\documentclass{article}
\usepackage{amsmath,geometry,physics,amssymb,mathtools,array,tabularx,amsfonts,cite,tcolorbox,subcaption,tocloft,mathrsfs,enumitem,textcomp,tikz,jheppub}

\newcommand{\zz}{\mathtt{z}}

\definecolor{DCviolet}{RGB}{140, 43, 226}

\begin{document}

\title{Holographic Spread Complexity at Fixed Charge: Routhians, Branes and Strings
}

\author[a]{Dimitrios Chatzis,} \author[a]{Madison Hammond,} \author[a]{Carlos Nunez,}\author[b]{Alfonso V. Ramallo }\author[a]{and  Ricardo T. Santamaria}
\affiliation[a]{Centre for Quantum Fields and Gravity, Department of Physics, Swansea University, Swansea SA2 8PP, United Kingdom}
\affiliation[b]{Departamento de F\'{\i}sica de Part\'{\i}culas, Universidade de Santiago de Compostela and Instituto Galego de
F\'{i}sica de Altas Enerx\'{\i}as (IGFAE). E-15782 Santiago de Compostela, Spain}
\emailAdd{dchatzis@proton.me}
\emailAdd{m.hammond.2412736@swansea.ac.uk}
\emailAdd{c.nunez@swansea.ac.uk}\emailAdd{alfonso.ramallo@usc.es}
\emailAdd{ricardo.sta2718@gmail.com}

\abstract
{Holographic spread (Krylov) complexity relates the growth of a boundary state's complexity to the proper radial momentum of a probe falling into the bulk.  Unitary evolution makes spread complexity an even function of time. We show that this requirement fails whenever a probe carries a conserved Noether charge and is described by its unreduced Lagrangian. The cure is simple and universal: passing to the Routhian of the fixed-charge sector restores the correct short-time behaviour of the complexity. We establish this prescription from first principles and test it across an extensive family of probes: charged particles, non-BPS D-branes with detuned tension and charge, branes excited along internal isometries, worldvolume gauge fields, a fluctuating D0-brane in AdS$_4\times \mathbb{CP}^3$
and fundamental strings combining winding with rotation in AdS$_5\times \mathrm{S}^5$
 complemented by further examples in AdS$_3\times \mathrm{S}^3\times T^4$,  ABJM, and the charged Anabal\'on-Ross background.  We then translate these results into Krylov-chain data, extracting Lanczos coefficients and Krylov-number correlators, and propose that complexity for charged, extended probes organises naturally into collective, fluctuation, charge and mixed contributions. This decomposition opens a concrete path toward a genuinely field-theoretic, multi-seed construction of holographic complexity.  }
\maketitle

\section{Introduction and General Idea of this Paper}\label{sec:introduction}
Understanding how the complexity of a quantum state or operator grows under time evolution has become a central theme connecting quantum information, many-body chaos and holography. Among the many diagnostics proposed to make ``complexity'' precise, those built from the Krylov subspace stand out for being essentially unique and free of the basis choice ambiguities that affect circuit complexity constructions. Given a Hamiltonian $H$ and a seed operator or state, the Lanczos algorithm generates an orthonormal Krylov basis $\{|K_n\rangle\}$ by repeated action of $H$ (or of the Liouvillian, for operator evolution) followed by Gram-Schmidt orthogonalisation. The dynamics reduces to hopping on a semi-infinite chain governed by the Lanczos coefficients $\{a_n,b_n\}$ \cite{Parker:2018yvk}. For operators, this defines Krylov complexity, whose asymptotic growth rate of $b_n$ bounds the Lyapunov exponent of chaotic theories  \cite{Parker:2018yvk}. For states, the analogous and widely used notion is spread complexity \cite{Balasubramanian:2022tpr}. Among all bases reached from a given seed state by allowed changes of basis, the Krylov basis is precisely the one that minimises the spread of the wavefunction at every instant of time. Hence,  spread complexity is a genuinely basis independent quantity, computable directly from the survival amplitude. Since the entire dynamics is packaged into a single sequence of real numbers $\{a_n,b_n\}$, together with a sharp characterisation of chaotic versus integrable dynamics \cite{Balasubramanian:2022tpr,Rabinovici:2023yex}, relations to circuit and geometric complexity \cite{Caputa:2021sib,Muck:2026top}, and extensions to quantum field theory \cite{Avdoshkin:2022xuw,Caputa:2024xkp}, have made Krylov and spread complexity among the most versatile complexity measures available today. See \cite{Nandy:2024evd,Rabinovici:2025otw,Baiguera:2025dkc} for recent reviews of the many facets of this programme.
\\
Let us elaborate the above in more detail: given a Hamiltonian $H$ and a normalised reference state $|K_0\rangle$, the Lanczos algorithm constructs the smallest subspace explored by the evolution $e^{-iHt}|K_0\rangle$.  In an orthonormal Krylov basis, the Hamiltonian is tridiagonal,
\begin{equation}\label{eq:intro_lanczos}
 H|K_n\rangle=b_{n+1}|K_{n+1}\rangle+a_n|K_n\rangle+b_n|K_{n-1}\rangle,
 \qquad b_0=0,
\end{equation}
and the time-evolved state can be written as $|\psi(t)\rangle=\sum_{n\geq 0}\phi_n(t)|K_n\rangle$.  Spread, or Krylov state, complexity is the mean position of this wave packet on the resulting semi-infinite chain,
\begin{equation}\label{eq:intro_spread_complexity}
 {\cal C}_{K}(t)=\sum_{n\geq0}n\,|\phi_n(t)|^2.
\end{equation}
It is therefore intrinsic to the pair $(H,|K_0\rangle)$, involves no externally chosen gate set, and packages the moments of the spectral measure (or equivalently, the derivatives of the return amplitude $\langle K_0|e^{-iHt}|K_0\rangle$) into the recursion data $\{a_n,b_n\}$ \cite{Balasubramanian:2022tpr,Caputa:2021sib,Nandy:2024evd,Rabinovici:2025otw}.  This construction has proved useful as a quantitative language for quantum information spreading, operator growth and chaos, while its geometric formulation interprets the Lanczos chain itself as the effective space in which the state moves \cite{Caputa:2021sib,Balasubramanian:2025xkj}. 

One elementary property is decisive in this work.  Since the Jacobi matrix in \eqref{eq:intro_lanczos} is real and Hermitian, $\phi_n(-t)=\phi_n(t)^*$ and hence ${\cal C}_K(t)={\cal C}_K(-t)$.  In particular,
\begin{equation}\label{eq:intro_evenness}
 {\cal C}_{K}(t)=b_1^2t^2+{\cal O}(t^4),
 \qquad
 \dot{\cal C}_{K}(t)=2b_1^2t+{\cal O}(t^3),
\end{equation}
so an analytic Krylov complexity has neither a linear term nor a non-zero initial velocity \cite{Huh:2023jxt, Fan:2022xaa}.

\subsection{Krylov spread complexity and holography}\label{sec:intro_holography}

The holographic route to \eqref{eq:intro_spread_complexity} grew out of the observation that the increase of boundary complexity is encoded in the momentum of an object falling into the bulk \cite{Susskind:2018tei,Susskind:2019ddc,Susskind:2020gnl,Barbon:2020uux}.  A direct bridge to Krylov dynamics was first exhibited in Jackiw--Teitelboim gravity, where the motion of the Krylov wave packet has an explicit bulk manifestation \cite{Rabinovici:2023yex}.  The sharp proposal of \cite{Caputa:2024sux} identifies the rate of spread complexity with the \emph{proper} radial momentum of a massive probe. The works \cite{Fan:2024iop,He:2024pox} clarified the dynamical assumptions and the range of the momentum--Krylov relation.  The construction was subsequently extended beyond the original $\mathfrak{sl}(2,\mathbb R)$ and JT setting in \cite{Heller:2024ldz}, while the planar holographic limit was studied directly in \cite{Das:2024tnw}.  In double-scaled SYK, operator Krylov complexity was identified with a bulk length in \cite{Ambrosini:2024sre}, the wormhole-velocity dictionary was developed in \cite{Fu:2025kkh}, and a near-horizon $\mathrm{AdS}_2$ interpretation of Krylov-subspace dynamics was given in \cite{Jeong:2026iac}.  The observer-based derivation of \cite{Li:2025observer} interprets spread complexity as a bulk-measured energy and its rate as radial momentum, whereas the quantum-probe analysis of \cite{Li:2026comments} isolates the coherent-state assumptions behind that classical relation and shows why a semiclassical limit by itself is insufficient.  The papers  \cite{Muck:2026top, Qu:2025lanczos} synthesise the orthogonal-polynomial, phase-space and holographic descriptions, the work \cite{Alfinito:2026vah} extends the dictionary from one-point spread complexity to exact multi-time Krylov correlators. Finally, \cite{Qu:2026dmv} studies properties of Lanczos coefficients and matrix models.

The first top-down applications already reveal that bulk geometry leaves information in the complete time dependence rather than only in a universal late-time growth law.  In $\mathcal N=4$ SYM, \cite{Fatemiabhari:2025cyy} relates motion in an $\mathrm{AdS}_3$ slice to an $SL(2)$ subsector and general motion to the full theory.  In the Anabal\'on-Ross soliton \cite{Anabalon:2021tua}, the paper \cite{Fatemiabhari:2025usn} obtains exact geodesics and finds oscillatory complexity generated jointly by the ultraviolet cutoff and the smooth confining end of space.  The comparison with the longitudinal field Ising chain in \cite{Jiang:2025wpj}, where oscillation frequencies track meson masses, provides an independent field-theory realisation of complexity and its relation with confinement and discrete spectra.  The systematic analysis of \cite{Fatemiabhari:2026goj} then establishes the same oscillatory pattern across several top-down confining backgrounds, with its frequency fixed by the confinement scale. A critical view on these results is presented in \cite{Nunez:2026vhw}, that established that the oscillations in the Krylov complexity are related to the presence of a discrete spectrum (not necessarily to the phenomenon of confinement). For conformal quiver theories, the work \cite{Fatemiabhari:2025poq} shows that motion along the quiver direction produces model-dependent early-time growth but is damped at late times, when the universal AdS behaviour is recovered.  The six-dimensional extension \cite{Fatemiabhari:2026six} adds both quiver motion and $SU(2)_R$ charge and finds the same separation between structured early dynamics and radial late-time growth.  Charged particles, composite objects and extended strings were treated in \cite{Nastase:2026lhz}. This  work showed that internal charge and spatial extent distinguish operators that would be indistinguishable to a purely radial point-particle prescription.  More recently, \cite{Nunez:2026rgflow} relates the local acceleration of spread complexity to holographic $c$-functions: it is inversely correlated with the covariant central function along fixed-dimensional flows but becomes co-monotonic for flows across dimensions.

These results have prompted a rapidly expanding set of tests and refinements.  The $\eta$-deformed backgrounds of \cite{Roychowdhury:2026eta} quantify how broken scale invariance alters the proper-momentum growth, while the Coulomb-branch analysis of \cite{Zoakos:2026coulomb} finds oscillations only for trajectories that avoid the interior singularity.  The orthogonal-polynomial correspondence of \cite{Qu:2025lanczos} identifies Lanczos and matrix-model recursion coefficients in the large-$N$ continuum limit.  Lin--Maldacena geometries were used in \cite{Roychowdhury:2026lin} to connect bulk probe momentum with BMN matrix-model operator growth and to initiate a direct calculation of its Lanczos data.  The reduced BMN model of \cite{Roychowdhury:2026bmnstate} computes state-complexity Lanczos coefficients at small and large deformation. The paper \cite{Roychowdhury:2026planewave} finds a universal linear dependence of both state and operator Lanczos coefficients on the plane-wave mass; and \cite{Roychowdhury:2026bmnspectral} supplements these results with moments, spectral densities, Krylov entropy and long-lived collective modes.  The boundary analysis of open strings attached to giant gravitons in \cite{Graef:2026probe} shows that protected few-body sectors have bounded Lanczos coefficients and therefore cannot by themselves test gravitational universality.  Likewise, \cite{Baume:2026chaos} studies orbifold SCFT spin chains and Krylov complexity.  On the gravity side, \cite{Fadafan:2026lifshitz} extends the proper-momentum prescription to Lifshitz and hyperscaling violating spacetimes, where the hyperscaling exponent controls the late time power law.  

Collectively, these works both support the momentum-spread correspondence and motivate the question addressed here.\\
{\underline{\bf Question}:\it  what is the correct momentum when the probe carries conserved internal quantum numbers or has genuine worldvolume structure?}
\\
This question was succinctly answered in the companion paper \cite{Chatzis:2026ekd}. The present work elaborates, makes clear, expands and shows different examples of the proposal in \cite{Chatzis:2026ekd}.
\subsection{General idea: fixed-charge dynamics and short-time evenness}\label{sec:general_idea}

The tension is already visible at the initial radial point (also at turning points).  The spread complexity of a unitary Krylov evolution obeys \eqref{eq:intro_evenness}; consequently, any bulk quantity identified with $\dot{\cal C}_K$ must vanish at $t=0$ and be odd at short times.  This requirement is automatic for a neutral particle released from rest, since its radial proper momentum vanishes.  It is not automatic for a probe described by a Lagrangian ${\cal L}(r,\dot r,\chi,\dot\chi)$ when $\chi$ is a cyclic internal coordinate and
\begin{equation}
 J=\frac{\partial {\cal L}}{\partial\dot\chi}
\end{equation}
is held fixed.  At the radial initial or turning point one has $\dot r(0)=0$. But $J\neq0$ generally requires $\dot\chi(0)\neq0$.  If the norm of the velocity in the \emph{unreduced} configuration space is used to define a generalised proper momentum, this internal motion leaves a charge-dependent constant in $\dot{\cal C}_K(t)$ and hence a linear term in the forward-time complexity.  The problem first became explicit for $R$-charged and extended probes in \cite{Nastase:2026lhz}; it is especially sharp when compared with the fixed-sector constructions of symmetry-resolved Krylov complexity \cite{Caputa:2025mii,Caputa:2025ozd}.

The resolution is to perform the fixed-charge reduction before defining the bulk complexity observable.  One solves the charge constraint for $\dot\chi$ and makes the partial Legendre transform
\begin{equation}\label{eq:intro_routhian}
 {\cal R}(r,\dot r;J)=J\dot\chi-{\cal L}(r,\dot r,\dot\chi)
 \big|_{\dot\chi=\dot\chi(r,\dot r;J)}.
\end{equation}
The Routhian governs the same radial trajectories in the sector of fixed $J$, but the cyclic motion now appears as a charge-dependent effective mass or potential.  The proper coordinate and its conjugate momentum must therefore be extracted from ${\cal R}$. In other words, from the reduced phase space appropriate to the boundary charge sector.  The resulting momentum is radial, vanishes at the initial and  turning points. It takes the form $P_y(t)=p_1t+{\cal O}(t^3)$; the holographic identification $\dot{\cal C}_K\propto P_y$ then gives ${\cal C}_K(t)\propto t^2+{\cal O}(t^4)$, as required.  This prescription applies to Noether charges, including angular momentum and electric displacement.  By contrast, a winding number or a fixed magnetic flux is a sector label rather than the momentum of a time-dependent cyclic variable; it modifies the effective tension or mass but requires no Routhian reduction (in other words, Noether charges require the Routhian treatment, but other conserved charges do not).  Section~\ref{sec:pointparticle} derives this mechanism in the simplest charged-particle example and fixes the prescription used throughout the paper.

The remainder of the paper is organised as follows.  Section~\ref{sec:pointparticle} isolates the fixed-charge issue for a particle moving in $\mathrm{AdS}_5\times \mathrm{S}^5$ and shows explicitly that the Routhian restores an odd proper momentum and an even short-time complexity.  Section~\ref{seccion3} develops the prescription for non-BPS branes: either detuning the DBI and Wess--Zumino couplings, exciting an internal isometry, or turning on magnetic or electric worldvolume fields cleanly separates effective parameters from genuine Noether charges.  Section~\ref{seccion4} studies a D0-brane in $\mathrm{AdS}_4\times\mathbb{CP}^3$, including its conserved internal momentum and transverse fluctuations.  Section~\ref{seccion5} treats fundamental strings in $\mathrm{AdS}_5\times \mathrm{S}^5$ in both Polyakov and Nambu--Goto formalisms and distinguishes winding from time-dependent rotation.  Section~\ref{seccion6} translates the bulk short-time expansions into combinations of Lanczos coefficients, discusses Krylov-number correlators and proposes a decomposition of complexity into collective, fluctuation, charge and mixed contributions.  Section~\ref{section7} summarizes the conclusions.  The appendices collect the worldvolume gauge-field equations, further F1, D1/D5, D5 and D3 examples in $\mathrm{AdS}_3\times \mathrm{S}^3\times T^4$ and $\mathrm{AdS}_5\times \mathrm{S}^5$, a D2-brane example in ABJM, and a charged-particle application to the Anabal\'on--Ross gauged-supergravity background and its eleven-dimensional interpretation.

\section{Motivation: Point particle}\label{sec:pointparticle}

After observing previously \cite{Nastase:2026lhz} that systems which contain conserved charges yield complexities whose derivative for early times is led by a charge dependent constant rather than the expected linear term \cite{Nastase:2026lhz}, we aim to resolve this issue and obtain the expected behaviour of the complexity \cite{Caputa:2024sux,Muck:2026top}, $\mathcal{C}\sim t^2$ via Legendre transforming to the Routhian.

Let us consider a motivating example. We consider a falling point particle in $\text{AdS}_5\times \mathrm{S}^5$ with angular momentum on the internal space, as is given in \cite{Nastase:2026lhz}.
The metric is written in the form, for $L$ the AdS radius
\begin{equation}\label{AdS5 sins}
    \begin{split}
        \dd s^2 &= \frac{r^2}{L^2}(-\dd t^2 + \dd x_1^2 + \dd x_2^2+ \dd x_3^2)+\frac{L^2 \dd r^2}{r^2}+L^2 \dd \Omega^2_5,\\
        \dd \Omega^2_5&= \dd \theta^2_1+\sin^2\theta_1 \dd \theta^2_2 +\sin^2\theta_1 \sin^2\theta_2 \dd \theta^2_3 + \sin^2 \theta_1 \sin^2 \theta _2 \sin^2 \theta_3 \dd\theta^2_4 + \sin^2 \theta _1 \sin^2 \theta_2\sin^2 \theta_3\sin^2\theta_4\dd\psi^2,
    \end{split}
\end{equation}
and the particle is embedded as
\begin{equation}
    r=r(t),  \quad \psi = \psi (t)  , \quad \theta_i = \frac{\pi}{2},
\end{equation}
with all other coordinates set to be constant. The Lagrangian reads
\begin{equation}
    {\cal L}_p = -m \sqrt{\frac{r^2}{L^2}-\frac{L^2 \dot{r}^2 }{r^2}-L^2 \dot{\psi}^2},
\end{equation}
with two conserved quantities
\begin{equation}
    {\cal J} = \frac{\partial {\cal L}_p}{\partial \dot{\psi}}\quad \& \quad {\cal H}=\frac{\partial {\cal L}_p}{\partial \dot{r}}\dot{r}+{\cal J}\dot{\psi} - {\cal L}_p,
\end{equation}
in terms of which we can solve for the profiles $r(t)$ and $\psi(t)$
\begin{equation}\label{r_and_psi_profiles_example}
    \begin{split}
        r(t) =  \frac{{\cal H}L^2}{\sqrt{{\cal H}^2 t^2 + {\cal J}^2+L^2 m^2}}\quad \& \quad \psi(t) = \frac{{\cal J}}{\sqrt{{\cal J}^2 + m^2 L^2}}\text{arctan}\left( \frac{{\cal H}t}{\sqrt{{\cal J}^2+m^2 L^2}}\right).
    \end{split}
\end{equation}

To describe the system in a sector of fixed angular momentum $\mathcal{J}$, we perform a  Legendre transform with respect to the cyclic coordinate $\psi$. The resulting quantity, the {\it Routhian}, provides the appropriate effective action for the dynamics at fixed conserved charge. In this procedure, $\dot{\psi}$ is eliminated in favour of the conserved quantity $\mathcal{J}$. Explicitly, in this case one substitutes 
\begin{equation}
    \dot{\psi}=\frac{{\cal J}\sqrt{r^4-L^4 \dot{r}^2}}{L^2 r \sqrt{L^2 m^2 + {\cal J}^2}},
\end{equation}
such that the Routhian does not depend on the coordinate $\dot{\psi}$,
\begin{equation}
    {\cal L}(r,\dot{r},\dot{\psi})\mapsto {\cal R}(r,\dot{r},{\cal J}) = {\cal J}\dot{\psi}(r,\dot{r},{\cal J}) - {\cal L}(r,\dot{r},{\cal J}).
\end{equation}

Indeed, we see that
\begin{equation}\label{eq:Ruthian_point_particle}
    {\cal R} _p= \frac{\sqrt{{\cal J}^2 + m^2 L^2 }}{L}\sqrt{ \frac{r^2}{L^2}-\frac{L^2 \dot{r}^2}{r^2}} = \sqrt{\hat{m}^2\left(\frac{r^2}{L^2}-\frac{L^2 \dot{r}^2}{r^2}\right) },\quad \hat{m}^2 = \frac{{\cal J}^2 + m^2 L^2 }{L^2}.
\end{equation}

This has a corresponding Hamiltonian given by

\begin{equation}\label{Ham_ruth}
     H=\frac{\partial {\cal R}_p}{\partial \dot{r}} \dot{r}-{\cal R}_p = - \frac{\sqrt{L^2 m^{2} +{\cal J}^2}r^3}{L^2 \sqrt{r^4 -L^4 \dot{r}^2}},
\end{equation}

by defining the energy to be $\mathcal{H} =- H$, we obtain
\begin{equation}
    \dot{r} = - \frac{r^2 \sqrt{ L^4 {\cal H}^2-(L^2 m^2 +{\cal J}^2)r^2}}{L^4{\cal H}}= -\frac{r^{2}}{L^{2}} \sqrt{1-\left(\frac{r}{r_{\text{UV}}}\right)^{2}},\quad r_{\text{UV}}=\frac{{\cal H} L^2}{\sqrt{{\cal J}^2 + L^2 m^2}} = \frac{L \mathcal{H}}{\hat{m}}.
\end{equation}
We can now proceed in calculating the proper momentum using the above instead of the Lagrangian. To do this we write\footnote{Notice that we chose $\dot{y} \sim \dot{r} \leq 0$. Furthermore, the Routhian unlike the Lagrangian does not have an overall minus sign as a prefactor of the square root, this is why one should define $P_{y} = -\frac{\partial \mathcal{R}}{\partial \dot{y}} \sim \dot{r}$, implying that $P_{y}\leq 0$. This agrees with the fact that as the particle falls the complexity grows, encapsulated in the relation $\dot{C}(t)\sim -P_y(t)$. \label{sign convention} }
\begin{equation}
    {\cal R}_p = \sqrt{\hat{m}^2\frac{r^2}{L^2}-\dot{y}^2},\quad \dot{y}=\hat{m}\frac{L}{r}\dot{r},
\end{equation}
which gives 
\begin{equation}
    P_y \equiv - \frac{\partial {\cal R}_p}{\partial \dot{y}}=\frac{\dot{y}}{\sqrt{\hat{m}^2\frac{r^2}{L^2}-\dot{y}^2}}=\frac{\dot{r}}{\sqrt{\frac{r^4}{L^{4}} - \dot{r}^2}}.
\end{equation}
Substituting the solution for $r(t)$ from \eqref{r_and_psi_profiles_example} in the above we get the exact result
\begin{equation}
   \dot{\cal C}\sim - P_y = \frac{{\cal H}}{L\, \hat{m}}t,
\end{equation}
which is the expected early time behaviour of the derivative of the complexity \cite{Caputa:2024sux,Muck:2026top}.
Note that the correct relation is $\dot{\cal C}\sim- P_y~$. Hence, when the particle falls $(\dot{y}<0)$  the complexity grows. Conversely, for situation in which the particle climbs-up the radial direction, the complexity decreases. See the discussion in \cite{Nunez:2026vhw}.
We also emphasise the fact that the information of the conserved momentum ${\cal J}$ is implicit through \eqref{Ham_ruth}. In the following sections, we  utilise the Routhian prescription to investigate more elaborate configurations possessing conserved charges. In such cases, the appropriate quantity to evaluate is the Routhian, rather than the Lagrangian, in order to recover the expected time dependence of the complexity.

\section{Complexity for non-BPS branes}\label{seccion3}

A BPS brane experiences no net force, as we have a cancellation between the gravitational attraction and the Ramond repulsion. So in order for the brane to fall, we must consider non-BPS branes. There are multiple ways in which we can do this, besides choosing a brane that is automatically non-BPS as will be seen in the case of the D5 brane falling in $\text{AdS}_5 \times \mathrm{S}^5$, as discussed in section \ref{sec: D5 AdS5xS5}. The ways in which we consider making branes non-BPS is via adding a gauge field, giving the brane motion in one of the internal directions of the geometry or preventing the mass and charge of the brane from being equal. 

\subsection{Detuning mass from charge}\label{section:detuning_mass_from_charge}

\begin{figure}
    \centering
    \includegraphics[width=0.45\linewidth]{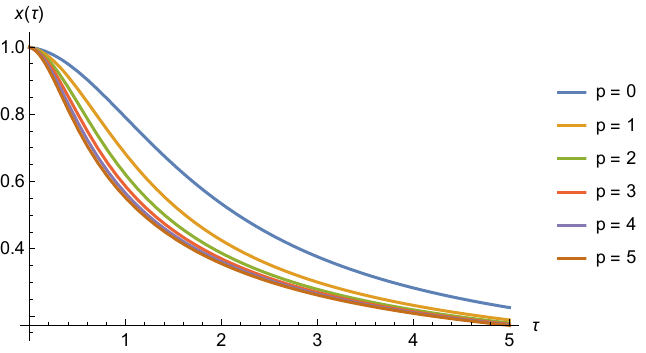}
    \includegraphics[width=0.45\linewidth]{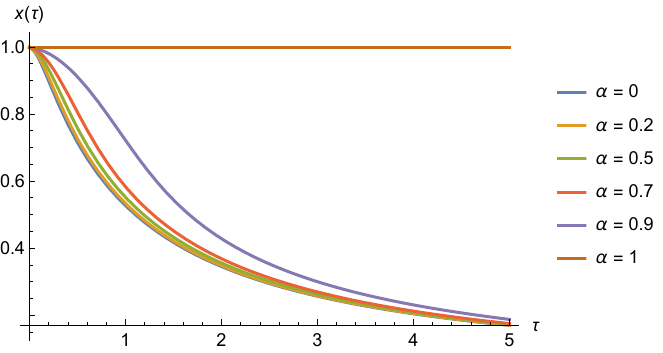}
    \caption{[Left] Trajectory for D$p$ branes for $\alpha=0.5$ and different values of $p$. [Right] Trajectory for a D5 brane with different values of $\alpha$, at $\alpha=1$ the brane is BPS and does not fall.}
    \label{fig:Dp alpha=0.5}
\end{figure}

Consider a probe D$p$-brane whose worldvolume is localised in a generic AdS$_d$ spacetime, with no extension along the internal manifold, where $d \geq p+2$. Its embedding is denoted by D$p[t,x_1,x_2,\ldots,x_p]$, and we allow for a non-trivial radial profile $r=r(t)$, with $(t,x_1,x_2,\ldots,x_p,r)$ the AdS coordinates. In this subsection, we refrain from specifying the underlying string theory background, so that the analysis remains applicable to a broad class of geometries. The brane is further coupled to a $(p+1)$-form potential, $C_{p+1}$, on its worldvolume, given by
\begin{eqnarray}\label{Cp+1}
    C_{p+1} = \left(\frac{r}{L}\right)^{p+1} \, \mathrm{d} t\wedge  \mathrm{d}x_{1} \wedge ... \wedge  \mathrm{d}x_{p} \, .
\end{eqnarray}
We write the action and Lagrangian from the induced metric, where we have contributions from the DBI part and WZ part, 
\begin{align}
    & \mathrm{d}s^2_{\text{ind}}=\left(\frac{L^2}{r^2}\dot{r}^2-\frac{r^2}{L^2}\right) \mathrm{d}t^2 + \frac{r^2}{L^2}\mathrm{d}x_1^2 + \ldots + \frac{r^2}{L^2}\mathrm{d}x_p^2  \, ,\\
    S &= -T_{\rm{D}p} \int \mathrm{d}^{p+1} \vec{x}\;  \sqrt{- \text{det}g_{\text{ind}}} \; +\; \mu_p \int \mathrm{d}^{p+1} \vec{x}\;P[C_{p+1}] \\
    &= T_{\rm{D}p}\, L_{x_1} \ldots L_{x_p} \Big(  -\int \mathrm{d}t  \; \frac{r^{p}}{L^{p}}\sqrt{\frac{r^{2}}{L^{2}}-\frac{L^{2}\dot{r}^{2}}{r^{2}}}+ \frac{\mu_{p}}{T_{\rm{D}p}} \int \mathrm{d}t \; \frac{r^{p+1}}{L^{p+1}} \Big) \, , \,
\end{align}
where we define $\nu = T_{\rm{D}p}  \int \dd x_{1} \ldots \int \dd x_{p}=T_{\rm{D}p}\,  L_{x_1} \ldots L_{x_p}$ and $\mu_{p}$ is the charge of the brane under the $C_{p+1}$ form. 
The quantities have the units $[T_{\rm{D}p}] =[\mu_p] = \text{length}^{-p-1}, \;[\nu] = \text{length}^{-1}$. In this way, the Lagrangian is defined to be 
\begin{equation}\label{lagrangian alpha case}
    \mathcal{L} =  \nu \left( -\frac{r^{p}}{L^{p}}\sqrt{\frac{r^{2}}{L^{2}}-\frac{L^{2}\dot{r}^{2}}{r^{2}}}+\alpha \frac{r^{p+1}}{L^{p+1}}  \right) \, ,
\end{equation}
where $\alpha = \frac{\mu_{p}}{T_{\text{D}p}}$. Then, the Hamiltonian is 
\begin{align}\label{hamilton}
    & \mathcal{H} = P_{r}\dot{r}-\mathcal{L} = \nu\frac{r^{p+1}}{L^{p+1}}\left( \frac{r}{L \tilde{\Delta}} -\alpha    \right) \, ,\\
    &\tilde{\Delta} = \sqrt{\frac{r^{2}}{L^{2}}-\frac{L^{2}\dot{r}^{2}}{r^{2}}}.
\end{align}
Since the Hamiltonian is conserved, it is fixed by the initial position and velocity of the brane, $r(t = 0) = r_{\text{UV}}$ and $\dot{r}(t= 0) = 0$, as the brane falls freely. This implies that
\begin{equation}
    \mathcal{H} =  \nu\frac{r_{\text{UV}}^{p+1}}{L^{p+1}} (1-\alpha) \, ,
\end{equation}
hence $\alpha$ \textit{parametrises} the BPS condition for the brane, meaning that for mass equals charge ($\alpha = 1$), the brane will not fall, as the gravitational attraction and the charge repulsion mutually cancel each other. As a consequence, the mass has to be greater than the charge, meaning $\alpha < 1$. \\
From \eqref{hamilton} we get the equation
\begin{equation}
    \dot{r} = - \frac{r^{2}}{L^{2}}\frac{\sqrt{L^{p+1}\mathcal{H}(L^{p+1}\mathcal{H}+2\alpha\nu r^{p+1})+(\alpha^{2}-1)\nu^{2}r^{2p+2}}}{L^{p+1}\mathcal{H}+\alpha\nu r^{p+1}} \, .
    \label{rdotalpha}
\end{equation}
If $\alpha = 1$, then the argument of the square root is always positive, meaning that the initial condition of $\dot{r}(t= 0) = 0$ is only possible when $\mathcal{H}= 0$, returning to the BPS case. Then, we can explicitly see how by detuning the mass and the charge, the brane will fall with zero initial velocity. \\
We now aim to solve the equation \eqref{rdotalpha}, define the dimensionless variables
\begin{equation}
    x(t) = \frac{r(t)}{r_{\text{UV}}}\, ,  \quad\quad  \tau = \frac{r_{\text{UV}}\, t}{L^{2}} \, ,
    \label{xandtau}
\end{equation}

such that $0<x\leq 1$, where $x \sim 0$ defines the IR and  $x \sim 1$ the UV. Then, equation \eqref{rdotalpha} is rewritten as
\begin{equation}
    \frac{\dd x}{\dd \tau}=- x^2\frac{ \sqrt{(\alpha -1) \left(x^{p+1}-1\right) \left[(1-\alpha )+(\alpha +1) x^{p+1}\right]}}{(1-\alpha )+\alpha  x^{p+1}} \, ,
    \label{xdotalpha}
\end{equation}
this equation \eqref{xdotalpha} was also studied numerically, and the solutions can be found in Figure \ref{fig:Dp alpha=0.5} for different  choices of $p$ and $\alpha$. Alternatively, one can try to solve the equation \eqref{xdotalpha} analytically, equivalent to 
\begin{equation}
    \tau-\tau_{0} = \int_{x}^{1} \dd z \,  \frac{1}{z^{2}} \frac{(1-\alpha )+\alpha  z^{p+1}}{ \sqrt{(\alpha -1) \left(z^{p+1}-1\right) \left[(1-\alpha )+(\alpha +1) z^{p+1}\right]}} \, ,
\end{equation}
whose solution is given in terms of Appell functions $F_{1} = \text{Appell}F_1$,
\begin{equation}
    \begin{split}
         \tau-\tau_{0} &= \frac{\alpha \, x^{p} }{(1-\alpha) p} \; F_1\left(\frac{p}{p+1}, \frac{1}{2}, \frac{1}{2} ; \frac{p}{p+1}+1; x^{p+1}, \frac{(\alpha +1) x^{p+1}}{\alpha -1}\right) \\
         &-\frac{1}{x} F_1\left(-\frac{1}{p+1}, \frac{1}{2},\frac{1}{2}; \frac{p}{p+1} ; x^{p+1}, \frac{(\alpha +1) x^{p+1}}{\alpha -1}\right) \, ,
    \end{split}
    \label{solalpha}
\end{equation}
The initial condition $x(\tau = 0) = 1$ sets the integration constant to be
\begin{equation*}
     \tau_{0} = \frac{\alpha  }{(1-\alpha ) p}F_1\left(\frac{p}{p+1}, \frac{1}{2}, \frac{1}{2} ; \frac{p}{p+1}+1 ; 1 , \frac{\alpha +1}{\alpha -1}\right) \\
         - F_1\left(-\frac{1}{p+1}, \frac{1}{2},\frac{1}{2}; \frac{p}{p+1} ; 1, \frac{\alpha +1 }{\alpha -1}\right) \, .
\end{equation*}
To study the asymptotic behaviour, one can expand the solution \eqref{solalpha} and then substitute the dimensionful variables, or equivalently propose a series expansion for \eqref{rdotalpha}. Expanding at $t\to 0$ for arbitrary $p$, so that $r \sim r_{\text{UV}}$ we obtain the following early time expansion,
\begin{equation} \label{ralphaexp}
    r(t) \approx r_{\text{UV}}- \frac{(p+1)(1-\alpha)}{2L^{4}}r_{\text{UV}}^{3}t^{2}-\frac{(p+1)^{2}(1-\alpha)^{2}(3\alpha(p+1)-2p-9)}{24 L^{8}}r_{\text{UV}}^{5}t^{4} +\mathcal{O}(t^{6}) \, .
\end{equation}
This expansion still reflects the non-BPS condition for the brane. In the next subsection, we use equation \eqref{ralphaexp} to obtain the early time behaviour for the complexity. 

\subsubsection{Holographic spread complexity}
In order to compute the complexity, we follow the prescription in \cite{Fatemiabhari:2025poq, Fatemiabhari:2025usn, Fatemiabhari:2026goj}, where the proper coordinate $y$ is such that $\dot{y} = \frac{L\dot{r}}{r}$. Then, the proper momentum is\footnote{Notice that for this case $P_{y} = \frac{\partial \mathcal{L}}{\partial \dot{y}}$, implying $P_{y} \leq 0$. This agrees with the fact that as the particle falls the complexity grows, encapsulated in the relation $\dot{C}(t)\sim -P_y(t)$.}
\begin{equation}
     P_{y} = \frac{\partial \mathcal{L}}{\partial \dot{y}}=  \frac{\partial \mathcal{L}}{\partial \dot{r}} \frac{\partial \dot{r}}{\partial \dot{y}}  = P_{r}\frac{\dot{r}}{\dot{y}} = P_{r}\frac{r}{L} \, ,
\end{equation}
which after substituting \eqref{rdotalpha} we get the expression
\begin{equation}
    P_{y} =- \frac{1}{L^{p}r}\sqrt{\mathcal{H}L^{p+1}(\mathcal{H}L^{p+1}+2\alpha\nu r^{p+1})+(\alpha^{2}-1)\nu^{2} r^{2p+2}} \, .
    \label{properalpha}
\end{equation}
Additionally, one can write \eqref{properalpha} in terms of a dimensionless variable $x(\tau)$ and parameter $\alpha$ by defining the rescaling $\tilde{P}_{y} = \left(\frac{
L}{r_{\text{UV}}}\right)^{p}\frac{P_{y}}{\nu}$ and $\tau = \frac{r_{\text{UV}}\, t}{L^{2}}$, where now $\tilde{P}_{y}$ is dimensionless

\begin{equation}
    \tilde{P}_{y} = -\frac{1}{x(\tau)}\sqrt{(1-\alpha)(1-x(\tau)^{p+1})\left[1-\alpha+(\alpha+1)x(\tau)^{p+1}   \right]} \, .
    \label{properalphadimless}
\end{equation}
Notice that for a BPS brane, meaning $\alpha = 1$, the proper momentum exactly vanishes, in agreement with the fact that the brane does not fall. 

\begin{figure}
    \centering
    \includegraphics[width=0.45\linewidth]{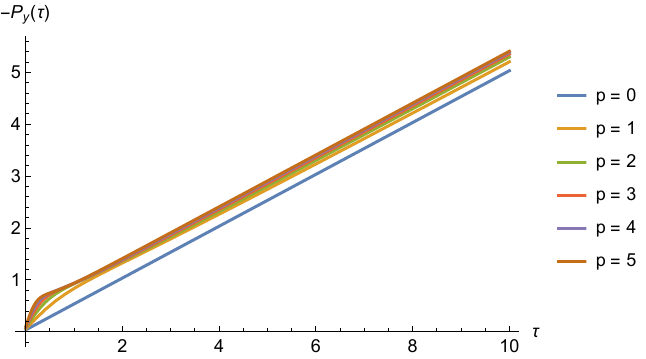}
    \includegraphics[width=0.45\linewidth]{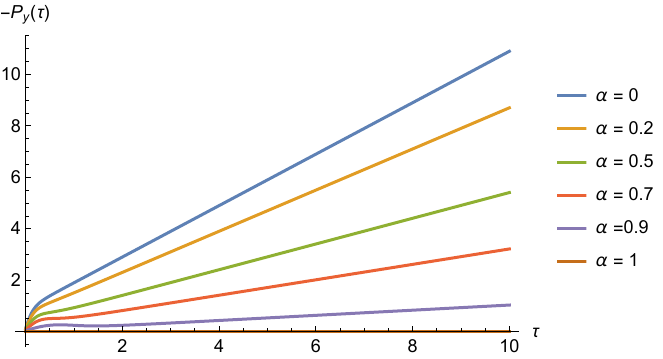}
    \caption{[Left]$P_y$ against $\tau$ for different values of $\alpha$ with $p=5$. [Right] $P_y$ against $\tau$ for $\alpha=0.5$ and different values of $p$. }
    \label{fig:Py alpha}
\end{figure}

We can analyse the behaviour of the derivative of the complexity by substituting \eqref{ralphaexp} in \eqref{properalpha}. For arbitrary $p$ at short times
\begin{equation}
    \dot{\mathcal{C}}(t) \propto P_{y} \approx \frac{(1-\alpha)(p+1)\nu}{L^{p+2}}r_{\text{UV}}^{p+1} t -\frac{1}{3}\frac{p(p+1)^{2}(1-\alpha)^{2}\nu}{L^{p+6}}r_{\text{UV}}^{p+3}t^{3} + \mathcal{O}(t^{5}) \quad \text{for} \quad t \to 0.\label{complexity-non-bps}
\end{equation}
We can also do the same procedure for late times and arbitrary $p$, getting
\begin{equation}
    P_{y} \approx \frac{(1-\alpha)\nu}{L^{p+2}}r_{\text{UV}}^{p+1}t+\alpha\nu \frac{L^{p}}{t^{p}} + \mathcal{O}\left(\frac{1}{t^{p+1}}\right) \quad \text{for} \quad  t \to \infty  \, .
    \label{pyalphalate}
\end{equation}
This is in agreement with the linear growth and implies that $\mathcal{C} \sim t^{2}$ matching the complexity being an even function of time. Additionally, Figure \ref{fig:Py alpha} shows numerical plots of $P_{y}(\tau)$ for different values of $p$ and $\alpha$, matching the expansions in \eqref{complexity-non-bps} and \eqref{pyalphalate}.

In Appendix \ref{sec: AdS3 x S3 x T4 examples} we study an F1[$t, x$] string in AdS$_{3} \times \mathrm{S}^{3} \times T^{4}$ corresponding to $p = 1$ and $ \alpha = 0$. Moreover, in Appendix \ref{sec: branes AdS5xS5} we analysed a D5[$t, x_{1}, x_{2}, \mathrm{S}^{3}$] brane wrapping the $\mathrm{S}^{3}\subset \mathrm{S}^{5}$ which is the case $p = 2$ and $\alpha = 0$. Finally, in Appendix \ref{sec: D2 ABJM} we study a D2[$t, x_{1}, x_{2}$] brane in ABJM which corresponds to the case $p = 2$ and $\alpha = \mu_{2}$.

In the next subsection, we study the complexity of branes with motion along an isometry direction of the internal space. 


\subsection{Adding an excitation within the internal space}\label{sec: internal space}
The probe branes considered in this subsection have mass equal to their charge, so $\alpha=1$. The way in which we make the branes non-BPS is by giving them an additional profile in one of the internal coordinates. We consider a probe $\text{D}p$ brane falling in $\text{AdS}_{d+1}\times \mathrm{S}^{9-d}$ supporting an RR flux $F_{p+2}$. We place the probe extended along D$p[t, x_{1}, x_{2}, \ldots, x_{p}]$,  with a profile in $r(t)$ and now also $\psi(t)$, where $\psi$ is  the equatorial angle of the sphere and is an isometry of the internal space $\mathrm{S}^{9-d}$.
See Appendix \ref{sec:Excited D3} for an explicit example in which we consider a D3 brane extended in $[t, x_{1}, x_{2}, x_{3}]$ in AdS$_{5} \times \mathrm{S}^{5}$, with $r = r(t)$ and $\psi = \psi(t)$, this corresponds to the case of $p=3$ below.

The metric is given by
\begin{eqnarray}
    \mathrm{d}s^2 = -\frac{r^2}{L^2} \mathrm{d}t^2 + \frac{r^2}{L^2}\mathrm{d}x_1^2 + \ldots + \frac{r^2}{L^2}\mathrm{d}x_{d}^2 + \frac{L^2}{r^2}\mathrm{d}r^2 + \lambda^2 \mathrm{d}\psi^2 + \ldots \, .
\end{eqnarray}
The DBI and WZ terms of the action of a brane with a profile in both $r(t)$ and $\psi(t)$ is given by,
\begin{equation}\label{L_brane_angularmomentum_generic}
    S = \nu \int \mathrm{d}t  \; \left(-\frac{r^{p}}{L^{p}}\sqrt{\frac{r^{2}}{L^{2}}-\frac{L^{2}\dot{r}^{2}}{r^{2}}-\lambda^{2} \dot{\psi}^{2}}+ \frac{r^{p+1}}{L^{p+1}}\right) \, ,
\end{equation}
where the parameter $\lambda$ is fixed by the geometry of the internal space. The units are $[r]=[L]=[\lambda]=\text{length}$, $[\nu]=\text{length}^{-1}$, $[\psi]=1$. In this case, $T_{\rm{D}p} = \mu_{p}$, meaning the tension is equal to the charge. However, the excitation adds mass to the brane, making the brane non-BPS. 
Define the Lagrangian to be
\begin{equation}
    \mathcal{L}= \nu \left(-\frac{r^{p}}{L^{p}}\sqrt{\frac{r^{2}}{L^{2}}-\frac{L^{2}\dot{r}^{2}}{r^{2}}-\lambda^{2} \dot{\psi}^{2}}+ \frac{r^{p+1}}{L^{p+1}}\right).
    \label{lagrangianJ}
\end{equation}

The Hamiltonian is then
\begin{align}
    & \mathcal{H} = P_{r}\dot{r}+P_{\psi}\dot{\psi}-\mathcal{L} = \nu\frac{r^{p+1}}{L^{p+1}}\left(\frac{r}{L\Delta } -1 \right) \, ,\\
    &\Delta = \sqrt{\frac{r^{2}}{L^{2}}-\frac{L^{2}\dot{r}^{2}}{r^{2}}-\lambda^{2} \dot{\psi}^{2}} \, .
\end{align}
The initial conditions for the probe are $\dot{\psi}(t=0)=\dot{\psi}_0$, $\; r(t=0)=r_{\text{UV}}$ and $\dot{r}(t = 0) = 0$. These fix the Hamiltonian to be
\begin{equation}
    \mathcal{H} = \nu\frac{r_{\text{UV}}^{p+1}}{L^{p+1}}\left(\frac{r_{\text{UV}}}{L\sqrt{\frac{r_{\text{UV}}^{2}}{L^{2}}-\lambda^{2} \dot{\psi}_{0}^{2}}} -1 \right) \, ,
\end{equation}
which is non-zero provided that $\dot{\psi} \neq 0$, such that the brane is non-BPS. \\
Due to the motion in $\psi$, there is a conserved quantity associated to $\psi$ being cyclic, we define this conserved quantity to be $P_{\psi} \equiv  {\cal J}$, this implies the equation for $\dot{\psi}$ to be
\begin{equation}
    {\cal J}= \frac{\partial {\cal L}}{\partial \dot{\psi}} , \quad \dot{\psi }(r, \dot{r}) =  \frac{L^{p-1}{\cal J}}{\lambda r}\sqrt{\frac{r^4 - L^4 \dot{r}^2}{L^{2p}{\cal J}^2 + \lambda^2 \nu^2 r^{2p}}},
    \label{psidotr}
\end{equation}
similarly, one can get the equation of motion for $\dot{r}$ from the Hamiltonian  
\begin{equation}
    \dot{r} =- \frac{r^{2}}{L^{2}}\frac{\sqrt{L^{p+1}\mathcal{H}(L^{p+1}\mathcal{H}+2\nu r^{p+1})-\lambda^{-2}L^{2p}{\cal J}^{2}r^{2}}}{L^{p+1}\mathcal{H}+\nu r^{p+1}} \, .
    \label{rdotj}
\end{equation}
After substituting $\dot{r}$ in \eqref{psidotr} we get
\begin{equation}
    \dot{\psi} = \frac{L^{p-1} {\cal J} r^{2} }{\lambda^{2}(L^{p+1}\mathcal{H}+\nu r^{p+1})} \, .
\end{equation}

 Notice that the role of the excitation in equation \eqref{rdotj} is to provide the argument of the square root with a real solution to the condition $\dot{r}(t = 0) = 0$, allowing the brane to fall freely. Else, if we set ${\cal J }= 0$ any $r_{\text{UV}}$ acts as a good initial condition and only $\mathcal{H} = 0$ is consistent with $\dot{r}(t = 0) = 0$ and we return to the BPS case. Let us analyse the relation between the energy $\mathcal{H}$ and the angular momentum $\mathcal{J}$, such that the brane falls with zero initial radial velocity. \\
Consider the square of equation \eqref{rdotj}
\begin{equation}
    \dot{r}^{2} = \underbrace{\frac{r^{4}}{L^{4}}\frac{1}{(L^{p+1}\mathcal{H}+\nu r^{p+1})^{2}}}_{A(r)} \, \underbrace{(L^{p+1}\mathcal{H}(L^{p+1}\mathcal{H}+2\nu r^{p+1})-\lambda^{-2}L^{2p}{\cal J}^{2}r^{2})}_{F(r)} \, .\label{julian}
\end{equation}
At $t = 0$:  $r(t = 0) = r_{\text{UV}}$ and $\dot{r}(t = 0) = 0$. Since $A(r)>0$, then it must hold that $F(r_{\text{UV}}) = 0$ which implies the relation
\begin{equation}
    (\mathcal{H}L^{p+1})^{2}+2\nu\mathcal{H}L^{p+1}r_{\text{UV}}^{p+1} = \frac{{\cal J}^{2}L^{2p}}{\lambda^{2}}r_{\text{UV}}^{2} \, .
    \label{Fcond}
\end{equation}
On the other hand, if we expand equation \eqref{julian} near $r_{\text{UV}}$, we get
\begin{equation}
    \dot{r}^{2} = A(r_{\text{UV}})F'(r_{\text{UV}})(r-r_{\text{UV}})+\mathcal{O}((r-r_{\text{UV}})^{2}) \, ,
\end{equation}
where we used the fact that $F(r_{\text{UV}}) = 0$. Notice that, since the probe is falling, then $r(t)\leq r_{\text{UV}}$, and given that $A(r)$ is positive, then $F'(r_{\text{UV}})< 0$, this condition is written by using equation \eqref{Fcond} as
\begin{equation}
    \mathcal{H} > \nu(p-1)\left(\frac{r_{\text{UV}}}{L}  \right)^{p+1} \, .
\end{equation}
Notice that for $p= 0,1$ this is always true, as we define $\mathcal{H}$ to be always positive. Moreover, the latter expression gives a lower bound to the energy for which the probe falls. Using equation \eqref{Fcond} we can rewrite the inequality in terms of the angular momentum as 
\begin{equation}
    \frac{{\cal J}^{2}}{\lambda^{2}\mathcal{H}^{2}} < \left(\frac{L}{r_{\text{UV}}}\right)^{2}\frac{p+1}{p-1}
    \label{jcond}
\end{equation}
this means that for $p>1$ the choice of ${\cal J}$ has to satisfy \eqref{jcond}, otherwise the probe will not fall. 

\begin{figure}[h!]
    \centering
    \includegraphics[width=0.45\linewidth]{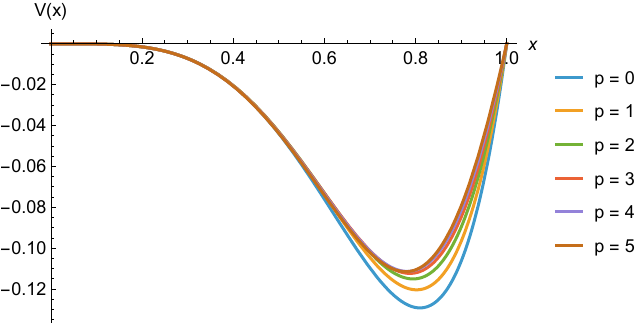}
    \quad \quad
        \includegraphics[width=0.45\linewidth]{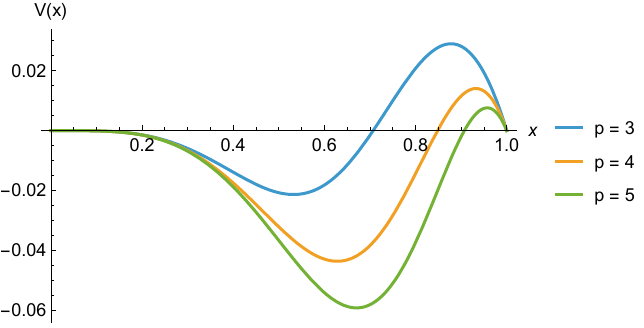}
    \caption{Plots of the effective potential $V_{\text{eff}}(x)$ where we chose $p+5 = \gamma > p-1$ (left) and $p-2=\gamma < p-1$ (right).}
    \label{Veff}
\end{figure}

\begin{figure}[h!]
    \centering
    \includegraphics[width=0.45\linewidth]{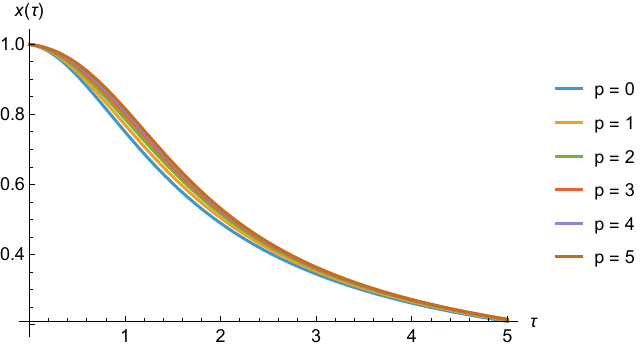}
     \includegraphics[width=0.45\linewidth]{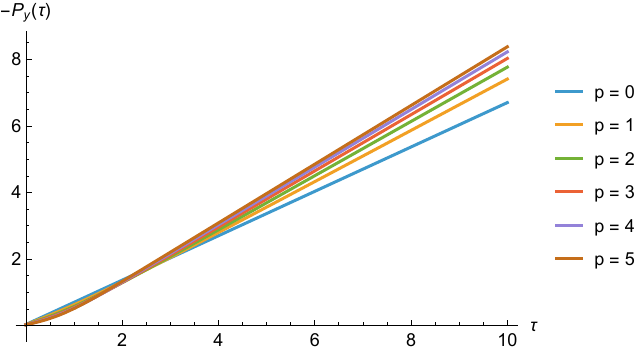}
    \caption{[Left] Trajectories for $x(\tau)$. [Right] Proper momentum $P_{y}(\tau)$ for different values of $p$.}
    \label{xtpyJ}
\end{figure}

We now aim to solve the equation \eqref{rdotj} and we do this via introducing the dimensionless variable defined in \eqref{xandtau} and the dimensionless parameters
\begin{equation}
   \gamma = \frac{\mathcal{H}}{\nu}\left(\frac{L}{r_{\text{UV}}}\right)^{p+1} \, , \quad \beta = \left(\frac{{\cal J}}{\lambda \mathcal{H}} \frac{r_{\text{UV}}}{L}  \right)^{2}
    \label{dimensionlessJ}
\end{equation}
such that $[\gamma] = [\beta]=  1$. In this way, by using the above definitions, equation \eqref{rdotj} reads
\begin{equation}
    \frac{\dd x}{\dd \tau} = -x^{2}\frac{\sqrt{\gamma^{2}+2\gamma x^{p+1}-\gamma^{2}\beta x^{2}}}{\gamma + x^{p+1}} \, .
    \label{xdotgamma}
\end{equation}
In this language, conditions \eqref{Fcond} and \eqref{jcond} are 
\begin{equation}
    \beta = \frac{\gamma + 2}{\gamma} \quad \text{and} \quad \gamma > p-1 \, ,
    \label{condgamma}
\end{equation}
respectively. Hence, equation \eqref{xdotgamma} is 
\begin{equation}
    \frac{\dd x}{\dd \tau} = - \frac{ x^{2}}{x^{p+1}+\gamma}\sqrt{\gamma^{2}-\gamma(\gamma+2)x^{2}+2\gamma x^{p+1}} \, .
    \label{xdotj}
\end{equation}
Equivalent to  
\begin{equation}
     \left(\frac{\dd x}{\dd \tau}\right)^{2}+  \frac{x^{4}}{(x^{p+1}+\gamma)^{2}}(\gamma(\gamma+2)x^{2}-\gamma^{2}-2\gamma x^{p+1}) = 0 \, ,
\end{equation}
which describes a particle with zero energy moving in an effective potential 
\begin{equation}
    V_{\text{eff}}(x) =  \frac{x^{4}}{(x^{p+1}+\gamma)^{2}}(\gamma(\gamma+2)x^{2}-\gamma^{2}-2\gamma x^{p+1})  \, .
    \label{veffgamma}
\end{equation}
In Figure \ref{Veff}, we can see that if $\gamma$ satisfies the conditions in \eqref{condgamma}, then the potential is such that the probe will fall from $x(0) =1$ ($r(0) = r_{\text{UV}}$). On the other hand, if those conditions are not satisfied, the probe will not move, as it will stay in an equilibrium position.\footnote{Another way of seeing this is to notice that  eq.(\ref{xdotj}) is $\frac{\dd x}{\dd \tau}=- F[x] \sqrt{-V_{\text{eff}}(x)}$, hence if $V_{\text{eff}}(x)\geq 0$ there cannot be motion. Note that $V_{\text{eff}}=0$ occurs for $x=0$ and $x = 1$, see eq. \eqref{veffgamma}.} Numerical solutions to equation \eqref{xdotj} for different values of $p$ were also studied, shown in Figure \ref{xtpyJ}. \\

One can also study the asymptotic behaviour of $x(\tau)$ by proposing a solution in series expansion for equation \eqref{xdotj} given by
\begin{equation}
    x(\tau) \approx 1-(\gamma+1-p)\left[ \frac{\gamma}{2(\gamma+1)^{2}}\tau^{2} + \frac{\gamma ^2  \big(p^{2}(\gamma -3) +(13 \gamma +9) p-3 \left(3 \gamma ^2+5 \gamma +2\right)\big)}{24 (\gamma +1)^5}\tau^{4}+\mathcal{O}(\tau^{6}) \right] \quad \text{for} \quad \tau \to 0 \, .
    \label{expxJ0}
\end{equation}
Notice that if $\gamma = p-1$, then the probe does not move, as this implies the BPS case. We can also consider the late time expansion
\begin{equation}
    \begin{split}
        & p=0: \quad x(\tau) \approx \frac{1}{\tau}-\frac{(\gamma+1)^{2}}{2\gamma^{2}}\frac{1}{\tau^{3}}+\mathcal{O}\left( \frac{1}{\tau^{4}}  \right) \quad \text{for} \quad \tau \to \infty  \\
        & 0<p\leq 5: \quad x(\tau) \approx \frac{1}{\tau}-\frac{\gamma+2}{2\gamma}\frac{1}{\tau^{3}} + \mathcal{O}\left( \frac{1}{\tau^{5}}  \right) \quad \text{for} \quad \tau \to \infty  \, . \\
    \end{split}
    \label{expxJI}
\end{equation}
These expansions will allow us to study the asymptotic behaviour of the holographic Krylov complexity in the next section. 

\subsubsection{Holographic spread complexity for an excited brane}

As explained in Section \ref{sec:pointparticle} in order to calculate the complexity for the probe described by the Lagrangian \eqref{lagrangianJ}, we need to utilise the Routhian which removes the dependence on $\dot{\psi}$. This is achieved by the Legendre transformation
\begin{equation}
    \mathcal{R}(r,\dot{r}) = {\cal J} \dot{\psi}(r, \dot{r})  - \mathcal{L}(r,\dot{r},\dot{\psi}) \, .
\end{equation}
The Routhian for this system is given by
\begin{equation}\label{ruthian_brane}
    {\cal R} = {\cal J}\dot{\psi}- {\cal L}= \sqrt{\frac{({\cal J}^2 L^{2p} + \lambda^2 \nu^2 r^{2p})}{\lambda^2 L^{2p}}\left(\frac{r^2}{L^2} - \frac{L^2 \dot{r}^2}{r^2}\right)} -\nu \frac{r^{p+1}}{L^{p+1}} \, .
\end{equation}
whose kinetic part is of the form \eqref{eq:Ruthian_point_particle}. The corresponding conserved Hamiltonian reads 

\begin{equation}
    \begin{split}\label{Hrouth excitation}
         H &= \frac{\partial{\cal R} }{\partial \dot{r}}\dot{r}-{\cal R} \\
        &=\nu\frac{r^{1+p}}{L^{1+p}}-\frac{r^3}{L^{1+p}\lambda}\sqrt{\frac{L^{2p}{\cal J}^2+\lambda^2\nu^2 r^{2p}}{r^4 -L^4 \dot{r}^2}},
    \end{split}
\end{equation}

by defining the energy from the Routhian $\mathcal{H} =- H $, \eqref{Hrouth excitation} gives the same equation as \eqref{rdotj}, here the dynamics encoded by the Routhian are the same as for the full Lagrangian in equation \eqref{lagrangianJ} with the only difference being that we are now working at fixed angular momentum, $\mathcal{J}$.






Defining the proper coordinate $y$ such that
 \begin{equation}
     \dot{y}^{2} =  \frac{({\cal J}^2 L^{2p} + \lambda^2 \nu^2 r^{2p})}{\lambda^2 L^{2p}}\frac{L^{2}}{r^{2}}\dot{r}^{2} \equiv g(r)^2 \frac{L^{2}}{r^{2}}\dot{r}^{2} \, .
     \label{properyJ}
 \end{equation}
 This definition of proper coordinate knows about the dynamics on $r$ but also about the conserved charge $\mathcal{J}$. Then, the proper momentum is defined as (see Footnote \ref{sign convention}) 
 \begin{equation}
     P_{y} \equiv - \frac{\partial {\cal R}}{\partial \dot{y}} = \frac{\dot{y}}{\sqrt{g(r)^2 \frac{r^{2}}{L^{2}}-\dot{y}^{2}}}= \frac{\dot{r}}{\sqrt{\frac{r^{4}}{L^{4}}-\dot{r}^{2}}} \, .
     \label{properJR}
 \end{equation}
 Note that in this way, $P_{y}$ is still negative as the expression for $\dot{r}$ has a minus sign. The dependence on $\cal J$ is implicit within $\dot{r}$. To explicitly see how the proper momentum depends on the conserved charge, we substitute the equation \eqref{rdotj} into \eqref{properJR} to get  
\begin{equation}
    P_y = -\frac{L^{\frac{p}{2}}\sqrt{L^{2+p} {\cal H}^2 \lambda^2-L^p{\cal J}^2 r^2+2L  {\cal H} \lambda^2\nu r^{p+1}}}{r \sqrt{L^{2p} {\cal J}^2 + \lambda^2 \nu^2 r^{2p}}} \, .
\end{equation}
Using the dimensionless parameters in equation \eqref{dimensionlessJ}, the proper momentum can be written as 
\begin{equation}
    P_{y}(\tau) = \frac{\dot{x}(\tau)}{\sqrt{x(\tau)^{4}-\dot{x}(\tau)^{2}}} \, ,
    \label{proper0}
\end{equation}
moreover, one can substitute equation \eqref{xdotj} to get
\begin{equation}
    P_{y}(\tau) =- \frac{1}{x(\tau)} \sqrt{\frac{\gamma^{2}-\gamma(\gamma+2)x(\tau)^{2}+2\gamma x(\tau)^{p+1}}{\gamma(\gamma+2)+x(\tau)^{2p}}} \, .
\end{equation}
 




We can analyse the early time behaviour of the proper momentum above by using the expansions in \eqref{expxJ0}, resulting in 
\begin{equation}
    P_{y}(\tau)\approx (p-1-\gamma)\left[\frac{\gamma }{(\gamma+1)^{2}}\tau + \frac{p \gamma^2 \left[(\gamma +1) (\gamma +3)+p\left(\gamma ^2+\gamma -3\right) \right]}{6 (\gamma +1)^6}\tau^{3}+\mathcal{O}(\tau^{5})\right] \quad \text{for} \quad \tau \to 0 \, .
    \label{expPyJ0}
\end{equation}
On the other hand, using  the expansion in \eqref{expxJI}, we can get the late time behaviour
\begin{equation}
    \begin{split}
         & p=0: \quad P_{y}(\tau) \approx -\frac{\gamma}{\gamma+1}\tau- \frac{1}{\gamma+1}+\mathcal{O}\left( \frac{1}{\tau^{3}}  \right) \quad \text{for} \quad \tau \to \infty  \\
        & 0<p\leq 5: \quad P_{y}(\tau) \approx -\sqrt{\frac{\gamma}{\gamma+2}}\tau+\mathcal{O}\left( \frac{1}{\tau^{3}}  \right) \quad \text{for} \quad \tau \to \infty  \, . \\
        \label{expPyJI}
    \end{split}
\end{equation}
The early-time expansion shows explicitly that the proper momentum vanishes for BPS branes, implying the absence of complexity growth. By contrast, in the late-time regime the proper momentum grows linearly, and does not depend on $p$ to this order.{ We would like to emphasise that if the calculation is done without the Routhian approach to fixed charge system, as in \cite{Fatemiabhari:2025poq, Fatemiabhari:2025usn, Fatemiabhari:2026goj}, one would get the early time behaviour $P_{y}(\tau) \sim k_{0} + k_{1} \tau^{2}$ where $k_{0}$ is proportional to the conserved charge and $k_{1}$ some constant. This constant $k_{0}$ in the early time behaviour originates from the fact that we cannot set the dynamical internal excitation to zero at $t = 0$, meaning in this case $\dot{\psi}(t = 0) \neq 0$ and the probe does not fall freely in all the coordinates, so this constant $k_{0}$ is the contribution from the non-zero velocity in the internal direction. Instead, what the Routhian prescription does is to implement the change $\dot{\psi} \to {\cal J}$ in such a way that it is no longer a dynamical degree of freedom and the problem no longer needs information about the initial condition of the internal coordinate. 

In Appendix \ref{sec: branes AdS5xS5} we study a D1[$t, x$] and $D5[t,x, T^{4}]$ wrapping the torus $T^{4}$, both on AdS$_{3} \times \mathrm{S}^{3} \times T^{4}$. These cases correspond with $p = 1$ and $\mathcal{J} \neq 0$. Another example is presented in Appendix \ref{sec: AdS3 x S3 x T4 examples} is that of a D3[$t, x_{1}, x_{2}, x_{3}, x_{4}$] in AdS$_{5} \times \mathrm{S}^{5}$ corresponding with $p = 3$ and $\mathcal{J} \neq 0$. 

In the next subsection we explore non-BPS branes by adding a gauge field excitation on the worldvolume.

\subsection{Adding a gauge field}\label{sec:Adding_a_gauge_field}

Another type of fluctuation that one can consider to make the brane non-BPS is a gauge field excitation in the worldvolume of the brane, for instance due to an F1 string ending on it. In this subsection, we consider a particular case which we choose to be the $p=3$ case of a D3 brane in AdS$_5 \times \mathrm{S}^5$, where we use the metric in terms of the left invariant $SU(2)$ forms $\omega_i$,
 \begin{align}\label{omegas}
       \omega^1&=\cos\psi\dd \vartheta+\sin\psi\sin\vartheta \dd \phi,\quad \omega^2 = \sin\psi\dd \vartheta - \cos\psi\sin\vartheta\dd \phi,\quad \omega^3 = \dd \psi + \cos\vartheta \dd \phi,\\
    \mathrm{d} s^2 & =\frac{r^2}{L^2}\left(-\mathrm{d} t^2+\mathrm{d} x_1^2+\mathrm{d} x_2^2+\mathrm{d} x_3^2\right)  + \frac{L^2}{r^2}\mathrm{d}r^2 \nonumber \\
    &\quad + L^2 \left[\mathrm{d}\theta^2 + \sin^2 \theta \, \mathrm{d}\phi^2 +\frac{\cos^2 \theta }{4}(\omega_1^2 + \omega_2^2 + \omega_3^2) \right].
\end{align}
The brane is extended in the AdS directions D3$[t, x_{1}, x_{2}, x_{3}]$. Then, its action is given by 
\begin{equation}
    S =- T_{D3} \int \dd^{4}\vec{x} \,  \sqrt{-\det\left(g_{\text{ind}}+ F \right)}+T_{D3} \int\;  C_{4} \, ,
\end{equation}
with $C_4$ given by the $p=3$ case of \eqref{Cp+1} and $F$ is the field strength associated to a constant electromagnetic field
\begin{equation}
    F = 
    \begin{pmatrix}
        0 & E & 0 & 0 \\
        -E & 0 & 0 & B \\
        0 & 0 & 0 & 0 \\
        0 & -B & 0 & 0  
    \end{pmatrix}
    \, .
\end{equation}
The action then takes the form
\begin{equation}
    S = \nu \int \dd t \, \left(-\sqrt{\underbrace{\left(B^{2} \frac{r^{2}}{L^{2}}+\frac{r^{6}}{L^{6}} \right)\left( \frac{r^{2}}{L^{2}}-\frac{L^{2}}{r^{2}}\dot{r}^{2}  \right)-E^{2}\frac{r^{4}}{L^{4}}}_{\bar \Delta}} +\frac{r^{4}}{L^{4}} \right) \, .
    \label{actionEB}
\end{equation} 
The Lagrangian for the case of both magnetic and electric fields turned on is given by
\begin{equation}
   {\cal L}=\nu \left(-\sqrt{\left(B^{2} \frac{r^{2}}{L^{2}}+\frac{r^{6}}{L^{6}} \right)\left( \frac{r^{2}}{L^{2}}-\frac{L^{2}}{r^{2}}\dot{r}^{2}  \right)-E^{2}\frac{r^{4}}{L^{4}}} + \frac{r^{4}}{L^{4}}\right) \, .
\end{equation}

The conjugate momenta are
\begin{equation}
     \begin{split}
         & \mathcal{P}_{E} \equiv \frac{\partial \mathcal{L}}{\partial E} = \frac{E \nu }{\sqrt{\bar \Delta}}\frac{r^{4}}{L^{4}} \, , \\
         & \mathcal{P}_{B} \equiv \frac{\partial \mathcal{L}}{\partial B} = \frac{\nu B}{\sqrt{\bar \Delta}}\left( \dot{r}^{2} -\frac{r^{4}}{L^{4}} \right) \, .
     \end{split}
\end{equation}
The Hamiltonian is 
\begin{equation}
   \mathcal{H} = P_{r} \dot{r} +\mathcal{P}_{E} E-\mathcal{L}
\end{equation}
which is conserved.  From the Hamiltonian one gets the equation
\begin{equation}
    \dot{r} = -\frac{r^{2}}{L^{4}\mathcal{H}+\nu r^{4}}\sqrt{\frac{L^8 \mathcal{H}^2 \left(B^2-E^2\right)+\nu  r^8 \left[2 \mathcal{H}-\nu  \left(B^2+E^2\right)\right]-L^4 r^4 \left(B^4 \nu ^2+2 \nu  \mathcal{H} \left(E^2-B^2\right)-\mathcal{H}^2\right)}{B^{2}L^{4}+r^{4}}} \, .
    \label{dotrEB}
\end{equation}
See Appendix \ref{sec:appendix EoM} for the equations of motion for the case with the gauge field.
The equations of motion for the magnetic components of $A$ are trivially satisfied provided that $B$ is constant. In the next sections we study the case of a constant magnetic field and a time-dependent electric field.  \\

\subsubsection{Constant magnetic field}
 Consider the case of $E = 0$, which simplifies equation \eqref{dotrEB} to 
\begin{equation}
    \dot{r} = -\frac{r^{2}}{L^{2}}\frac{\sqrt{L^{4}\mathcal{H}(L^{4}\mathcal{H}+2\nu r^{4})-L^{4}B^{2}\nu^{2}r^{4}}}{L^{4}\mathcal{H}+\nu r^{4}} \, ,
    \label{dotrB}
\end{equation}
 one can choose $r(t= 0) = r_{\text{UV}}$ such that the argument of the square root vanishes, i.e.
\begin{equation}
    r_{\text{UV}} = \frac{L\sqrt{\mathcal{H}}}{(B^{2} \nu^{2}-2\mathcal{H}\nu)^{1/4}} \quad \text{equivalently} \quad \mathcal{H} = \frac{\nu r_{\text{UV}}^{2}}{L^{4}}\left(\sqrt{r_{\text{UV}}^{4}+B^{2}L^{4}}-r_{\text{UV}}^{2}\right)\, ,
    \label{ruvB}
\end{equation}
in order for $r_{\text{UV}}$ to be well defined, the condition $B^{2}  \nu > 2 \mathcal{H}$ has to be satisfied. \\
As in the previous sections, we solve equation \eqref{dotrB} by defining the dimensionless quantities
\begin{equation}
   \gamma = \left(\frac{L}{r_{\text{UV}}}\right)^{4}\frac{\mathcal{H}}{\nu}, \quad \beta_B = \frac{\nu^{2}}{\mathcal{H}^{2}}\left(\frac{r_{\text{UV}}}{L}  \right)^{4}B^{2}
\end{equation}
where $\gamma +2 = \gamma\, \beta_B$. This allows, using \eqref{xandtau}, to rewrite equation \eqref{dotrB} as
\begin{equation}
    \frac{\dd x }{\dd\tau } = -\gamma x^{2}\frac{ \sqrt{1-x^{4}}}{x^{4}+\gamma} \, .
    \label{xdotB}
\end{equation}
This equation describes a particle with zero energy moving in the effective potential
\begin{equation}
    V_{\text{eff}}(x) = -\gamma^{2} x^{4} \frac{1-x^{4}}{(\gamma + x^{4})^{2}} \, .
\end{equation}
Plots of the effective potential are shown in Figure \ref{xVB}, where we can see that for the initial condition $x(\tau = 0) = 1$ the probe will fall with zero initial velocity. Moreover, numerical solutions of equation \eqref{xdotB} were obtained and are shown in Figure \ref{xVB}.

\begin{figure}[h!]
    \centering
    \includegraphics[width=0.45\linewidth]{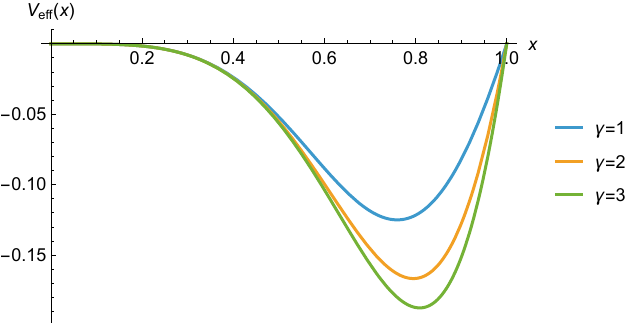}
    \quad \quad
        \includegraphics[width=0.45\linewidth]{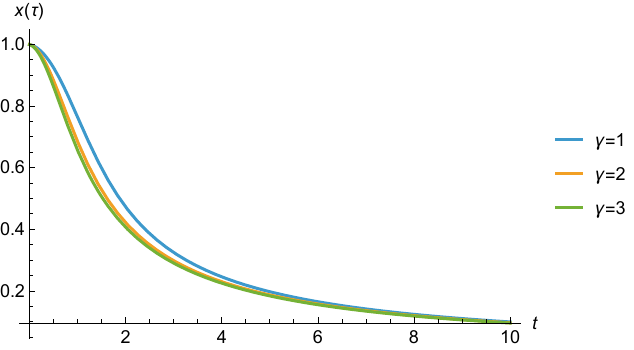}
    \caption{[Left] Plot of the effective potential $V_{\text{eff}}(x)$ and [Right] plot of $x(\tau)$ for different values of $\gamma$.}
    \label{xVB}
\end{figure}

Alternatively, one can solve equation \eqref{xdotB} by series expansion getting
\begin{equation}
    \begin{split}
        & x(\tau) \approx 1 - \frac{\gamma^{2}}{(\gamma+1)^{2}}\tau^{2}+\frac{\gamma ^4 (11 \gamma -5)}{6 (\gamma +1)^5}\tau^{4} + \mathcal{O}(\tau^{6}) \quad \text{for} \quad \tau \to 0 \, , \\
        & x(\tau) \approx \frac{1}{\tau} -\frac{\gamma + 2}{6 \gamma}\frac{1}{\tau^{5}} + \mathcal{O}\left(\frac{1}{\tau^{6}}\right) \quad \text{for} \quad \tau \to \infty \, .
        \label{expB}
    \end{split}
\end{equation}

Notice that the equations of motion, see Appendix \ref{sec:appendix EoM}, do not define a conserved quantity associated to the magnetic field, this implies that we do not need to use the Routhian prescription, and we can define the proper coordinate as usual, meaning $\dot{y}^{2} = \frac{L^{2}}{r^{2}}\dot{r}^{2}$ and the proper momentum is defined from the Lagrangian $P_{y} = \frac{\partial \mathcal{L}}{\partial \dot{y}}$. Then, after substituting \eqref{dotrB} we get the explicit expression 
\begin{equation}
    P_{y} = -\frac{1}{L^{3}r}\sqrt{L^{4}\mathcal{H}(L^{4}\mathcal{H}+2\nu r^{4})- L^{4} B^{2}\nu^{2}r^{4}} \, .
\end{equation}
Or in terms of the dimensionless variables
\begin{equation}
    \tilde{P}_{y}(\tau) = -\gamma\frac{\sqrt{1-x(\tau)^{4}}}{x(\tau)} \, ,
\end{equation}
where $\tilde{P}_{y} = \frac{P_{y}}{\nu}$. Plots of the proper momentum for the numerical solutions are shown in Figure \ref{PYBP}. On the other hand, using the expansions in \eqref{expB} we get the asymptotic behaviour for the proper momentum
\begin{equation}
    \begin{split}
        & \tilde{P}_{y}(\tau) \approx -\frac{2\gamma^{2}}{\gamma+1}\tau +\frac{4 (\gamma -1) \gamma ^4 }{3 (\gamma +1)^4}\tau^{3}  + \mathcal{O}(\tau^{4}) \quad \text{for} \quad \tau \to 0 \, , \\
        & \tilde{P}_{y}(\tau) \approx \gamma \tau +\frac{1-\gamma}{3}\frac{1}{\tau^{3}} + \mathcal{O}\left(\frac{1}{\tau^{4}}\right) \quad \text{for} \quad \tau \to \infty \, .
        \label{expPyB}
    \end{split}
\end{equation}
Note that in this case, we obtain the desired behaviour, meaning linear at early times. In the next section we will see that for a time-dependent electric field, the Routhian prescription is necessary, as there is a conserved charge reflecting the difference of the complexity in the presence of magnetic and electric fields.


\begin{figure}[h!]
    \centering
        \includegraphics[width=0.45\linewidth]{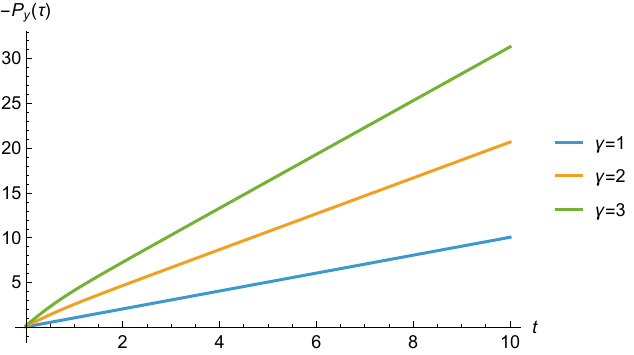}
    \caption{Plot of the proper momentum for different values of $\gamma$.}
    \label{PYBP}
\end{figure}

\subsubsection{Time dependent electric field}

Consider now the case of zero magnetic field, $B = 0$, and a non-zero time-dependent electric field $F_{01} = E(t)$, such that $A_{1} = \int_{0}^{t} \dd \tau \, E(\tau) $ and $A_{0}=0$. Following the same procedure as above, the action is given by equation \eqref{actionEB}, in this case
\begin{equation}\label{L time dep E}
    S = \nu \int \dd t \, \left(-\frac{r^{3}}{L^{3}}\sqrt{\frac{r^{2}}{L^{2}}-\frac{L^{2}}{r^{2}}(\dot{r}^{2}+E^{2})}  +\frac{r^{4}}{L^{4}} \right) \, .
\end{equation}
Consistency with the equations of motion requires that the quantity $\mathcal{P}_{E}$ is conserved, meaning that we can express the electric field in terms of $\mathcal{P}_{E}$, that is
\begin{equation}\label{E rdot}
    E = \mathcal{P}_{E}\sqrt{\frac{r^{4}-L^{4}\dot{r}^{2}}{L^{4}\mathcal{P}_{E}^{2}+\nu^{2}r^{4}}} \, .
\end{equation}
Since there is a conserved Noether charge, we compute the Routhian such that we can remove the dependence on the coordinate associated to the conserved charge,
\begin{equation}
    \mathcal{R} = \mathcal{P}_{E} E - \mathcal{L} = \sqrt{\frac{r^{2}(L^{4}\mathcal{P}_{E}^{2}+\nu^{2}r^{4})}{L^{6}}\left(\frac{r^{2}}{L^{2}}-\frac{L^{2}}{r^{2}}\dot{r}^{2}  \right)} - \nu\frac{r^{4}}{L^{4}}\, .
\end{equation}
Then, by computing the Hamiltonian from the Routhian, which we denote as $H$ and defining the energy to be positive, meaning $\mathcal{H}=-H$, we get the equation
\begin{equation}
    \dot{r} = -\frac{r^{2}}{L^{2}}\frac{\sqrt{L^{4}\mathcal{H}(L^{4}\mathcal{H}+2\nu r^{4})-L^{4}\mathcal{P}_{E}^{2} r^{4}}}{L^{4}\mathcal{H}+\nu r^{4}} \, .
    \label{mexico}
\end{equation}
The initial conditions $\dot{r}(t = 0) = 0,\; \; r(t=0)=r_{\text{UV}}$ give the constraint
\begin{equation}
    \left(\frac{L}{r_{\text{UV}}}\right)^{4}\frac{\mathcal{H}}{\nu}+2 = \frac{\mathcal{P}_{E}^{2}}{\mathcal{H}\nu} \, ,
\end{equation}
and by defining the dimensionless parameters 
\begin{equation}\label{dimless E}
\gamma = \left(\frac{L}{r_{\text{UV}}}\right)^{4}\frac{\mathcal{H}}{\nu}, \qquad \beta_{E} = \frac{\mathcal{P}_{E}^{2}}{\mathcal{H}\nu},
\end{equation}
the condition reads $\gamma+2 = \beta_{E}$. Moreover, using the definitions in \eqref{xandtau}, equation \eqref{mexico} along with the constraint gives
\begin{equation}
    \dot{x}(\tau) = -\gamma \, x(\tau)^{2}\,\frac{\sqrt{1-x(\tau)^{4}}}{\gamma + x(\tau)^{4}} \, ,
    \label{dotxE}
\end{equation}
which is the same equation as \eqref{xdotB}, and then the expansions in \eqref{expB} are also valid.

We can substitute \eqref{mexico} into \eqref{E rdot} to obtain 
\begin{align}
    E = \frac{\mathcal{P}_E \, r^4}{L^4 \mathcal{H} + \nu \, r^4} \, ,
\end{align}
which in the UV, in terms of \eqref{dimless E} gives $E_{\rm{UV}}$ from which we can define the electric field $\mathcal{E}(t) = \frac{L^{2}}{r_{\text{UV}}^{2}}E(t)$ in terms of the dimensionless variable and parameters
\begin{equation}
    E_{\rm{UV}} = \frac{r^2_{\rm{UV}} \sqrt{ \gamma (\gamma +2 )}}{L^2 (1+\gamma)}, \qquad \mathcal{E}(\tau) = \sqrt{\gamma (\gamma +2)}\,  \frac{x(\tau)^4}{x(\tau)^4 + \gamma }.
\end{equation}

This result shows that the electric field vanishes at late times. This behaviour is desirable, since an electric field that increases along the brane trajectory could cause the Lagrangian \eqref{L time dep E} to become imaginary. This phenomenon is familiar from flavour-brane embeddings extending along the radial direction, where the DBI action develops a worldvolume pseudo-horizon even at zero temperature \cite{Karch:2007pd,OBannon:2007cex}. In that case, the electric field depends on the radial coordinate (which is part of the brane worldvolume), and there is a singular shell in the IR, which can be interpreted as an effective temperature in the open-string metric.

In the present case, the argument of the square root remains positive throughout the brane trajectory. It is also noteworthy that, despite the existence of a conserved Noether charge and the corresponding need to work with the Routhian, no analogue of the centrifugal barrier discussed in Sections (\ref{sec: internal space}),(\ref{sec:Excited D3}) arises. Consequently, there is no dynamical mechanism that bounds the electric field from above, although the DBI action nevertheless remains real along the entire trajectory.

We now proceed to calculating the complexity, first by defining the proper coordinate from the Routhian as 
\begin{equation}
    \dot{y}^{2} = \frac{r^{2}(L^{4}\mathcal{P}_{E}^{2}+\nu^{2}r^{4})}{L^{6}} \frac{L^{2}}{r^{2}}\dot{r}^{2} \equiv g(r)^{2}\frac{L^{2}}{r^{2}}\dot{r}^{2} \, ,
\end{equation}
then the proper momentum is given by
\begin{equation}
    P_{y}(t) = -\frac{1}{r^{2}}\sqrt{\frac{L^{4}\mathcal{H}(L^{4}\mathcal{H}+2\nu r^{4})-L^{4}\mathcal{P}_{E}^{2}r^{4}}{L^{4}\mathcal{P}_{E}^{2}+\nu^{2}r^{4}}} \, . 
\end{equation}
The definition of the proper coordinate produces the same expression for the proper momentum as in \eqref{proper0}, which after using \eqref{dotxE} we get
\begin{equation}
    P_{y}(\tau) = -\frac{\gamma}{x(\tau)^{2}}\sqrt{\frac{1-x(\tau)^{4}}{\gamma(\gamma+2)+x(\tau)^{4}}} \, .
\end{equation}
This can be expanded as
\begin{equation}
    \begin{split}
        & P_{y}(\tau) \approx -\frac{2\gamma^{2}}{(\gamma+1)^{2}}\tau +\mathcal{O}(\tau^{3}) \quad \text{for} \quad \tau \to 0 \, , \\
        & P_{y}(\tau) \approx -\sqrt{\frac{\gamma}{\gamma+2}} \tau^{2} + \mathcal{O}\left(\frac{1}{\tau^{2}}\right) \quad \text{for} \quad \tau \to \infty \, .
        \label{expE}
    \end{split}
\end{equation}
Notice that the leading order early time behaviour for the electric and magnetic field are the same, compare equations \eqref{expE} with \eqref{expPyB}, however for late times the proper momentum grows quadratically for the electric field instead of linear in time as it does for the magnetic field, see Figure \ref{xandPyE}. Therefore, we can visualise how the presence of a conserved charge could change the late time behaviour of the complexity. 

\begin{figure}[h!]
    \centering
    \includegraphics[width=0.45\linewidth]{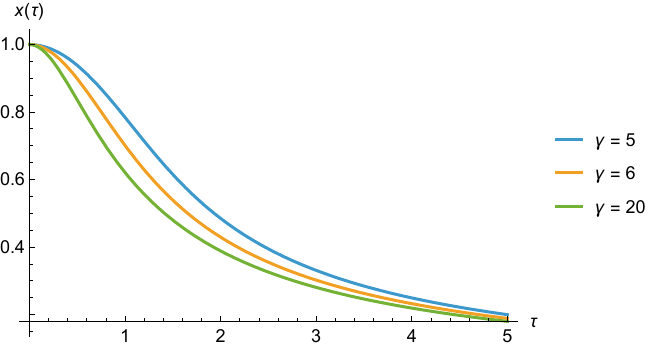}
    \quad \quad
        \includegraphics[width=0.45\linewidth]{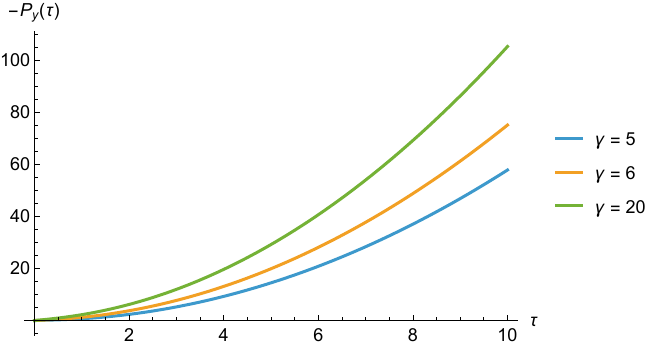}
    \caption{[Left] Plots of the trajectory $x(\tau)$.  [Right] Plot of the proper momentum $P_{y}(\tau)$ for the time-dependent electric field case.}
    \label{xandPyE}
\end{figure}

\section{ABJM}\label{sec:ABJM}\label{seccion4}
In this section we study a  D0-brane in $\mathrm{AdS}_4\times\mathbb{CP}^3$, including its conserved internal momentum and transverse fluctuations, and calculate the complexity generated by the evolution of this state. We start with a brief review of the background to set the notation.

\subsection{Review of the ABJM background}

We dimensionally reduce M-theory on $\mathrm{AdS}_4\times \mathrm{S}^7/\mathbb{Z}_k$ to the ABJM type IIA solution \cite{Aharony:2008ug,Benincasa:2011zu}, given by the metric in string frame and $\alpha'=1$,
\begin{equation}\label{AdS4xCP3}
    \begin{split}
        &\dd s^2_{10}=L^2_{\text{ABJM}}(\dd s^2_{\text{AdS}_4}+4 \dd s^2_{\mathbb{CP}^3}),\\
        &\dd s^2_{\text{AdS}_4}=r^2 (-\dd t^2+\dd x_1^2+\dd x_2^2)+\frac{\dd r^2}{r^2},\\
        &\dd s^2_{\mathbb{CP}^3}=g_{ij}(\zz_l,\bar{\zz}_l)\dd \zz^i \dd \bar{\zz}^j ,
    \end{split}
\end{equation}
We use the Fubini-Study metric of $\mathbb{CP}^3$,
\begin{align}
    \dd s^{2}_{\mathbb{CP}^{3}} =& \frac{1}{4} \Bigl[
        \dd\alpha^{2} + \cos^{2}\frac{\alpha}{2} \left( \dd\theta_{1}^{2} + \sin^{2}\theta_{1}\,\dd\varphi_{1}^{2} \right) + \sin^{2}\frac{\alpha}{2} \left( \dd\theta_{2}^{2} + \sin^{2}\theta_{2}\,\dd\varphi_{2}^{2} \right) \nonumber\\
        &\quad + \sin^{2}\frac{\alpha}{2}\cos^{2}\frac{\alpha}{2}  \left( \dd\chi + \cos\theta_{1}\,\dd\varphi_{1} + \cos\theta_{2}\,\dd\varphi_{2} \right)^{2} \Bigr],
\end{align}
with the angles parametrised as $\, \alpha, \theta_1,\theta_2 \in [0,\pi], \;\; \varphi_1,\varphi_2 \in [0,2\pi], \;\; \chi \in [0,4\pi)$. 

The radius of the space, $L_{\rm{ABJM}}$ is related to the integers $N,k$, which are the rank of the gauge group and the Chern-Simons level in the dual gauge theory respectively, 
\begin{equation}
    L_{\text{ABJM}}^4=2\pi^2 \frac{N}{k}.
\end{equation}
There are RR-fields $F_2$ and $F_4$ given by
\begin{align}
    F_2 &= \frac{k}{4}\dd A \\
    A &= \cos\alpha\,\dd\chi + 2\cos^{2}\frac{\alpha}{2}\cos\theta_{1}\,\dd\varphi_{1}- 2\sin^{2}\frac{\alpha}{2}\cos\theta_{2}\,\dd\varphi_{2} \, \\
    F_4 &= \frac{3}{2}k L^2_{\text{ABJM}} \, \operatorname{vol}(\text{AdS}_4),\\
    \operatorname{vol}({\text{AdS}_4)}& =r^2\dd t\wedge \dd x_1 \wedge \dd x_2\wedge \dd r.
\end{align}

We will also consider the potentials for these fluxes, which can be written in the form
    \begin{equation}\label{RR_potetials}
        C_1=\frac{k}{2}A,\quad C_3 = -\frac{k}{2} L^2_{\text{ABJM}} r^3 \dd t\wedge \dd x_1 \wedge \dd x_2.
    \end{equation}
 The background also contains a constant dilaton,
\begin{equation}
    e^\Phi= \frac{2}{k}L_{\text{ABJM}}=2\sqrt{\pi} \left( \frac{2N}{k^5}\right)^{1/4}.
\end{equation}

We will study the motion of a D0 and, in Appendix \ref{sec: D2 ABJM}, a D2 brane in this background.

\subsection{D0-brane in \texorpdfstring{$\mathrm{AdS}_4\times\mathbb{CP}^3$}{AdS4xCP3}}}

We first consider a D0-brane with motion in $r(t), \, \alpha(t), \, \chi(t)$, and find the classical solutions for the trajectory of the brane in this embedding. After finding the classical solutions, we will then consider fluctuations about the $\alpha, \chi$ angles contained within $\mathbb{CP}^3$. The induced metric is given by
\begin{align}
    \dd s^2_{\text{ind}} = L^2_{\rm{ABJM}} \Big[  -r^2  + \frac{\dot{r}^2}{r^2} + \dot{\alpha}^2 + \frac{\sin^2 \alpha }{4} \dot{\chi}^2 \Big]\dd t^2\, .
\end{align}

The DBI part of the Lagrangian is given by
\begin{equation}
    \mathcal{L}_{\rm{DBI}} = e^{-\Phi}\sqrt{- \operatorname{det}g_{\rm{ind}}} = L_{\rm{ABJM}} \; e^{-\Phi} \, \sqrt{ r^2 - \frac{\dot{r}^2}{r^2} - \dot{\alpha}^2 - \frac{\sin^2 \alpha}{4}\dot{\chi}^2}
\end{equation}
The full DBI-WZ action then reads
\begin{align}
    S &= T_{D0}\;    \left( \, -  \int \dd t \; \mathcal{L}_{\rm{DBI}} + \mu_0 \int P[C_1] \right)\\
    &=T_{D0}\;   \,  \int \dd t \, \left( -e^{-\Phi}L_{\rm{ABJM}}\sqrt{ r^2 - \frac{\dot{r}^2}{r^2} - \dot{\alpha}^2 - \frac{\sin^2 \alpha}{4}\dot{\chi}^2} \; + \mu_0\,  \cos\alpha\, \dot{\chi} \right)
\end{align}
where $\mu_0$ is a parameter related to the charge of the brane under the WZ term, $T_{D0}$ is the tension and $P[C_1]$ is the pull-back of the $C_1$ form on the worldvolume of the D0.

There is a conserved charge associated to the $\chi$ coordinate,
\begin{align}
  \mathcal{P}_\chi = \frac{\partial\mathcal{L}}{\partial\dot{\chi}} =    \frac{e^{-\Phi}\, L_{\text{ABJM}}\sin ^2\alpha \; \dot{\chi}}{4 \sqrt{-\frac{\dot{r}^2}{r^2}+r^2-\dot{\alpha}^2-\frac{1}{4} \sin ^2\alpha \;  \dot{\chi}^2}}+\mu_0\cos \alpha \, .
\end{align}
We write the Routhian, and substitute the velocity $\dot{\chi}$ using conserved momenta above,
\begin{align}\label{routh D0}
    \mathcal{R} &= \mathcal{P}_\chi \dot{\chi} - \mathcal{L}= 
\sqrt{e^{-2\Phi}L_{\text{ABJM}}^2+4(\mu_0 \cot \alpha-{\cal P}_\chi \csc \alpha)^2}\sqrt{r^2 - \frac{\dot{r}^2}{r^2} - \dot{\alpha}^2}=M(\alpha) \sqrt{r^2 - \frac{\dot{r}^2}{r^2} - \dot{\alpha}^2},\\
&\text{with}\qquad M(\alpha):= \sqrt{e^{-2\Phi}L_{\text{ABJM}}^2+4(\mu_0 \cot \alpha-{\cal P}_\chi \csc \alpha)^2}.
\end{align}
We define the momenta along $\alpha$ and $r$ as well as the conserved Hamiltonian in the Routhian formalism
\begin{align}
    &\mathcal{P}_\alpha = \frac{\partial\mathcal{R}}{\partial\dot{\alpha }} = - \frac{M(\alpha )\,  \dot{\alpha}}{\sqrt{r^2 - \frac{\dot{r}^2}{r^2} - \dot{\alpha}^2}}, \qquad \mathcal{P}_r = \frac{\partial\mathcal{R}}{\partial\dot{r}}\\
    &\mathcal{H} = \mathcal{P}_r \dot{r} + \mathcal{P}_\alpha \dot{\alpha} - \mathcal{R} = \frac{M(\alpha) r^2}{\sqrt{r^2 - \frac{\dot{r}^2}{r^2} - \dot{\alpha}^2}}\, .
\end{align}
Since the momentum associated to $\alpha$ is not conserved, we construct another conserved quantity
\begin{align}
    K = \mathcal{P}_\alpha^2 + M(\alpha)^2.
\end{align}
It can be shown that $\dot{K}=0$ via the equation of motion of $\alpha$ using the Routhian.

We obtain the classical solution
\begin{align}\label{rcl}
    &\dot{r} = -\frac{r^2 \sqrt{\mathcal{H}^2 - K r^2}}{\mathcal{H}}\Rightarrow r(t)\equiv r_{\rm{cl}}(t) = \frac{\mathcal{H}}{\sqrt{K + \mathcal{H}^2 t^2}},\\
    & \dot{\alpha} = \frac{r^2 \sqrt{K - M(\alpha )^2}}{\mathcal{H}}= \frac{{\cal H}\sqrt{K-M(\alpha)^2}}{K+{\cal H}^{2}t^2}.
\end{align}

The $\chi$ and $\alpha$ equations of motion in this embedding are
\begin{align}
    &\frac{\dd }{\dd t} \left( e^{-\Phi}L_{\text{ABJM}}\frac{\sin^2 \alpha \; \dot{\chi}}{4 \, \sqrt{\Xi}} +\mu_0\cos \alpha\right) =0,\\
    &e^{-\Phi}L_{\text{ABJM}}\frac{\dd }{\dd t}\left( \frac{\dot{\alpha}}{\sqrt{\Xi}}\right) = -\mu_0\sin \alpha \; \dot{\chi} + e^{-\Phi}L_{\text{ABJM}}\frac{\sin\alpha \, \cos \alpha\,  \dot{\chi}^2}{4 {\sqrt{\Xi}}},\\
    &\Xi := r^2 - \frac{\dot{r}^2}{r^2} - \dot{\alpha}^2 - \frac{\sin^2 \alpha}{4}\dot{\chi}^2,
\end{align}
which means that $\alpha = \frac{\pi}{2},\;  \dot{\chi}=0$ is a consistent solution, so we can take $\chi = \chi_0$ a constant. Therefore we consider fluctuations of the D0-brane around 
\begin{equation}
    r(t) = r_{\rm{cl}}(t), \qquad \alpha(t) = \frac{\pi}{2} + \epsilon \, Z(t), \qquad \chi(t) = \chi_0 + \epsilon\, Y(t),\quad |\epsilon|\ll 1.
\end{equation}
\begin{equation}
    \mathcal{L}= -e^{-\Phi}L_{\text{ABJM}}\sqrt{r^2-\frac{\dot{r}^2}{r^2}-\frac{1}{4} \epsilon ^2 \left(\dot{Y}(t)^2 \cos ^2(\epsilon  Z(t))+4 \dot{Z}(t)^2\right)}-\epsilon \mu_0 \dot{Y}(t) \sin (\epsilon  Z(t))\,.
\end{equation}

Expanding the Lagrangian in powers of $\epsilon$ gives
\begin{align}
    &\mathcal{L} = \mathcal{L}_{0} + \epsilon^2 \, \mathcal{L}_2 + \mathcal{O}(\epsilon^4) \\
    &\mathcal{L}_0 = -\frac{e^{-\Phi}L_{\text{ABJM}}}{\mathcal{M}(t)}, \quad \mathcal{L}_2 = e^{-\Phi}L_{\text{ABJM}}\frac{\dot{Y}(t)^2 + 4 \dot{Z}(t)^2}{8 }\mathcal{M}(t) -\mu_0\,\dot{Y}(t)\, Z(t),\\
    &\mathcal{M}(t)=\left(\sqrt{r_{\rm cl}(t)^2 - \frac{\dot{r}_{\rm cl}(t)^2}{r_{\rm cl}(t)^2}}\right)^{-1}.
\end{align}
The $Z,Y$ equations of motion from the second order Lagrangian are
\begin{align}
    &\frac{\dd }{\dd t } \Bigg( e^{-\Phi}L_{\text{ABJM}}\dot{Z}(t)\mathcal{M}(t) +\mu_0\, Y(t)\Bigg) =0,\\
    &\frac{\dd}{\dd t }\Bigg(  \frac{e^{-\Phi}L_{\text{ABJM}}}{4}\dot{Y}(t) \mathcal{M}(t) -\mu_0\, Z(t) \Bigg) =0,
\end{align}
with solutions after a change of variable $\dd u = \frac{1}{\mathcal{M}(t)} \dd t$ and using \eqref{rcl}
 \begin{align}\label{sol Z,Y,u}
     &Z(t) = A\cos(\omega u) + B\sin(\omega u), \qquad Y(t) = 2A \sin(\omega u) - 2B \cos(\omega u), \\
     &u =\int \dd t \sqrt{r_{\rm{cl}}^2 - \frac{\dot{r}^2_{\rm{cl}}}{r^2_{\rm{cl}}}} = \arctan \left( \frac{t\mathcal{H}}{\sqrt{K}}\right),\quad \omega^2 = \frac{4e^{2\Phi}\mu_0^2}{L_{\text{ABJM}}^2}.
 \end{align}

In order to compute the complexity, we define the proper coordinate from the Routhian \eqref{routh D0}, such that $\mathcal{R} = \sqrt{M(\alpha)^2 r^2 - \dot{y}^2}$, then
\begin{align}
    \dot{y}^2 = M(\alpha)^2 \left( \frac{\dot{r}^2}{r^2} + \epsilon ^2 \dot{Z}^2(t) \right)\, .
\end{align}
The rate of change of complexity is given by the proper momentum, which for small fluctuations $\epsilon \ll 1$ gives
\begin{align}
    &\dot{\mathcal{C}} \propto P_y  = -\sqrt{\frac{r^4}{r^4-\dot{r}^2 - \epsilon ^2\,  r^2 \,\dot{Z}(t)^2  }-1} \, ,\\
    &P_y \sim \frac{\mathcal{H} t}{\sqrt{K}}+  \frac{ \dot{Z}(t)^2 \left(K+\mathcal{H}^2 t^2\right)^3}{2 K^{\frac{3}{2}} \mathcal{H}^3 t} \,\epsilon ^2 + \mathcal{O}(\epsilon^3) \, .
\end{align}
Setting $B=0$ (otherwise one obtains a constant zeroth order term in $t$ in $P_y$) in \eqref{sol Z,Y,u} before expanding in $t$, in which case we then obtain the early times expansion
\begin{align}
    P_y \sim \frac{ \mathcal{H} \sqrt{\omega^4 A^2 \epsilon ^2+1}}{\sqrt{K}} \, t + \mathcal{O}(t^2) \, .
\end{align}

Further expanding in $\epsilon$ gives us the correction
\begin{equation}
    P_y\approx -\frac{{\cal H}}{\sqrt{K}}\left(1+\frac{A^2\omega^4\epsilon^2}{2}\right)t + {\cal O}(t^2).
\end{equation}


In the next section we study an explicit example of an extended object, a fundamental string winding around directions in the internal $\mathrm{S}^5$ space whilst falling in $\rm{AdS}_5$. This example shows how a wrapped string can be recast as an effective massive particle. We study the distinction between cases with a conserved Noether charge studied above, which require choosing a fixed-charge sector and using the Routhian, compared to examples with winding data such as the fundamental string studied below, in which it acts effectively like the massive particle falling in AdS.
\\
\section{F1 String falling and rotating in  {$\rm{AdS}_5\times \mathrm{S}^5$}}\label{seccion5}
In Section \ref{sec:ABJM} we studied an excitation in the field theory represented by a D0-brane. In this section we discuss an excitation in $\mathcal{N}=4$ SYM. The holographic dual of the excitation  is a fundamental (F1) string propagating along the radial direction while wrapping three internal directions of $\rm{AdS}_5 \times \mathrm{S}^5$, which in this example are within $\mathrm{S}^3 \subset \mathrm{S}^5$ of $\rm{AdS}_5 \times \mathrm{S}^5$. Although the fundamental string is an extended object, its dynamics reduce to those of an effective massive particle. This example illustrates the distinction between conserved Noether charges, which necessitates working within a fixed charge sector via Legendre transforming to the Routhian, and winding data of the string encoded within $\phi_i = m_i \sigma$, which instead enters through the effective mass of the particle. 

We let $(\tau,\sigma)$ denote the worldsheet coordinates of the fundamental string. Consider the F1 string falling in one spatial direction $r$ and rotating within three angular directions of the internal space $\phi_i(\tau,\sigma), \; i=1,2,3$.
We now write the metric of $\rm{AdS}_5 \times \mathrm{S}^5$ in the form 
\begin{align}\label{metricaads5s5}
    \mathrm{d}s^2 =& -\frac{r^2}{L^2} \mathrm{d}t^2 + \frac{r^2}{L^2} \mathrm{d}x^2_1 + \frac{r^2}{L^2} \mathrm{d}x_2^2 + \frac{r^2}{L^2} \mathrm{d}x_3^2 + \frac{L^2}{r^2}\mathrm{d}r^2 \\
    &+ L^2 \Big[\mathrm{d}\theta^2
+ \cos^2\theta\, \mathrm{d}\psi^2
+ \cos^2\theta\, \sin^2\psi\, \mathrm{d}\phi_1^2
+ \cos^2\theta\, \cos^2\psi\, \mathrm{d}\phi_2^2
+ \sin^2\theta\, \mathrm{d}\phi_3^2
\Big].
\end{align}
with $\theta,\psi \in [0,\frac{\pi}{2}], \; \phi_i \in [0,2\pi]$ and define the following in which we will use below
\begin{equation}\label{mus}
    \mu_1 = \cos\theta \sin\psi , \qquad \mu_2 = \cos\theta\cos\psi, \qquad \mu_3 = \sin\theta\, .
\end{equation}

We will study the F1 in the Polyakov and Nambu-Goto formalisms, in the next subsection we begin with the Polyakov action for the most generic string embedding in the above background. 

\subsection{Polyakov action}

We consider a general embedding before truncating to ensure consistency, for $i,j=1,2,3$,
\begin{equation}\label{generalembedding}
    t=t(\tau,\sigma) , \quad r=r(\tau,\sigma), \quad x_j=x_j(\tau,\sigma), \quad \phi_i= \phi_i(\tau,\sigma), \quad \theta = \theta(\tau,\sigma), \quad \psi= \psi(\tau,\sigma)
\end{equation}
one can check that the equations of motion are satisfied for this embedding, we check that the equations of motion are consistent in the Polyakov formalism. We let $\dot{X}^\mu=\partial_\tau X^\mu$ and $(X^\mu)'=\partial_\sigma X^\mu$ and the worldsheet metric is $\eta^{\alpha\beta}=\text{diag}(-1,1)$, The Polyakov action is given by 
\begin{align}
    &S_{P} = T_{F1} \int \mathrm{d}\sigma \mathrm{d}\tau ~{\cal L}_P=T_{F1}\int\mathrm{d}\sigma \mathrm{d}\tau \; \eta^{\alpha\beta}G_{\mu\nu} \partial_\alpha X^\mu \partial_\beta X^\nu \\
    &\mathcal{L}_P = G_{tt} (t^{\prime \, 2 } - \dot{t}^2) +  G_{rr}(r^{\prime \, 2} - \dot{r}^2) +\sum_{j=1}^3 G_{x_j x_j}( x_j^{\prime \, 2} -  \dot{x}_j^2)\nonumber \\
    & \qquad \quad + G_{\theta\theta}(\theta^{\prime \, 2} -\dot{\theta}^2 ) + G_{\psi\psi}(\psi^{\prime \, 2}-\dot{\psi}^2 ) + \sum_{i=1}^3 G_{\phi_i \phi_i}(\phi_i^{\prime \, 2} - \dot{\phi}_i^2)\, .
\end{align}
Along with the Virasoro constraints,
\begin{align}
    &T_{\tau\tau} = T_{\sigma \sigma}=\frac{1}{2}\Bigg( G_{tt} (t^{\prime \, 2} + \dot{t}^2) + G_{rr}( r^{\prime \, 2} + \dot{r}^2 ) + \sum_{j=1}^3 G_{x_j x_j}( x_j^{\prime \, 2} +  \dot{x}_j^2) \nonumber \\
    & \qquad \qquad \qquad \quad + G_{\theta\theta} (\dot{\theta}^2 + \theta^{\prime \, 2} ) + G_{\psi\psi}( \dot{\psi}^2 +\psi^{\prime \, 2} ) + \sum_{i=1}^3 G_{\phi_i \phi_i}(\phi_i^{\prime \, 2} + \dot{\phi}_i^2) \Bigg) =0 \label{constraint1}\\
    &T_{\tau \sigma }= T_{\sigma \tau } = G_{tt}\dot{t}t^{\prime } + G_{rr}\dot{r}r^\prime  + \sum_{j=1}^3 G_{x_j x_j}\dot{x}_jx_j^\prime  +  G_{\theta\theta} \dot{\theta}  \theta^{\prime }  + G_{\psi\psi} \dot{\psi} \psi^{\prime }  + \sum_{i=1}^3 G_{\phi_i \phi_i} \dot{\phi}_i \phi_i^{\prime } =0.
\end{align}
The equations of motion in the general embedding are
\begin{align}
    &- \partial_\tau (G_{tt}\dot{t}) + \partial_\sigma ( G_{tt }t^{\prime} ) =0,\label{eqtt}\\
    &- 2 \partial_\tau (G_{rr}\partial_\tau r) + 2 \partial_\sigma (G_{rr} \partial_\sigma r) = \frac{\partial G_{tt}}{\partial r} \eta^{\alpha \beta }\partial_\alpha t \partial_\beta t +~~~ \frac{\partial G_{rr}}{\partial r} \eta^{\alpha \beta }\partial_\alpha \, r \partial_\beta \, r \nonumber\\
    & ~~+ \sum_{j=1}^3  \frac{\partial G_{x_j x_j}}{\partial r} \eta^{\alpha \beta }\partial_\alpha \, x_j \partial_\beta \, x_j       ,\label{eqrr}\\
    &- \partial_\tau (G_{x_j x_j}\dot{x}_j) + \partial_\sigma ( G_{x_j x_j }x_j^{\prime} ) =0, \label{eqxx}\\
    & -2 \partial_\tau (G_{\theta \theta} \partial_\tau \theta) + 2 \partial_\sigma (G_{\theta \theta }\partial_\sigma \theta ) = \frac{\partial G_{\psi \psi }}{\partial\theta} (\psi^{\prime \, 2} - \dot{\psi}^2) + \sum_{i=1}^3 \frac{\partial G_{\phi_i \phi_i}}{\partial \theta }(\phi_i^{\prime \, 2} - \dot{\phi}_i^2),\label{eqthetatheta}\\
    & -2 \partial_\tau (G_{\psi \psi} \partial_\tau \psi ) + 2 \partial_\sigma (G_{\psi \psi } \partial_\sigma \psi ) = \sum_{i=1}^3 \frac{\partial G_{\phi_i \phi_i}}{\partial \psi }(\phi_i'^2 -\dot{\phi}_i^2),\label{eqpsipsi}\\
    &-\partial_\tau (G_{\phi_i \phi_i } \partial_\tau \phi_i ) + \partial_\sigma (G_{\phi_i \phi_i } \partial_\sigma \phi_i) = 0\label{eqphiphi}
\end{align}
the components of the spacetime metric $G_{\mu\nu}$ are read from eq.(\ref{metricaads5s5}).
We now focus on a consistent truncation of the above embedding, leading to a solution in which the string propagates along the radial direction while wrapping the angular coordinates $\phi_i$,
\begin{align}\label{truncationpol}
    &t=t(\tau) , \quad r=r(\tau),\quad  x_j  =x_0 = \text{constant}, \quad \phi_1= \phi_1(\tau,\sigma), \quad \phi_2 = \phi_2(\tau,\sigma), \quad \phi_3= \phi_3(\tau,\sigma), \nonumber\\
    &\theta=\theta_0= \text{constant}, \qquad \psi=\psi_0 = \text{constant}.
\end{align}
Now the Polyakov Lagrangian simplifies to
\begin{equation}
    \mathcal{L}_P = -G_{tt} \dot{t}^2 - G_{rr}\dot{r}^2 + \sum_{i=1}^3 G_{\phi_i \phi_i}(\phi_i^{\prime \, 2} - \dot{\phi}_i^2)\, .
\end{equation}
The equations of motion in this truncated embedding are satisfied if the following equations hold,
\begin{align}
    &\ddot{r} - \frac{\dot{r}^2}{r} + \frac{r^3}{L^4} \dot{t}^2 =0 \label{reom},  \\
    &\dot{t} = \hat{\mathcal{E}} \frac{L^2}{r^2},\quad \hat{\cal E}=\text{const.}\label{hatE} \\
    & (\phi_1^{\prime \, 2} - \dot{\phi}_1^2)\sin^2\psi_0 + (\phi_2^{\prime \, 2} - \dot{\phi}_2^2)\cos^2 \psi_0 = \phi_3^{\prime \, 2} - \dot{\phi}_3^2 \\
    &\phi_1^{\prime \, 2} - \phi_2^{\prime \, 2} = \dot{\phi}_1^2 - \dot{\phi}_2^2, \qquad \qquad \phi_i^{\prime \prime} -\ddot{\phi}_i=0,  \qquad i=1,2,3.
\end{align}
The Virasoro constraints reduce to 
\begin{align}
    &T_{\tau \tau } = T_{\sigma \sigma} = G_{tt}\dot{t}^2 + G_{rr} \dot{r}^2 + \sum_{i=1}^3 G_{\phi_i \phi_i } (\dot{\phi}_i^2 + \phi_i^{\prime \, 2}) = 0\\
    &T_{\sigma \tau } = \sum_{i=1}^{3} G_{\phi_i \phi_i} \dot{\phi}_i \phi_i^{\prime} = 0.
\end{align}
The equations of motion for $\theta, \psi$ imply they can be taken as constants if $\dot{\phi}_i^2 = \phi_i^{\prime \, 2}$. Alternatively, $\dot{\phi}_i=0$ is also allowed, see \cite{Chatzis:2026ekd}.

For this truncation, the equations of motion allow for the ansatz $\phi_i = w_i \tau + m_i \sigma,$ in which case the Virasoro constraint from $T_{\tau\tau} =0$ gives
\begin{align}
    &\frac{r^2}{L^2} \dot{t}^2 - \Omega^2 = \frac{L^2}{r^2}\dot{r}^2 \label{constraint}\\
    &\Omega^2 = \mu_1^2 (  m_1^2 +  w_1^2)  + \mu_2^2 ( m_2^2 +  w_2^2) + \mu_3^2 (  m_3^2+ w_3^2 )
\end{align}
with the $\mu_i$ defined as in \eqref{mus}. Together with the Virasoro constraint \eqref{constraint}, one can obtain a constraint on the parameters $w_i,m_i$ given by
\begin{align}
    &m_3 w_3 \sin^2\theta + \cos^2 \theta (m_2 w_2 \cos^2\psi + m_1 w_1 \sin^2\psi) = 0\\
    &\Rightarrow \mu_1^2 \,  m_1 w_1 + \mu_2^2\,  m_2 w_2 + \mu_3^2\,  m_3 w_3 =0 .
\end{align}
Which means that $|m_i| =|w_i|$, but not such that all of $m_i,w_i$ are equal. Substituting the constraint \eqref{constraint} into \eqref{reom} gives the equation 
\begin{equation}
    \ddot{r}(\tau) + \frac{r(\tau)}{L^2} \Omega^2 = 0\, .
\end{equation}
Proceeding similarly, we consider a string which starts falling from the position $r=r_\text{UV}$ with zero initial velocity $\dot{r}(\tau=0) = 0$, the integration constants give the solution to the equations of motion as
\begin{equation}
    r(\tau) = r_{\text{UV}} \, \cos \left( \frac{\Omega}{L}\tau\right).
\end{equation}
Substituting the solution for $r(\tau)$ into the equation of motion for $t$ and then inverting gives the solution
\begin{align}
    &t(\tau) = \xi\,  \tan \left( \frac{\Omega}{L}\tau\right), \qquad \xi = \frac{\hat{\mathcal{E}} \, L^3}{r_{\text{UV}}^2\,  \Omega }\\
    &r(t) = \frac{r_{\text{UV}} \, \xi}{\sqrt{ t^2  + \xi^2  }}.
\end{align}

In this example we do not require treatment using the Routhian described in the previous sections, as it is possible to have vanishing initial velocities and we do not have a Noether conserved quantity which needs to be fixed. Our conserved quantity in this example is the winding number $m$, which is not associated to an initial velocity. Different string solutions to eqs.(\ref{constraint1})-(\ref{eqphiphi}) give rise to Noether conserved quantities and hence would then require the Routhian procedure.

We now study the same setup in the Nambu-Goto formalism. See \cite{Chatzis:2026ekd} for a similar example, with the string embedded in $\rm{AdS}_5 \times \mathrm{S}^5$ with $\phi_i = m_i \sigma$.

\subsection{Nambu-Goto action}

We begin by checking that the equations of motion are satisfied in the general embedding \eqref{generalembedding} above for the Nambu-Goto action, before truncating further. 
The Nambu-Goto action is given by
\begin{align}
    &S_{\mathrm{NG}} = - T_{F1} \int \mathrm{d}{\sigma}\mathrm{d}\tau\;  \sqrt{-\operatorname{det}g_{\text{ind}} }.
\end{align}
The induced metric in worldsheet coordinates,
\begin{align}
    \mathrm{d}s^2_{\text{ind}} = g_{\tau \tau} \mathrm{d}\tau^2 + g_{\sigma \sigma } \mathrm{d}\sigma^2 + 2\,  g_{\sigma \tau }\mathrm{d}\sigma \,\mathrm{d}\tau
\end{align}
where we used the same notation as above. We have
\begin{align}
    &g_{\tau\tau} = G_{\mu\nu}\partial_\tau X^\mu \partial_\tau X^\nu =  \frac{L^2 \dot{r}^2}{r^2} - \frac{r^2 \dot{t}^2}{L^2} + L^2 \cos^2\theta_0\,  \sin^2 \psi_0\, \dot{\phi}_1^2 + L^2 \cos^2 \theta_0 \, \cos^2\psi_0 \, \dot{\phi}_2^2+ L^2 \sin^2 \theta_0 \, \dot{\phi}_3^2 \\
    &g_{\sigma\tau} = g_{\tau\sigma} = \frac{L^2 r^{\prime} \dot{r}}{r^2}-\frac{r^2 t^{\prime} \dot{t}}{L^2}+ L^2  \, \Bigg( \cos^2 \theta_0 \, \sin^2 \psi_0 \,  \dot{\phi}_1 \phi_1^{\prime} + \cos^2\theta_0\, \cos^2\psi_0 \,  \dot{\phi}_2 \phi_2^{\prime}  + \sin^2 \theta_0 \, \,  \dot{\phi}_3 \phi_3^{\prime} \Bigg) \\
    &g_{\sigma\sigma} =\frac{L^2 r^{\prime 2}}{r^2}-\frac{r^2t^{\prime 2}}{L^2}+ L^2 \, \Bigg( \cos^2 \theta_0 \, \sin^2 \psi_0 \, \phi_1^{\prime \, 2} + \cos^2 \theta_0 \, \cos^2 \psi_0 \, \phi_2^{\prime \, 2} +  \sin^2 \theta_0 \, \phi_3^{\prime \, 2} \Bigg)
\end{align}

The $\theta$ and $\psi$ equations of motion are satisfied if $\dot{\phi}_i^2 = \phi_i^{\prime \, 2}$. 

We consider the following consistent truncation of the configuration \eqref{truncationpol}
\begin{equation}
\begin{split}
        &t=t(\tau),\quad r=r(\tau),\quad \phi_1 = \phi(\tau)+m \sigma,\quad \phi_2 = \phi(\tau) -m\sigma ,\quad \phi_3 =0,\\
        &\theta = \theta_0,\quad \psi_0=\frac{\pi}{4},\quad x_{i}=x_{i,0}\quad (i=1,2,3).
\end{split}
\end{equation}
This yields the following induced metric and Nambu-Goto action
\begin{equation}
    \dd s^2 _{\text{ind}} = L^2m^2 \cos^2\theta_0\dd \sigma^2 + \left[ \frac{L^2}{r(\tau)^2}\dot{r}(\tau)^2-\frac{r(\tau)^2}{L^2}\dot{t}(\tau)^2+L^2\cos^2\theta_0 \dot{\phi}(\tau)^2\right]\dd \tau^2,
\end{equation}
\begin{equation}
S_{\text{NG}}=T_{\text{F}1}mL L_\sigma \cos\theta_0  \int \dd \tau \sqrt{\frac{r(\tau)^2}{L^2}\dot{t}(\tau)^2 - \frac{L^2}{r(\tau)^2}\dot{r}(\tau)^2-L^2\cos^2\theta_0 \dot{\phi}(\tau)^2}.
\end{equation}
This is effectively describing the dynamics of a falling massive particle which (in addition to the one found in \cite{Chatzis:2026ekd}) has conserved angular momentum, with the winding number $m$ and the constant value of the angle $\theta_0$ contributing to the mass. 

The Lagrangian is
\begin{equation}
    {\cal L} =\widehat{\mu} ~\sqrt{\frac{r(\tau)^2}{L^2}\dot{t}(\tau)^2-\frac{L^2}{r(\tau)^2}\dot{r}(\tau)^2-L^2 \cos^2\theta_0 \dot{\phi}(\tau)^2},
\end{equation}
{where} ~$\widehat{\mu}=T_{\text{F}1}mL L_\sigma \cos\theta_0$. The conserved conjugate momentum of $\phi(\tau)$
\begin{equation}
    {\cal P}_\phi = \frac{\partial {\cal L}}{\partial \dot{\phi}}=-\frac{\widehat{\mu}L^2\cos^2\theta_0 \dot{\phi}}{\sqrt{\frac{r^2}{L^2}-\frac{L^2}{r^2}\dot{r}^2 -L^2 \cos^2\theta_0 \dot{\phi}^2}}\Rightarrow \dot{\phi}=\frac{{\cal P}_\phi \sec\theta_0}{L\sqrt{{\cal P}_\phi^2+\widehat{\mu}^2L^2\cos^2\theta_0}}\sqrt{\frac{r^2}{L^2}\dot{t}^2-\frac{L^2}{r^2}\dot{r}^2},
\end{equation}
to apply the Routhian prescription with respect to $\phi(\tau)$. The resulting expression takes the form
\begin{equation}
    {\cal R} = \frac{\partial {\cal L}}{\partial \dot{\phi}}\dot{\phi}-{\cal L} = -\sqrt{\widehat{\mu}^2+\frac{{\cal P}_\phi^2 \sec^2\theta_0}{L^2}}\, \sqrt{\frac{r^2}{L^2}\dot{t}^2-\frac{L^2\dot{r}^2}{r^2}},
\end{equation}
matching in a particular gauge the dynamics of the falling point particle described by \eqref{eq:Ruthian_point_particle}.





\section{Some  comments on field theory and complexity}\label{seccion6}

The purpose of this section is twofold.  We first extract concrete boundary
information from the bulk results obtained above. The short-time proper
momentum constrains the first Lanczos coefficients. The
$\mathrm{AdS}_{3}$ examples give conditional predictions for
correlators of the Krylov speed.  After this, we formulate a
field theory interpretation of the fixed charge reduction and of the
additional structure carried by branes and strings.  We separate 
the statements that follow directly from the
short-time expansions in Sections \ref{sec:lanczos-data} and \ref{sec:krylov-correlators} from the more
conjectural ones about organisation of collective, charge and fluctuation sectors in Sections \ref{sec:fixed-charge-boundary} and \ref{sec:collective-krylov}.

\subsection{From proper momentum to Lanczos data}\label{sec:lanczos-data}

Consider the hopping equation on the Krylov chain,
\begin{equation}\label{eq:krylov-hopping-section6}
 i\,\partial_t\phi_n(t)=a_n\phi_n(t)+b_{n+1}\phi_{n+1}(t)
 +b_n\phi_{n-1}(t),
 \qquad \phi_n(0)=\delta_{n0},
\end{equation}
where $a_n\in\mathbb{R}$, $b_n\geq0$ and $b_0=0$.  The short-time
expansion of spread complexity is \cite{Balasubramanian:2022tpr,
Balasubramanian:2025xkj,Caputa:2025ozd}
\begin{align}
 {\cal C}(t)
 &=b_1^2t^2+b_1^2\left[
 \frac{b_2^2}{6}-\frac{b_1^2}{3}
 -\frac{(a_1-a_0)^2}{12}\right]t^4+{\cal O}(t^6),
 \label{eq:C-short-section6}\\
 \dot{\cal C}(t)
 &=2b_1^2t+4b_1^2\left[
 \frac{b_2^2}{6}-\frac{b_1^2}{3}
 -\frac{(a_1-a_0)^2}{12}\right]t^3+{\cal O}(t^5).
 \label{eq:Cdot-short-section6}
\end{align}
It is convenient to introduce the combination
\begin{equation}\label{eq:Xi-section6}
 \Xi_2:=\frac{b_2^2}{6}-\frac{(a_1-a_0)^2}{12}.
\end{equation}
Suppose that, in a fixed normalisation of the bulk--boundary dictionary, the
proper-momentum expansion takes the form
\begin{equation}\label{eq:Py-general-section6}
 P_y(t)=p_1t+p_3t^3+{\cal O}(t^5),
 \qquad \dot{\cal C}(t)=\Lambda P_y(t)\,,\quad  \text{with}\quad \Lambda <0.
\end{equation}
Matching \eqref{eq:Py-general-section6} with
\eqref{eq:Cdot-short-section6} gives the model-independent relations
\begin{equation}\label{eq:matching-section6}
 b_1^2=\frac{\Lambda p_1}{2},
 \qquad
 \Xi_2=\frac{b_1^2}{3}+\frac{p_3}{2p_1}.
\end{equation}
The sign convention in $\Lambda P_y$ is chosen so that an infalling probe has
$\dot{\cal C}>0$.  Equation \eqref{eq:matching-section6} also makes precise
the information content of the bulk calculation.  The coefficient $p_1$
fixes $b_1^2$, which is the energy variance of the seed state, whereas $p_3$
fixes only the combination \eqref{eq:Xi-section6}.  Separating $b_2$ from
$a_1-a_0$ requires additional boundary information, for example the first
few moments of the survival amplitude \cite{Nandy:2024evd,Muck:2026top}.
Only when the relevant spectral measure is known to give $a_1=a_0$ may one
set $b_2^2=6\Xi_2$.


We now apply \eqref{eq:matching-section6} to three of the calculations in the
preceding sections.  For the point particle of
Section~\ref{sec:pointparticle}, the Routhian
\eqref{eq:Ruthian_point_particle} gives
$P_y=\frac{{\cal H}t}{L\, \hat m}$ exactly. Therefore
\begin{equation}\label{eq:lanczos-particle-section6}
 b_1^2=\frac{\Lambda{\cal H}}{2L\hat m},
 \qquad
 \Xi_2=\frac{\Lambda{\cal H}}{6L\hat m}.
\end{equation}
The absence of a cubic term in $P_y$ does not imply $b_2=0$. It imposes
the second relation in \eqref{eq:lanczos-particle-section6}.

For the detuned non-BPS D$p$ branes of
Section~\ref{section:detuning_mass_from_charge}, equation
\eqref{complexity-non-bps} yields
\begin{align}
 p_1&=(1-\alpha)(p+1)\nu
 \frac{r_{\rm UV}^{p+1}}{L^{p+2}},\nonumber\\[-1mm]
 p_3&=-\frac{p(p+1)^2(1-\alpha)^2\nu}{3}
 \frac{r_{\rm UV}^{p+3}}{L^{p+6}}.
\end{align}
Consequently,
\begin{align}
 b_1^2
 &=\frac{\Lambda}{2}(1-\alpha)(p+1)\nu
 \frac{r_{\rm UV}^{p+1}}{L^{p+2}},
 \label{eq:b1-nonbps-section6}\\
 \Xi_2
 &=\frac{\Lambda}{6}(1-\alpha)(p+1)\nu
 \frac{r_{\rm UV}^{p+1}}{L^{p+2}}
 -\frac{p(p+1)(1-\alpha)}{6}
 \frac{r_{\rm UV}^{2}}{L^{4}}.
 \label{eq:Xi-nonbps-section6}
\end{align}

In particular, $b_1$ vanishes in the BPS limit $\alpha\to1$, consistently
with the absence of radial motion and with the short-time evenness condition.

Finally, consider the internally rotating D$p$-branes of
Section~\ref{sec: internal space}.  Define
\begin{equation}
 {\cal Q}_p(\gamma):=(1+\gamma)(3+\gamma)
 +p(\gamma^2+\gamma-3).
\end{equation}
Using the short-time expansion following \eqref{proper0}, together with
$\tau=r_{\rm UV}t/L^2$, gives
\begin{align}
 p_1&=\frac{\gamma(\gamma+1-p)}{(\gamma+1)^2}
 \frac{r_{\rm UV}}{L^2},\nonumber\\
 p_3&=\frac{p\gamma^2(\gamma+1-p){\cal Q}_p(\gamma)}
 {6(\gamma+1)^6}
 \left(\frac{r_{\rm UV}}{L^2}\right)^3.
\end{align}
The corresponding Lanczos constraints are
\begin{align}
 b_1^2
 &=\frac{\Lambda\gamma(\gamma+1-p)}{2(\gamma+1)^2}
 \frac{r_{\rm UV}}{L^2},
 \label{eq:b1-rotating-section6}\\
 \Xi_2
 &=\frac{\Lambda\gamma(\gamma+1-p)}{6(\gamma+1)^2}
 \frac{r_{\rm UV}}{L^2}
 +\frac{p\gamma{\cal Q}_p(\gamma)}{12(\gamma+1)^4}
 \frac{r_{\rm UV}^2}{L^4}.
 \label{eq:Xi-rotating-section6}
\end{align}
The threshold $\gamma=p-1$ therefore has a direct Krylov interpretation:
at this point, $b_1=0$ and the seed does not leave the first Krylov site within
the semiclassical approximation.  The same matching procedure applies to the
magnetic and electric-displacement results \eqref{expPyB} and
\eqref{expE}, and to the fluctuation-corrected D0-brane result of
Section~\ref{sec:ABJM}.  In all cases, the Routhian prescription is what makes
$P_y$ odd at short times and hence permits a consistent identification with
\eqref{eq:Cdot-short-section6}.

\subsection{Krylov-speed correlators}\label{sec:krylov-correlators}

The preceding matching uses a one-point function.  Indeed, if
$N|K_n\rangle=n|K_n\rangle$ and
$N(t)=e^{iHt}Ne^{-iHt}$, then
\begin{equation}
 {\cal C}(t)=\langle K_0|N(t)|K_0\rangle,
 \qquad
 \dot{\cal C}(t)=\langle K_0|\dot N(t)|K_0\rangle,
 \qquad \dot N(t)=i[H,N(t)].
\end{equation}
Reference~\cite{Alfinito:2026vah} extended this statement to multi-time
Krylov correlators for the $\mathfrak{sl}(2,\mathbb R)$ chain associated with
an excited thermofield-double state in $\mathrm{AdS}_3/\mathrm{CFT}_2$.  In
the large-dimension, semiclassical regime, its two-speed commutator obeys
\begin{equation}\label{eq:speed-correlator-section6}
 \left\langle[\dot N(t_1),\dot N(t_2)]\right\rangle
 =\frac{i}{\varepsilon_{\rm UV}^{2}}
 \big[P_y(t_1)-P_y(t_2)\big]+{\cal O}(\Delta^{-2}),
 \qquad \varepsilon_{\rm UV}=\frac{L^2}{r_{\rm UV}}.
\end{equation}
The r.h.s. being imaginary, agrees with the antisymmetry of the hermitian conjugate of the commutator. Equation \eqref{eq:speed-correlator-section6} is not a universal identity for arbitrary probes: it relies on the
algebraic closure of the $\mathfrak{sl}(2,\mathbb R)$ model and on a specific
normalisation of the boundary state.  We may nevertheless use it as a
sharply stated ansatz for the collective sector of the top-down
$\mathrm{AdS}_3\times \mathrm{S}^3\times T^4$ examples in
Appendix~\ref{sec: AdS3 x S3 x T4 examples}.  The following equations should
therefore be read as predictions to be tested once the corresponding boundary
seeds are identified, rather than as consequences of the bulk trajectory
alone.

For the F1 string, let
$\Pi_{\rm F1}:=P_y/{\cal N}_{\rm F1}=(r_{\rm UV}/L)\widetilde P_y$.
Using \eqref{properF1ads3} in \eqref{eq:speed-correlator-section6} gives
\begin{align}
 \left\langle[\dot N(t_1),\dot N(t_2)]\right\rangle
 &\simeq 2i\frac{r_{\rm UV}^4}{L^7}(t_1-t_2)
 +{\cal O}(t_1^3,t_2^3,\Delta^{-2}),
 && t_1,t_2\to0,
 \label{eq:F1-correlator-early-section6}\\
 \left\langle[\dot N(t_1),\dot N(t_2)]\right\rangle
 &\simeq i\frac{r_{\rm UV}^4}{L^7}(t_1-t_2)
 +{\cal O}(t_1^{-3},t_2^{-3},\Delta^{-2}),
 && t_1,t_2\to\infty.
 \label{eq:F1-correlator-late-section6}
\end{align}
For the D1/D5 probe, one instead finds
\begin{align}
 \left\langle[\dot N(t_1),\dot N(t_2)]\right\rangle
 &\simeq i\frac{r_{\rm UV}^3}{L^6}
 \frac{\gamma^2}{(\gamma+1)^2}(t_1-t_2)
 +{\cal O}(t_1^3,t_2^3,\Delta^{-2}),
 && t_1,t_2\to0,
 \label{eq:D1-correlator-early-section6}\\
 \left\langle[\dot N(t_1),\dot N(t_2)]\right\rangle
 &\simeq i\frac{r_{\rm UV}^3}{L^6}
 \sqrt{\frac{\gamma}{\gamma+2}}(t_1-t_2)
 +{\cal O}(t_1^{-1},t_2^{-1},\Delta^{-2}),
 && t_1,t_2\to\infty.
 \label{eq:D1-correlator-late-section6}
\end{align}
The common linear late-time behaviour of several probes in
Section~\ref{sec: internal space} is consistent with the progressive loss of
probe-specific information in the radial collective motion.  By itself,
however, this similarity neither proves a universal $\mathrm{AdS}_2$
description nor establishes \eqref{eq:speed-correlator-section6} outside its
original $\mathfrak{sl}(2,\mathbb R)$ setting.

\subsection{Fixed-charge sectors and the boundary seed}\label{sec:fixed-charge-boundary}

We now turn from the explicit Krylov data to their field-theory
interpretation. We propose a way to understand the bulk Legendre transform in field theory.  In the construction of \cite{Caputa:2024sux}, the bulk point
particle is the semiclassical saddle associated with a heavy boundary
excitation.  In a top-down compactification, such a pointlike description
should be understood as a collective-coordinate limit of the corresponding
string or brane state, in which internal and worldvolume modes have been
frozen.  Restoring those modes changes the boundary seed and therefore changes
its spectral measure; it is not merely a kinematical correction to a fixed
Krylov chain \cite{Das:2024tnw,Li:2026comments}.

The role of the Routhian becomes particularly transparent in this language.
Let $Q$ be a conserved field-theory charge, $[H,Q]=0$, so that
\begin{equation}\label{eq:charge-decomposition-section6}
 {\cal H}_{\rm QFT}=\bigoplus_q{\cal H}_q,
 \qquad
 |\Psi_{0;q}\rangle=
 \frac{\Pi_q|\Psi_0\rangle}
 {\sqrt{\langle\Psi_0|\Pi_q|\Psi_0\rangle}},
\end{equation}
where $\Pi_q$ projects onto the sector of charge $q$.  The relevant survival
amplitude and moments are
\begin{equation}\label{eq:fixed-charge-survival-section6}
 S_q(t)=\langle\Psi_{0;q}|e^{-iHt}|\Psi_{0;q}\rangle,
 \qquad
 \mu_k^{(q)}=\langle\Psi_{0;q}|H^k|\Psi_{0;q}\rangle.
\end{equation}
They generate a charge-resolved Lanczos sequence and a charge-resolved spread
complexity \cite{Caputa:2025mii,Caputa:2025ozd}.  {\it The partial Legendre
transform in the bulk is the semiclassical counterpart of making this sector
choice before defining the collective observable}.  The charge is no longer a
velocity in the reduced phase space, but it remains present parametrically in
the effective mass, potential and hence in every moment in
\eqref{eq:fixed-charge-survival-section6}. Thus, the Routhian does not erase
charge dependence; it removes an inappropriate dynamical direction after the
charge has been fixed \footnote{As an aside, one may ask what is the holographic equivalent of the unresolved complexity, in particular wonder about its late time behaviour, as in \cite{Das:2026gko}.}.

This distinction organises the examples of the paper.  The angular momentum
${\cal J}$ in \eqref{psidotr}, the electric displacement ${\cal P}_E$ in
\eqref{E rdot}, the D0-brane momentum ${\cal P}_\chi$ in
\eqref{routh D0}, and the time-dependent string rotation of
Section~\ref{seccion5} are Noether charges.  They require a fixed-sector
Routhian such as \eqref{ruthian_brane}. By contrast, the detuning parameter
$\alpha$ in Section~\ref{section:detuning_mass_from_charge}, the constant
magnetic flux in Section~\ref{sec:Adding_a_gauge_field}, and the winding
numbers $m_i$ entering \eqref{constraint} label the probe or its topological
sector without being conjugate to a time-dependent cyclic coordinate.  They
modify the effective Hamiltonian but do not call for an additional partial
Legendre transform.  This is the boundary counterpart of the distinction
between selecting a value of the Noether charge $q$ (a superselection sector) and exciting a dynamical mode (like the winding $m_i$).

Worldvolume or internal fluctuations change the seed within a chosen charge
sector.  Schematically, a family of such seeds may be written as
\begin{equation}\label{eq:excited-seed-section6}
 |\Psi_{\boldsymbol n;q}\rangle=
 \frac{1}{\sqrt{{\cal Z}_{\boldsymbol n,q}}}\,
 \Pi_q\prod_\alpha(c_\alpha^\dagger)^{n_\alpha}|\Psi_0\rangle,
\end{equation}
where $c_\alpha^\dagger$ creates a normal-mode excitation and
${\cal Z}_{\boldsymbol n,q}$ normalises the state.  Its survival amplitude
$S_{\boldsymbol n,q}(t)$ defines a new spectral measure and, in general, a new
set of all the $a_n$ and $b_n$. The fluctuation correction to the D0-brane
proper momentum following \eqref{sol Z,Y,u} is a concrete bulk manifestation
of this fact: already the coefficient $p_1$, and hence the energy variance
$b_1^2$, depends on the fluctuation amplitude.

\subsection{Collective motion, fluctuations and multi-seed Krylov space}
\label{sec:collective-krylov}

It is important to distinguish two related statements.  For any single seed
and Hermitian Hamiltonian, the ordinary Lanczos algorithm always produces a
one-dimensional tridiagonal chain; internal structure is encoded in its
coefficients.  A higher-dimensional Krylov lattice becomes useful when one
retains several algebraic directions or several seeds before reducing to a
single path.  This is precisely the setting of the higher-dimensional Krylov
paths for semiclassical string states in \cite{Das:2024tnw} and of the
block-Lanczos, multi-seed construction in \cite{Craps:2024suj}.

With this qualification, the bulk calculations motivate the schematic
bookkeeping
\begin{equation}\label{diegoar}
 {\cal C}(t)={\cal C}_{\rm collective}(t)
 +{\cal C}_{\rm fluctuations}(t)
 +{\cal C}_{\rm charges}(t)
 +{\cal C}_{\rm mix}(t).
\end{equation}
Equation \eqref{diegoar} should not be regarded as an automatic additive
identity.  It becomes literal only when the effective Hilbert space, seed and
Krylov number operator factorise sufficiently well, for example when
$N\simeq N_{\rm collective}\otimes\mathbf 1+\mathbf 1\otimes
N_{\rm fluctuations}$.  Interactions between these factors define the mixed
term.  At strictly fixed charge, the charge is a sector label rather than an
independent dynamical coordinate; its contribution is most cleanly measured
by comparing ${\cal C}_q(t)$ between sectors, or by a symmetry-resolved
decomposition.  A separate ${\cal C}_{\rm charges}$ is therefore useful as
bookkeeping across sectors, but should not be double-counted within a single
fixed-$q$ chain.

This formulation embeds, rather than replaces, the proper-momentum proposal.
The reduced $P_y$ computed in Sections~\ref{sec:pointparticle}--\ref{seccion5}
captures the collective contribution in the appropriate fixed-charge sector.
A complete boundary calculation must additionally identify the seed, compute
\eqref{eq:fixed-charge-survival-section6}, and determine whether the remaining
modes factorise or require a block-Krylov description.  The proposal is then
testable: it must reproduce the neutral point-particle limit, match
\eqref{eq:matching-section6} from a boundary survival amplitude, distinguish
probes with identical radial motion but different charge or fluctuation data,
and suppress $b_1$ in the BPS limits found above.  Agreement of the conditional
correlator predictions \eqref{eq:F1-correlator-early-section6}--
\eqref{eq:D1-correlator-late-section6} would provide a substantially stronger
test, because it probes multi-time information not contained in the
one-point complexity alone.

\section{Conclusions and Future Directions}\label{section7}

In this work we have developed a fixed-charge formulation of holographic
Krylov spread complexity for probes whose dynamics contains cyclic degrees of
freedom.  The starting point was a simple consistency condition:\\
{\it for unitary evolution from a normalised state, spread complexity is an even
function of time and therefore begins as $\mathcal{C}(t)=b_1^2t^2+O(t^4)$}.\\  A bulk
prescription that predicts a non-vanishing complexity velocity at $t=0$ cannot
represent this evolution in the chosen sector.  We traced precisely this
problem to the use of an unreduced Lagrangian after the value of a conserved
Noether charge had already been fixed.  The appropriate object is instead the
Routhian.  It implements the partial Legendre transform and the reduction to
the fixed-charge phase space before the radial motion is solved.  Applied to
the proper-momentum proposal of \cite{Caputa:2024sux}, this replacement restores the short-time parity required by the
Krylov construction.

The same logic was then tested across a  broad collection of
holographic probes.  We first analysed a charged particle in
$\mathrm{AdS}_5\times \mathrm{S}^5$ and non-BPS branes for which the tension and Wess--Zumino
charge are detuned.  We next introduced internal angular momentum and
worldvolume gauge fields, separating a magnetic flux, which is a fixed
parameter of the effective problem, from an electric displacement, which is a
conserved canonical charge and hence requires a Routhian.  The D0-brane in
$\mathrm{AdS}_4\times\mathbb{CP}^3$ provided a non-trivial test with internal
motion, while the fundamental string examples showed that winding and a
time dependent internal rotation must  be treated differently. Indeed,
winding changes the effective mass, whereas a fixed angular momentum reduces
the phase space.  The additional examples in the appendices demonstrate that
this distinction persists for higher-dimensional branes and in the charged
Anabal\'on-Ross geometry. \\ In conclusion: {\it the Routhian is not an ad hoc repair, but the
mechanism by which the bulk observable is made compatible with the quantum
sector whose complexity it is meant to approximate}.

Finally, we translated the near-boundary probe data into information about the
Krylov chain.  The leading proper-momentum coefficient fixes $b_1$, while the
first correction determines the combination of $b_1$, $b_2$ and the diagonal
Lanczos coefficients that controls the $t^4$ term.  We also discussed Krylov
correlators and argued that an extended or charged probe should not in general
be represented by a single one-dimensional chain.  Its collective coordinate,
fixed charges and fluctuation modes naturally suggest the decomposition in
eq.~\eqref{diegoar}, and ultimately a multi-seed or block-Krylov description.
This statement also delimits the present result. In fact, the proper radial momentum
captures a controlled collective contribution, but a complete boundary
derivation must reconstruct the full spectral measure of the state or
operator dual to the probe.

\paragraph{Outlook.}
One immediate problem is therefore to perform the above mentioned reconstruction.  For
each particle, brane or string considered here, one should identify a boundary
seed with the same global charges, project it onto the corresponding
selection sector, and compute its survival amplitude.  The resulting
moments would determine the Lanczos coefficients independently of the bulk
trajectory and provide a sharp test of the proper-momentum dictionary beyond
$b_1$.  Worldvolume normal modes and Kaluza--Klein excitations supply a natural
family of seeds, so the multi-seed construction of \cite{Craps:2024suj}, the
planar semiclassical analysis of \cite{Das:2024tnw}, and symmetry-resolved
complexity \cite{Caputa:2025mii,Caputa:2025ozd,Balasubramanian:2024ghv} should fit together particularly
well.  A block-Lanczos recursion may then turn the schematic separation into
collective, charge, fluctuation and mixed terms into an operational boundary
definition.

An especially fertile laboratory is supplied by the Anabal\'on--Ross family.
The original solutions exhibit smooth supersymmetric solitons and competing
bulk saddles at fixed boundary data \cite{Anabalon:2021tua}; their gauged
$\mathcal N=8$ extensions and M-theory relatives enlarge the set of charges and
regular infrared endings \cite{Anabalon:2022aig,Anabalon:2024qhf}.  Related
$\mathcal N=4$ constructions interpolate among discrete spectra, continuum with
a gap and gapless continuum \cite{Anabalon:2024che}.  These are precisely the
data against which Krylov recurrence should be tested.   It is important not to identify oscillatory complexity with
confinement by itself, as confinement, screening and spectral discreteness are
logically distinct, see the paper \cite{Nunez:2026vhw}.  The useful question is  whether recurrence times
and their charge dependence can be derived from the glueball spectrum and the
infrared geometry.

There is already a top-down setting in which this programme can be made
quantitative.  The uplifted-soliton constructions of
\cite{Fatemiabhari:2024aua,Chatzis:2024top,Chatzis:2024kdu,Macpherson:2024qfi, Macpherson:2025pqi, Nunez:2023nnl, Nunez:2023xgl} connect conformal
theories to gapped or confining compactifications and make a large class of
Wilson loops, entanglement observables and central functions accessible.  The
subsequent universal-observable and supersymmetric-flow analyses
\cite{Chatzis:2025dnu,Chatzis:2025hek} provide families in which the ultraviolet
SCFT data factorise from the flow dependence.  It would be striking to learn
whether fixed-sector Lanczos coefficients, or suitable Krylov correlators,
obey an analogous factorisation.  The moduli-space construction of
\cite{Anabalon:2026moduli}, with several current sources and $U(1)^3$ charges,
offers a sharper test: the Routhian prescription predicts how to compare
complexity along the moduli space.  This could also be confronted with the proposed relations
between Krylov growth, confinement and universality
\cite{Fatemiabhari:2025usn,Fatemiabhari:2026goj} and between Krylov complexity
and holographic $c$-functions along RG flows \cite{Nunez:2026rgflow}.

A second direction is to go beyond the probe and collective-coordinate
approximations.  Quantising the quadratic brane and string fluctuations would
test whether the Routhian reduction commutes with the semiclassical expansion,
and would determine the mixed term in eq.~\eqref{diegoar}.  Including
backreaction would then reveal how the fixed-charge prescription is encoded in
the geometry itself, especially near soliton/black-hole saddle changes or
finite-temperature deconfinement.  Extended observables provide further
targets: Wilson and 't~Hooft loops, baryon vertices, giant gravitons and defect
operators all carry charges or worldvolume fluxes for which the distinction
established here is unavoidable.  Their charge-resolved Krylov evolution could
be compared with conventional tests of screening, confinement and stability,
and with the glueball towers of the same background. Studying the problems in Sections \ref{seccion3},\ref{seccion4},\ref{seccion5} in the presence of black holes, is also a natural problem. Similarly, studying the case of non-relativistic QFTs and their dual backgrounds, using the developments of \cite{Imani:2025etp} is a possible natural next step.

The broader prospect is that fixed-charge reduction may become a useful
organising principle for holographic complexity rather than a correction tied
to one observable.  The decisive next step is a boundary calculation that
simultaneously resolves charge sectors and spectral channels and reproduces
the bulk short-time coefficients.  Agreement would promote the proper-momentum
rule from a successful semiclassical diagnostic to a quantitative dictionary;
disagreement would be equally informative, because it would isolate the
fluctuation or mixing sectors that the collective trajectory leaves behind.

\section*{Acknowledgments} We want to thank various colleagues for their input that improved the contents and presentation of this work. In particular we thank: Ali Fatemiabhari, Horatiu Nastase, Dibakar Roychowdhury. D.C. has been supported by the STFC consolidated grand ST/Y509644-1. M.H. has been supported by the STFC consolidated grant ST/Y509644/1. 
CN is supported by  STFC’s grants UKRI4243, ST/Y509644- 1, ST/X000648/1 and ST/T000813/1. The work of R.T. has been supported by EPSRC Grant EP/Z535175/1 and STFC grant UKRI1787.
The work of AVR has received financial support from the Xunta de Galicia (CIGUS
Network of Research Centres and grant ED431C-2021/14), the European Union, the
Mar\'\i a de Maeztu grant CEX2023-001318-M funded by MICIU/AEI/10.13039/501100011033
and the Spanish Research State Agency (grant PID2023-152148NB-I00).


\appendix


\section{Example: Branes in $\rm{AdS}_{3} \times \mathrm{S}^{3} \times T^{4}$}\label{sec: AdS3 x S3 x T4 examples}

In this section, we consider an example of \ref{sec: internal space} in AdS$_{3} \times \mathrm{S}^{3} \times$ T$^{4}$. Consider the solution to Type IIB \cite{Thompson:2019ipl,Sfetsos:2010uq,Lozano:2019ywa} given by
\begin{equation}
    \dd s_{10}^{2} = \dd s^{2}(\text{AdS}_{3})+L^{2}\dd s^{2}(S^{3})+ M^{2} \dd s^{2}(\text{T}^{4})
\end{equation}
where\footnote{Note that \cite{Lozano:2019ywa, Lozano:2019emq, Lozano:2019zvg} is written in the democratic formalism, however we choose to use $F_3$ and implement $F_7 = -\star F_3$ along the conventions of \cite{Sfetsos:2010uq}.}
\begin{equation}
    \begin{split}
        & \dd s^{2}(\text{AdS}_{3}) = \frac{r^{2}}{L^{2}}(-\dd t^{2}+\dd x^{2}) + L^{2}\frac{\dd r^{2}}{r^{2}} \\
        & \dd s^{2}(S^{3}) = \frac{1}{4}(\omega_{1}^{2}+\omega_{2}^{2}+\omega_{3}^{2}) \\
        & \dd s^{2}(\text{T}^{4}) = \dd y_{1}^{2}+\dd y_{2}^{2}+ \dd y_{3}^{2}+ \dd y_{4}^{2}
    \end{split}
    \label{ads3}
\end{equation}
with the coordinates on the torus are defined that $y_{i} \sim y_{i}+2\pi, \; i=1,2,3,4$. The left-invariant SU(2) forms are given in \ref{omegas}.

The background has a 3-form given by
\begin{equation}
     F_{3} = \frac{2}{L} \Big( \text{vol}(S^{3})+\text{vol}(\text{AdS}_{3}) \Big) \, ,
    \label{3form}
\end{equation}
where $\text{vol}(\text{AdS}_{3}) = \frac{r}{L} \,  \dd t\wedge \dd x\wedge \dd r$.

\subsection{F1}\label{sec: F1 AdS3 x S3 x T4}

Firstly we consider an F1 fundamental string extended along the $[t, x]$ directions with profile in $r = r(t)$. This example follows the method as in \cite{Nastase:2026lhz}, as there is not an associated Noether charge we do not require the Routhian prescription here. We begin by writing the induced metric, 
\begin{equation}
    \dd s^2_{\text{ind}} =  \left(-\frac{r^{2}}{L^{2}}+\frac{L^{2}}{r^{2}}\dot{r}^{2}\right)\dd t^{2}+\frac{r^{2}}{L^{2}} \dd x^{2} \, .
\end{equation}
The action includes only the DBI term, as $H_3 = \dd B_2 =0$ in the background,
\begin{equation}
    S_{\text{F1}} = -T_{\text{F1}}L_{x} \int \dd t \, \frac{r}{L}\sqrt{\frac{r^{2}}{L^{2}}-\frac{L^{2}}{r^{2}}\dot{r}^{2}} \, .
\end{equation}
We define $\mathcal{N}_{\text{F1}} := T_{\text{F1}}L_{x}$ for the $\text{F1}$. This corresponds with the Lagrangian in \eqref{lagrangian alpha case} with $p = 1$ and $\alpha = 0$. This case is non-BPS as there is no WZ term in the action, and the string free falls radially. 


This case corresponds to equation \eqref{xdotalpha}, which be solved exactly
\begin{equation}
    \tau = \frac{\sqrt{\pi}\Gamma\left(\frac{3}{4}  \right)}{\Gamma\left(\frac{1}{4}  \right)}-\frac{1}{x} \, {}_2 F_{1}\left[-\frac{1}{4}, \frac{1}{2};\frac{3}{4}; x^{4}\right] \, .
\end{equation}
where $ {}_2 F_{1}$ is the hypergeometric function and we imposed $x( \tau = 0) = 1$ and $\dot{x}(\tau = 0) = 0$. This solution correspond with the series expansions 
\begin{equation}
    \begin{split}
        & x(\tau) \approx 1-\tau^{2} +\frac{11}{6}\tau^{4} \quad \text{for}\quad \tau \to 0 \, ,\\
        & x(\tau) \approx \frac{1}{\tau}-\frac{1}{6\tau^{5}} + \mathcal{O}\left( \frac{1}{\tau^{6}}\right) \quad \text{for}\quad \tau \to \infty \, . 
    \end{split}
    \label{expF1A}
\end{equation}
Numerical solutions for the trajectory are shown in Figure \ref{fig:Dp alpha=0.5}.

The proper momentum is given by \eqref{properalphadimless}. Using the expansions in \eqref{expF1A} we can analyse the asymptotics for the proper momentum
\begin{equation}
    \begin{split}
        & \tilde{P}_{y}(\tau) \approx 2\tau + \frac{4}{3}\tau^{3}+21 \tau^{5} + \mathcal{O}(\tau^{7})   \quad \text{for}\quad \tau \to 0 \, ,\\
        &\tilde{P}_{y}(\tau) \approx \tau- \frac{1}{3 \tau^{3}} + \frac{11}{27 \tau^{7}} +\mathcal{O}\left( \frac{1}{\tau^{8}}\right)   \quad \text{for}\quad \tau \to \infty \, . 
    \end{split}
    \label{properF1ads3}
\end{equation}
The numerical plots in Figure \ref{fig:Py alpha} match the latter asymptotic behaviour.





\subsection{D1/D5}

We begin by studying a D1 brane in the background described by \eqref{ads3}, with worldvolume extending in the $t,x$ directions. Additionally, it has a profile on $r = r(t)$ and $\psi = \psi(t)$, where $\psi$ is the equatorial angle of the sphere. Therefore, the induced metric is 
\begin{equation}
    \dd s^2_{\text{ind}} = \left(-\frac{r^{2}}{L^{2}}+\frac{L^{2}}{r^{2}}\dot{r}^{2}+L^{2}\dot{\psi}^{2}   \right)\dd t^{2}+ \frac{r^{2}}{L^{2}} \dd x^{2} \, .
\end{equation}
On the other hand, the background 3-form in \eqref{3form} can be obtained from a 2-form potential, such that its pull-back to the worldvolume of the D1 brane is given by
\begin{equation}
    P[C_{2}] = \frac{r^{2}}{L^{2}}  \,  \dd t\wedge \dd x \, .
\end{equation}
Therefore, the action is given by
\begin{align}
    S &= T_{\rm{D}1}\left(- \int \mathrm{d}^{2} \vec{x}\;  \sqrt{- \text{det}g_{\text{ind}}} \; +\;  \int \,P[C_{2}] \right) \\
    &= T_{\rm{D}1}\, L_{x} \int \mathrm{d}t \left(  - \frac{r}{L}\sqrt{\frac{r^{2}}{L^{2}}-\frac{L^{2}\dot{r}^{2}}{r^{2}}-L^{2}\dot{\psi}^{2}}+  \frac{r^{2}}{L^{2}} \right) \, . \,
\end{align}
Define $\nu = T_{\rm{D}1}\, L_{x_1} $ such that $[\nu] = \frac{1}{\text{length}}$, hence define the Lagrangian to be 
\begin{equation}
    \mathcal{L} = \nu  \left(-\frac{r}{L}\sqrt{\frac{r^{2}}{L^{2}}-\frac{L^{2}}{r^{2}}\dot{r}^{2}-L^{2}\dot{\psi}^{2}}+ \frac{r^{2}}{L^{2}}\right) \, .
    \label{lagraniand1}
\end{equation}
Notice that this Lagrangian is the same as \eqref{lagrangianJ} for $p = 1$. \\
Before studying the complexity for the D1 brane, we first show that a probe D5 brane with the same profile $r = r(t)$, $\psi = \psi(t)$, and whose worldvolume is localized  as D5$[t,x,\text{T}^{4}]$ has the same (up to a constant) Lagrangian as \eqref{lagraniand1}.  The induced metric for this D5 probe is given by
\begin{equation}
    \dd s^2_{\text{ind}} = \left[\left(-\frac{r^{2}}{L^{2}}+\frac{L^{2}}{r^{2}}\dot{r}^{2}+L^{2}\dot{\psi}^{2}\right)\dd t^{2}+\frac{r^{2}}{L^{2}} \dd x^{2}  \right]+M^{2}(\dd y_{1}^{2}+ \dd y_{2}^{2}+ \dd y_{3}^{2}+ \dd y_{4}^{2}) \, .
\end{equation}
Notice that $F_3$ in \eqref{3form} contains a $\rm{vol}(\mathrm{S}^3)$ term, meaning that $\star F_{3} =- F_7$ has a $\rm{vol}(\rm{AdS}_3) \wedge \rm{vol}(T^4)$ term. This implies there is a $C_6$ which will couple with this D5 brane. The pull-back of such 6-form to the worldvolume is given by
\begin{equation}
    P[C_{6}] = \frac{ M^{4} r^{2}}{L^{2}} \,   \dd t\wedge \dd x \wedge  \dd y_{1}\wedge \dd y_{2} \wedge \dd y_{3}\wedge \dd y_{4} \, ,  
\end{equation}
which contributes to the corresponding Wess-Zumino term. As a consequence, the action is 
\begin{align}
    S &= T_{\rm{D}5}\left(- \int \mathrm{d}^{6} \vec{x}\;  \sqrt{- \text{det}g_{\text{ind}}} \; +\;  \int \,P[C_{6}] \right) \\
    &= (2\pi M)^{4} T_{\rm{D}5}\, L_{x} \int \mathrm{d}t \left(  - \frac{r}{L}\sqrt{\frac{r^{2}}{L^{2}}-\frac{L^{2}\dot{r}^{2}}{r^{2}}-L^{2}\dot{\psi}^{2}}+  \frac{r^{2}}{L^{2}} \right) \, . \,
\end{align}
After defining $\kappa =  (2\pi M)^{4} T_{\rm{D}5}\, L_{x}$, such that $[\kappa] = \frac{1}{\text{length}} $. Then the corresponding Lagrangian is  
\begin{equation}
    \mathcal{L} = \kappa  \left(-\frac{r}{L}\sqrt{\frac{r^{2}}{L^{2}}-\frac{L^{2}}{r^{2}}\dot{r}^{2}-L^{2}\dot{\psi}^{2}}+ \frac{r^{2}}{L^{2}}\right) \, .
\end{equation}
This Lagrangian is the same as in \eqref{lagraniand1} which corresponds effectively to equation \eqref{lagrangianJ} with $p = 1$, as from the field theory perspective it is a two dimensional object. \\
For the rest of the section and without lost of generality we study the Lagrangian in \eqref{lagraniand1} which can correspond to a D1/D5 brane. This case contains a conserved charge associated to the motion in $\psi$, hence the Routhian prescription is implemented.

For this setup, equation \eqref{xdotj} for $p = 1$ can be solved, leading to
\begin{equation}
    \tau(x) = \frac{\pi}{2\gamma}-\frac{\sqrt{1-x^{2}}}{x}-\frac{2}{\gamma}\arctan\left( \frac{1+\sqrt{1-x^{2}}}{x}  \right) \, .
\end{equation}
We can expand this solution near $x \sim 1 $ obtaining
\begin{equation}
    \tau(x) \approx \frac{\gamma+1}{\gamma}\sqrt{2(1-x)}+\mathcal{O}\left[(x-1)^{3/2}\right]\,, \quad x \to 1 \, ,
\end{equation}
which can be inverted to give the expressions in \eqref{expxJ0} and \eqref{expxJI}. By defining the proper coordinate as in \eqref{properyJ} then, the proper momentum is given by the expression in \eqref{proper0} whose asymptotic behaviour is given by \eqref{expPyJ0} and \eqref{expPyJI} for $p = 1$. Numerical plots for this case are shown in Figures \ref{Veff} and \ref{xtpyJ}.


\section{Example: Branes in {$\rm{AdS}_5\times \mathrm{S}^5$}}\label{sec: branes AdS5xS5}

\subsection{D5}\label{sec: D5 AdS5xS5}

We now specialise the general discussion to explicit probes in $\rm{AdS}_5 \times \mathrm{S}^5$. In this example, we consider a D5-brane wrapping the directions $(t,x_1,x_2)$ in $\rm{AdS}_5$ and an $\mathrm{S}^3$ inside $\mathrm{S}^5$. In addition, we allow the transverse radial position and two transverse internal coordinates to depend on time, $r = r(t)$, $\theta= \theta(t)$ and $\phi = \phi(t)$. After deriving the general reduced DBI system, we will consistently truncate to the fixed-charge sector $\mathcal{J} = 0$ with $\theta = 0$ and $\dot{\phi}= 0$. In this sector the radial dynamics coincides with the $p = 2$, $\alpha= 0$ case of Section \ref{section:detuning_mass_from_charge}, with the volume of the wrapped $\mathrm{S}^3$ absorbed into the effective normalisation ${\cal N}$.

We use the metric in terms of the left invariant $SU(2)$ forms $\omega_i$ defined in \eqref{omegas}, and from this we write the induced metric as
\begin{align}
   \mathrm{d}s^2_{\text{ind}} =  \frac{r^2}{L^2} (\mathrm{d} x_1^2 + \mathrm{d}x_2^2) + \left(-\frac{r^2}{L^2} + \frac{L^2 \dot{r}^2}{r^2} + L^2 \dot{\theta}^2 + L^2 \sin^2 \theta \dot{\phi}^2\,  \right)\, \mathrm{d}t^2 + L^2 \cos^2\theta\, \mathrm{d}\Omega_3^2\,,
\end{align}
where $\mathrm{d} \Omega_3^2$ is the metric of the unit 3-sphere given by
\begin{align}
    \mathrm{d}\Omega_3^2 = \frac{1}{4}(\omega_1^2 + \omega_2^2 + \omega_3^2).
\end{align}
The D5 brane is non-BPS as here we have $S_{\text{WZ}}=0$, which means we consider only the DBI term of the action,
\begin{align}
    &S_{\text{DBI}} = - \mathcal{N} \int \mathrm{d} t \cos^3\theta\; r^2 \sqrt{\Delta} \\
    &\mathcal{N} = T_{\text{D5}}\,  L_{x_1} L_{x_2} \, L \;  \mathrm{vol}\Omega_3  , \quad \Delta = \frac{r^2}{L^2} - \frac{L^2 \dot{r}^2}{r^2} - L^2 \dot{\theta}^2 - L^2 \sin^2 \theta \dot{\phi}^2\,.
\end{align}
We have the Lagrangian and Hamiltonian given by
\begin{align}\label{L_and_H_of_D5}
    &\mathcal{L} = - \mathcal{N} \cos^3 \theta \; r^2 \sqrt{\Delta}\\
    &{\cal H} = P_r \dot{r} + P_\theta \dot{\theta} + P_\phi \dot{\phi} - \mathcal{L} =  \frac{\mathcal{N}\cos^3 \theta \; r^4}{L^2 \sqrt{\Delta}}\, .
\end{align}
The coordinates $(t,x_1,x_2,r,y)$ and the parameters $L_{x_1}, L_{x_2}, L,r_{\text{UV}}$ have units of length. The angles, $\Delta$ and complexity are dimensionless. $T_{D5}$ has units length$^{-6}$. $\mathcal{N}$ has units length$^{-3}$. $\mathcal{L}, {\cal H},{\cal R}$ and the momenta have units of inverse length. 

The momenta are defined as
\begin{align}
    &P_\phi = \mathcal{J} = -\frac{L^2 \mathcal{N} \dot{\phi}\, r^2 \sin ^2\theta \cos^3\theta}{\sqrt{\Delta}}\,,\quad P_r = - \frac{L^2   \mathcal{N}\dot{r}\, \cos^3\theta }{\sqrt{\Delta}}\\
    &P_\theta = - \frac{L^2 \mathcal{N} r^2 \dot{\theta} \cos^3 \theta}{\sqrt{\Delta}}\nonumber
\end{align}
with
\begin{align}
    &\dot{\phi} = - \frac{\mathcal{J} \, r^2}{H \, L^4 \sin^2 \theta} = \pm \frac{\mathcal{J} \, \sqrt{r^4 - L^4 \dot{r}^2 - r^2 L^4 \, \dot{\theta}^2}}{L^2 \, r \sin\theta \, \sqrt{L^2 \cos^6 \theta \sin^2 \theta \, \mathcal{N}^2 \,r^4 + \mathcal{J}^2}}\, .
\end{align}
We utilise the Routhian as we have a conserved charge $\mathcal{J}$ associated to the coordinate $\phi$, Legendre transforming to the Routhian allows us to eliminate $\dot{\phi}$ and restrict to a fixed charge plane. We write the Routhian as
\begin{align}
    &\mathcal{R} = \mathcal{J} \dot{\phi} - \mathcal{L} =  M(r,\theta) \sqrt{\frac{r^2}{L^2} - \frac{L^2 \dot{r}^2}{r^2} - L^2 \dot{\theta}^2}\\
    & M(r,\theta) =- \frac{\sqrt{\mathcal{J}^2 + L^2 \mathcal{N}^2 \cos^6\theta \, \sin^2 \theta \, r^4}}{L  \sin\theta} \, .
\end{align}

Since $\phi$ is cyclic, $P_\phi ={\cal J}$ is conserved. The Routhian implements the fixed-${\cal J}$ description. For non-zero ${\cal J}$, the coordinate patch is singular at $\theta=0$ as the $\phi$-circle shrinks at this point. In the sector used below we first set ${\cal J}=0$ and only then can we take the consistent truncation $\theta=\dot{\theta}=\dot{\phi}=0$. In this order, the Routhian has the regular limit $M(r, \theta)\to -{\cal N} r^2$ and the dynamics is equivalent, up to the overall sign convention, to the radial DBI Lagrangian. The truncated Lagrangian and Hamiltonian are then given by
\begin{align}
    \mathcal{L} = \mathcal{N} r^2 \sqrt{ \frac{r^2}{L^2}- \frac{L^2 \dot{r}^2}{r^2}}, \qquad  \qquad {\cal H} =  \frac{\mathcal{N}r^4}{L^2 \sqrt{ \frac{r^2}{L^2}- \frac{L^2 \dot{r}^2}{r^2}}}\, .
\end{align}
This leads to, where we take the negative sign solution such that the brane falls
\begin{align}
    \dot{r} = - \frac{r^2}{L^3 {\cal H} }\sqrt{L^2 {\cal H}^2 - \mathcal{N}^2 r^6},
\end{align}
with $r_{\text{UV}}^3 = \frac{L \, {\cal H} }{\mathcal{N}}$. We can analytically solve the above equation for $t(r)$ as

\begin{align}\label{t(r) D5}
   & \frac{1}{r} \; {}_2 F_1\left(-\frac{1}{6},\frac{1}{2}; \frac{5}{6}; \frac{r^6}{r_{\text{UV}}^6}\right) - \frac{1}{r_{\text{UV}}} \; {}_2 F_1\left(-\frac{1}{6},\frac{1}{2};\frac{5}{6}; 1 \right) = \frac{t}{L^2}\,\\
   &\Rightarrow t(r) = \frac{L^2}{r} {}_2 F_1\left(-\frac{1}{6},\frac{1}{2};\frac{5}{6}; \frac{r^6}{r_{\text{UV}}^6}\right) - \frac{L^2}{r_{\text{UV}}}  \frac{\sqrt{\pi} \; \Gamma\left( \frac{5}{6}\right)}{ \Gamma\left(\frac{1}{3} \right)},\label{lionel}
\end{align}
where ${}_2 F_1(a,b;c;z)$ is the hypergeometric function.
The proper coordinate is defined as
\begin{equation}\label{D5 y}
    \mathrm{d}y^{2} = \frac{L^{2}}{r^{2}} \mathrm{d}r^{2} , 
\end{equation}
which gives the proper momentum and following the prescription of \cite{Caputa:2024sux} the time derivative of the complexity,
\begin{align}
    P_{y} &= P_{r} \frac{\partial \dot{r}}{\partial \dot{y}} = -\frac{1}{r} \sqrt{L^2 {\cal H}^2- {\cal N}^2 r^6}\\
    &\dot{\mathcal{C}}(t) \sim -P_y \, .\label{diegoa}
\end{align}

\begin{figure}[t!]
    \centering
    \includegraphics[width=0.5\linewidth]{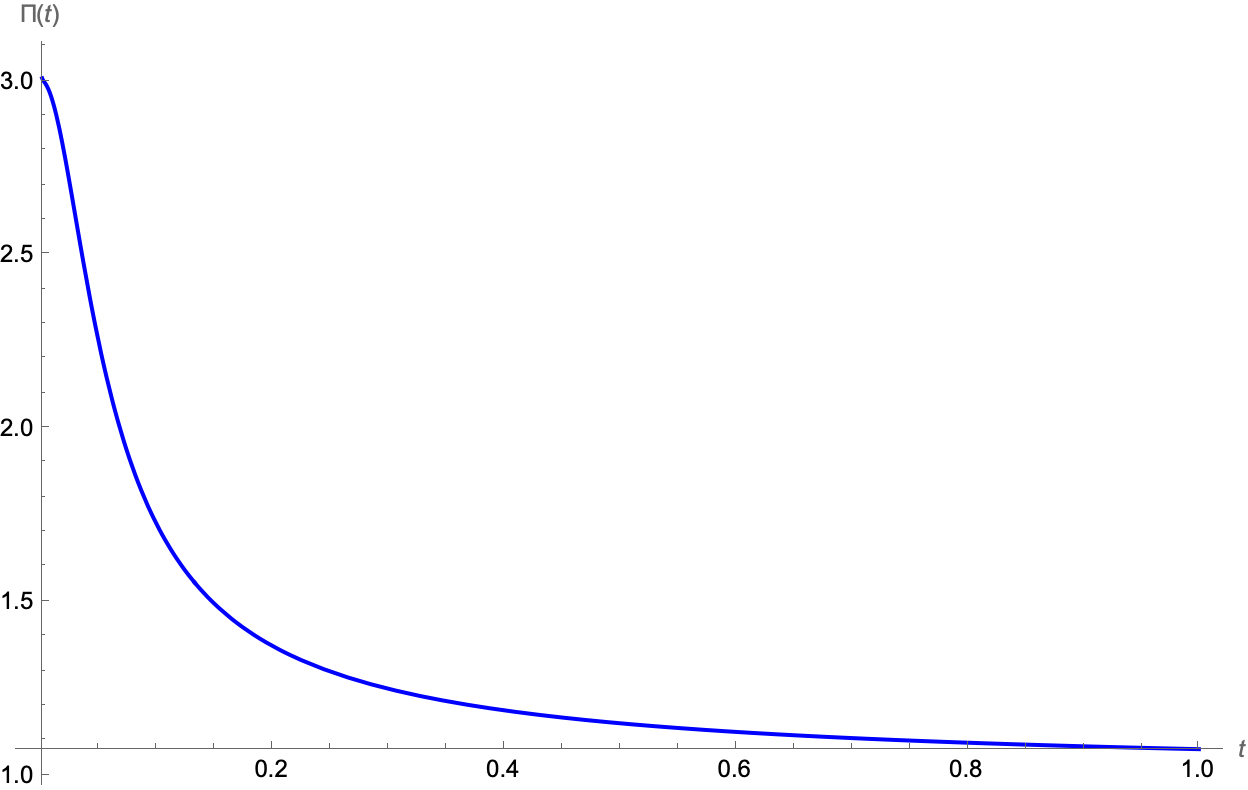}
    \caption{Plot of $\Pi(t) = \frac{\mathcal{\dot{C}} (t)}{\mathcal{H}t}$, the ratio of the first time derivatives of the complexity of the D5 to the particle. The D5 complexity is truncated in $\mathcal{J}=0, \; \theta=0, \; \dot{\psi}=0.$ Where $\mathcal{\dot{C}}_{\text{particle}}=\mathcal{H}t $ is the time derivative of the complexity of the particle. We see that initially the complexity of the D5 grows faster than in the particle case but then the ratio plateaus such that the complexity grows at the same rate for larger $t$.}
    \label{fig:d5 ratio}
\end{figure}

It is now within this truncation only that we can match with the case of $p=2$ of Section \ref{section:detuning_mass_from_charge}, as we have the D5 brane extending in two spatial AdS coordinates, $x_1,x_2$. We find the same Lagrangian as \eqref{lagrangian alpha case} with $\alpha=0.$ 

Which then leads to the same small time expansion when expanding \eqref{t(r) D5} close to $r\sim r_{\text{UV}}$, 
\begin{align}\label{small t D5}
   &\frac{ r_{\text{UV}} \, t}{L^2} \approx\sqrt{\frac{2}{3}} \sqrt{1-\frac{r}{r_{\text{UV}}}} - \sqrt{\pi}\; \frac{\Gamma\left( \frac{5}{6}\right) }{\Gamma \left( \frac{1}{3}\right)}\left( 1- \frac{r}{r_{\text{UV}}}\right) + \mathcal{O}\left( \left( 1-\frac{r}{r_{\text{UV}}}\right)^{\frac{3}{2}}\right)\\
   &r \approx r_{\text{UV}} \left( 1 - \frac{ 3 \, r_{\text{UV}}^2 }{2 \, L^4}t^2 + \frac{39 r_{\text{UV}}^4}{8 L^8}t^4 \right) \qquad \text{ as } \; t\rightarrow 0,
\end{align}
In this case, for small times we find the expansion of the complexity to be
\begin{equation}
    \dot{\mathcal{C}} \sim  |P_y | \sim -\frac{8L^8 \sqrt{L^2 {\cal H}^2-\mathcal{N}^2 \left( r_{\text{UV}}- \frac{ 3 \, r_{\text{UV}}^3 }{2 \, L^4}t^2 + \frac{39 r_{\text{UV}}^5}{8 L^8}t^4 \right)^6}}{8L^8 r_{\text{UV}} -12 L^4r_{\text{UV}}^3 t^2 + 39 r_{\text{UV}}^5 t^4}\approx \frac{3 {\cal H}}{L} t - \frac{6 {\cal H}^{\frac{5}{3}}}{ L^{\frac{13}{3}} \mathcal{N}^{\frac{2}{3}}} t^3 +\ldots 
\end{equation}
\\
Now, expanding around $\frac{r}{r_{\text{UV}}} \sim 0^+$,

\begin{align}
    & \frac{r_{\text{UV}}}{L^2}(t+t_0) \approx  \frac{r_{\text{UV}}}{r}  - \frac{1}{10} \left( \frac{r}{r_{\text{UV}}}\right)^5 - \frac{3}{88} \left( \frac{r}{r_{\text{UV}}}\right)^{11} + \ldots  \\
    &t_0 = \frac{L^2}{r_{\text{UV}}}\sqrt{\pi} \frac{\Gamma\left( \frac{5}{6}\right)}{\Gamma\left( \frac{1}{3}\right)}\, .
\end{align}

Inverting, for $t \gg t_0$, 
\begin{equation*}
  r(t) \approx \frac{L^2}{t+t_0} - \frac{L^{14}}{10 r_{\text{UV}}^6 (t+t_0)^7} + \mathcal{O}\left( \frac{1}{(t+t_0)^8} \right)\, .
\end{equation*}
From this we obtain
\begin{equation}
\dot{\mathcal{C}}\sim |P_y| \sim \frac{{\cal H}}{L}(t+t_0)  - \frac{2 {\cal H} L^{11}}{ 5 \, r_{\text{UV}}^6 (t+t_0)^5} + \ldots    
\end{equation}

for late times, the complexity is quadratic in $t$, consistent with the proposal in \cite{Caputa:2024sux}.

\subsection{D3}\label{sec:Excited D3}

The next example we will consider is a D3 brane extended on D3$[t, x_{1}, x_{2}, x_{3}]$ in AdS$_{5} \times \mathrm{S}^{5}$, with a profile in $r = r(t)$ and $\psi = \psi(t)$. Since there is a conserved charge, This example  falls in that of Section \ref{sec: internal space}, for $p=3$.
The action which includes a DBI and WZ term is written as
\begin{equation}
    S = \nu \int \dd t \, \left(-\sqrt{\left(\frac{r^{2}}{L^{2}} -\frac{L^{2}}{r^{2}}\dot{r}^{2}-L^{2}\dot{\psi}^{2}\right)\frac{r^{6}}{L^{6}}}+ \frac{r^{4}}{L^{4}}\right) \, .
\end{equation}

We can find an exact solution to the equation \eqref{xdotj} for $ p = 3$, subject to the initial condition $x(\tau=0)=1$, where $\tau_0$ is the integration constant. We write this solution in terms of the elliptic functions of the first kind (denoted EllipticF) and second kind (denoted EllipticE),
\begin{eqnarray}
& & \tau=-\int \dd x \frac{x^4+\gamma}{x^2\sqrt{-\gamma(\gamma+2)x^2 +\gamma^2+2\gamma x^4}} \nonumber\\
& & =\frac{\sqrt{\gamma(\gamma-2 x^2)(1-x^2)}}{\gamma~ x}+\frac{3}{2}\left(\text{EllipticE}[\arcsin x,\frac{2}{\gamma}] -\text{EllipticF}[\arcsin x,\frac{2}{\gamma}] \right)   + \tau_0.
\end{eqnarray}
This solution can be inverted to get \eqref{expxJ0} and \eqref{expxJI} for $p = 3$. Furthermore, the proper momentum is defined from the corresponding Routhian in \eqref{ruthian_brane} and is given by equation \eqref{proper0}. Numerical plots for this setup are shown in Figures \ref{Veff} and \ref{xtpyJ}.

\section{Example: D2 brane in ABJM}\label{sec: D2 ABJM}


Let us now consider a D2 brane in the background \eqref{AdS4xCP3} which is extended in $(t,x_1,x_2)$ and falls radially, $r = r (t)$. This is the $p=2$ example of Section \ref{section:detuning_mass_from_charge}. 

This object will couple to the three-form RR potential $C_3$ written in \eqref{RR_potetials}. The induced metric on the brane reads
\begin{equation}
    \dd s^2_{\text{ind}}=L^2_{\text{ABJM}} \left[ \frac{r^2}{L^2}(-\dd t^2+\dd x_1^2+\dd x_2^2)+\frac{L^2 \dot{r}^2 }{r^2}\dd t^2\right]
\end{equation}
and the Lagrangian density 
\begin{equation}\label{Lagrangian_D2}
    {\cal L}_{\text{D2}} = T_{\text{D2}}\Big(-\sqrt{-e^{-2\Phi}\text{det}(g_{\text{ind}})} + \mu _2P[C_3]\Big)= \nu \left( -\frac{r^{2}}{L^{2}}\sqrt{\frac{r^{2}}{L^{2}}-\frac{L^{2}\dot{r}^2}{r^{2}}}+\mu _2 \, \frac{r^{3}}{L^{3}}\right),
\end{equation}

where $P[C_3]$ is the pull-back of the three-form on the worldvolume of the D2 and $\mu_2$ is a parameter describing the charge of the brane under the WZ term, we also defined $\nu = T_{\text{D2}}\frac{k}{2}L_{\text{ABJM}}^2$. The system is BPS exactly when $\mu_2=1$ and therefore can fall for any value of $\mu_2<1$.

The corresponding equation is given by \eqref{xdotalpha} for $p = 2$ and $ \alpha = \mu_{2}$. We obtain an analytic solution for the function $t(x)$ in terms of $F_1(a;b_1,b_2;c;x,y)$, the Appell hypergeometric function of two variables
\begin{equation}
\begin{split}
 & \qquad  \tau(x)-\tau_0 = {\cal Y}(x),
      \end{split}
\end{equation}
where
\begin{equation}
  \begin{split}
        {\cal Y}(x)&= \frac{x^2 \mu_2}{\sqrt{\mu_2-1}}F_1 \left(\frac{2}{3}; \frac{1}{2},\frac{1}{2};\frac{5}{3};x^3,\frac{\mu_2+1}{\mu_2-1}x^3\right) - \frac{2 (\mu_2+1) x^5}{5\sqrt{\mu_2-1}}F_1 \left(\frac{5}{3};\frac{1}{2},\frac{1}{2};\frac{8}{3};x^3, \frac{\mu_2+1}{\mu_2-1}x^3 \right)\\
        &+\frac{\sqrt{ x^6-1 + (x^3 -1)^2 \mu_2}}{x}.
  \end{split}
\end{equation}

The constant $\tau_0$ is found from the initial condition $x(t=0)=1$. In terms of the hypergeometric function $_{2} F_1$
\begin{equation}
    \tau_0 = \frac{\sqrt{\pi}\,\Gamma\left( \frac{5}{3}\right)\left[ -7 \mu_2 \,_2 F_1\left( \frac{1}{2},\frac{2}{3};\frac{7}{6};\frac{1+\mu_2}{1-\mu_2}\right)+4(1+\mu_2)\, _2 F_1 \left(\frac{1}{2},\frac{5}{3};\frac{13}{6};\frac{1+\mu_2}{1-\mu_2} \right)\right]}{6\sqrt{\mu_2-1} \,  \, \Gamma\left( \frac{13}{6}\right)}.
\end{equation}

We obtain the late time expansion by inverting $t(x)$ near $x=0$, to get the following behaviour for the function $r(t)$ in series, which are in agreement with the early time expansion of \eqref{ralphaexp}, for the case in which $p=2$, and $\alpha= \mu_2$. Moreover, the proper momentum is given by \eqref{properalphadimless} and the corresponding expansions in \eqref{complexity-non-bps} and \eqref{pyalphalate}. Numerical solutions for this case are shown in Figures \ref{fig:Dp alpha=0.5} and \ref{fig:Py alpha}.

\section{Equations of motion for worldvolume gauge field}\label{sec:appendix EoM}

In this section we check the equations of motion in Section \ref{sec:Adding_a_gauge_field}. We consider a D3 brane with electric and magnetic worldvolume gauge field components. We begin by writing the DBI action for the D3 brane, in string frame units
\begin{equation}
\mathcal{L}_{\text{DBI}} =- T_{D{3}}\sqrt{-\det(g_{\text{ind}} + F)}
\end{equation}
where we absorbed the factors $e^{-\Phi_{0}}(2\pi\alpha')$ into the definition of the gauge field $F$ and the tension $T_{D3}$. We define the quantity $M =g_{\text{ind}}+F$. Then,  varying $A_{a} \to A_{a}+\delta A_{a}$, it follows that
\begin{equation}
    \delta \sqrt{-\det M} =\underbrace{ -\frac{1}{2}\sqrt{-\det M}(M^{-1})^{a b}}_{S^{ab}}\delta F_{a b} 
\end{equation}
gives the variation of the action as
\begin{equation}
    \begin{split}
        \delta S_{\text{DBI}} &=  T_{D{3}}\int \dd^{4}x \, S^{ab}(\partial_{a}\delta A_{b}-\partial_{b}\delta A_{a})  \\
        & =T_{D{3}}\int 
        \dd^{4}x \, \partial_{a}(S^{ba}-S^{ab})\delta A_{b} + \text{boundary terms}  \, ,
    \end{split}
\end{equation}
then, the corresponding equations of motion with anti-symmetrised indices are given by 
\begin{equation}
    \begin{split}
        & \partial_{a}\left( \sqrt{-\det M}(M^{-1})^{[a b]} \right) = 0 \, , \\
        & \partial_{[a}F_{bc]}=0 \, .
    \end{split}
\end{equation}
Where for the choice of $F$ in section \ref{sec:Adding_a_gauge_field}, we have
\begin{equation}
    -\det M = \left(\frac{r^{4}}{L^{4}} -\dot{r}^{2}  \right)\left(\frac{r^{4}}{L^{4}} +B^{2}\right)-\frac{r^{4}}{L^{4}} E^{2} \equiv \Theta \, .
\end{equation}
Define $\Pi^{ab} =\sqrt{-\det M}(M^{-1})^{[a b]} $, then the terms that contribute to the equations of motion are 
\begin{equation}
    \Pi^{tx} = \frac{r^{4}}{L^{4}}\frac{E}{\sqrt{\Theta}} \, , \quad 
     \Pi^{xz} = -\frac{B}{\sqrt{\Theta}}\left(\frac{r^{4}}{L^{4}}-\dot{r}^{2}\right) \, .
\end{equation}
Then, the non-trivial equations of motion are 
\begin{equation}
    \begin{split}
        & \partial_{t} \Pi^{tx} + \partial_{z}\Pi^{zx} = 0 \\
        & \partial_{x}\Pi^{xz} = 0 \, ,
    \end{split}
\end{equation}
the remaining components automatically satisfy the equations of motion after we impose the homogeneous ansatz $E = E(t)$ and the magnetic field $B$ is constant.

\section{Anabalon-Ross 4d gauged supergravity}

In this section, we consider a charged particle moving in the background described in \cite{Anabalon:2021tua}, which is a solution to $d = 4\;,$ $\mathcal{N} = 2$  minimally gauged supergravity. This is given by the metric
\begin{align}\label{AR4}
    & \dd s^{2}_{4} = \frac{r^{2}}{L^{2}}(- \dd t^{2} + \dd x^{2})+\frac{\dd r^{2}}{f(r)}+f(r) \dd \phi^{2} \, , \\
    & f(r) = \frac{r^{2}}{L^{2}}-\frac{\mu}{r}-\frac{Q^{2}}{r^{2}} \, . \nonumber
\end{align}
Where $\mu$ and $Q$ are free parameters. Additionally, there is a background gauge field given by
\begin{equation}
    A = A(r) \dd \phi =2Q\left(\frac{1}{r}-\frac{1}{r_{0}}    \right) \dd \phi \, .
\end{equation}
The solution is supersymmetric if $\mu = 0$, which implies that the end of the space is given by $r_{0} = \sqrt{Q L}$. In order for the solution to be smooth, the coordinate $\phi$ has period
\begin{equation}
    \Delta\phi = \frac{4\pi L^{2}r_{0}^{3}}{3r_{0}^{4}+Q^{2} L^{2}}.
\end{equation}
Consider a charged particle falling radially in the background described by (\ref{AR4}). In order for the particle to be charged under the background gauge field, it must have a profile the $\phi$ direction. Hence, the action of such particle is  
\begin{equation}
    S = -m\int \dd t \; \underbrace{\sqrt{\frac{r^{2}}{L^{2}}-\frac{\dot{r}^{2}}{f(r)}-f(r)\dot{\phi}^{2}}}_{\Lambda}+q \int \dd t \, A(r) \dot{\phi} \, ,
    \label{action}
\end{equation}
where $q$ is the charge of the particle. The corresponding equations of motion are
\begin{equation}
    \begin{split}
        & r: \quad \frac{\dd}{\dd t}\left(\frac{1}{f(r)}\frac{\dot{r}}{\Lambda}   \right) = -\frac{1}{2\Lambda}\left[\frac{2r}{L^{2}}+f'(r)\left(\frac{\dot{r}^{2}}{f(r)^{2}} -\dot{\phi}^{2}  \right) \right]+\frac{q}{m} A'(r)\dot{\phi} \, , \\
        & \phi: \quad \frac{\dd}{\dd t}\left(\frac{f(r)}{\Lambda}\dot{\phi}  \right) =-\frac{q}{m} \frac{\dd A(r)}{\dd t} \, .
    \end{split}
    \label{eom4}
\end{equation}
Notice that the second equation, defines a conserved quantity
\begin{equation}
     P_{\phi} \equiv \mathcal{J} = \frac{m \, f(r)}{\Lambda}\dot{\phi}+ q A(r) \,.
    \label{J}
\end{equation}
An important point to stress is that $\mathcal{J}$ and $q$ are not independent. Before we notice that in order for $q \neq 0$ there must exist motion on $\phi$, implying $\mathcal{J} \neq 0$. However $\mathcal{J} \neq 0$ does not imply that we should take $q \neq 0$, because in principle we can have a non-charged particle falling radially and rotating simultaneously. We now explain why this cannot be possible by considering the uplift of this soliton to $D = 11$ and how it is needed to embed the particle in a consistent way.

\subsection{Interpreting the charge $q$ of the particle}

In order to get an interpretation of the charge $q$, consider the uplift of the soliton in \eqref{AR4} to $D =11$ over a deformed $\mathrm{S}^7$ given by \cite{Anabalon:2021tua}. The complexity in this background was studied in \cite{Fatemiabhari:2026goj}, we do not calculate complexity here in $11d$ and instead we see how we can associate the charge $q$ of the particle in the $4d$ theory to the R-symmetry charge in the $11d$ theory. We then calculate complexity in the $4d$ gauged supergravity using the Routhian. The metric of the deformed $\rm{AdS}_4 \times \mathrm{S}^7$ is given by
\begin{equation}
    \dd s^{2}_{11} = \dd s_{4}^{2}+4L^{2}\sum_{i=1}^{4} \Bigg[\dd\mu_{i}^{2} + \mu_{i}^{2}\left(\dd\varphi_{i}-\frac{A}{4L}  \right)^{2}\Bigg] \, ,
\end{equation}
where \footnote{Note that these $\mu_i$ where $i=1,2,3,4$ are different to those defined in \eqref{mus}}
\begin{equation}
    \mu_1 = \sin \theta , \qquad \mu_2 = \cos\theta \sin \gamma , \qquad \mu_3 = \cos\theta \cos \gamma \sin \psi , \qquad \mu_4 = \cos\theta \cos \gamma \cos\psi.
\end{equation}
Consider a particle falling with a profile $r = r(t)$, $\phi = \phi(t)$ and $\theta = \gamma = \psi = 0$. As a result of the motion in $\phi$, consistency of its equations of motion implies motion in $\varphi_{4} =\varphi_{4}(t) $, due to the fibration in the geometry. Therefore, the action is given by
\begin{align}\label{action11d}
S &=-m\int \dd t \, \mathcal{L} = -m \int \dd t \,
\sqrt{
\frac{r^2}{L^2}
- \frac{\dot{r}^2}{ f(r)} 
- f(r) \dot{\phi}^2
- 4L^2 \left(\dot{\varphi}_4 - \frac{A(r)}{4L}\dot{\phi} \right)^2
} \, .
\end{align}
The corresponding equations of motion are 
\begin{equation}
\begin{split}
 \frac{\dd}{\dd t}\left(\frac{1}{f(r)}\frac{\dot{r}}{\mathcal{L}}   \right)= -\frac{1}{2\mathcal{L}}\left[ \frac{2r}{L^{2}}+f'(r)\left(\frac{\dot{r}^{2}}{f(r)^{2}}-\dot{\phi}^{2} \right) + 2L A'(r)\dot{\phi}\left( \dot{\varphi}_4 - \frac{A(r)}{4L}\dot{\phi}   \right)  \right] \, ,
 \label{eomr}
\end{split}
\end{equation}
\begin{equation}
    \frac{\dd}{\dd t} \left[
\frac{
\dot{\varphi}_4 - \frac{A(r)}{4L}\dot{\phi}
}{\mathcal{L}}
\right] =0  \, , 
\label{eomvarphi}
\end{equation}
\begin{equation}
    \frac{\dd}{\dd t} \left[
\frac{
f(r) \dot{\phi}
+ \frac{A^2}{4} \dot{\phi}
- L A \dot{\varphi}_4
}{\mathcal{L}}
\right] = 0 \, .
\label{eomphi}
\end{equation}
Notice that the last two equations define the conserved charges
\begin{align}
    & P_{\varphi_4} \equiv  \mathcal{J}_4 = \frac{ 4 L^{2} m \,\left( \dot{\varphi}_4 - \frac{A(r)}{4L}\dot{\phi} \right) }{\mathcal{L}} \, , \\
    & P_{\phi} \equiv  \mathcal{J} = \frac{m(f(r) \dot{\phi}+ \frac{A^2}{4} \dot{\phi}- L A \dot{\varphi}_4)}{\mathcal{L}} \, . \nonumber
\end{align}
In this way, equations (\ref{eomr}) and (\ref{eomphi}) can be rewritten as
\begin{equation}
    \frac{\dd}{\dd t}\left(\frac{1}{f(r)}\frac{\dot{r}}{\mathcal{L}}   \right) = -\frac{1}{2\mathcal{L}}\left[\frac{2r}{L^{2}}+f'(r)\left(\frac{\dot{r}^{2}}{f(r)^{2}}-\dot{\phi}^{2}   \right)  \right]- \frac{\mathcal{J}_{4}}{4 L m} A'(r)\dot{\phi}
\end{equation}
\begin{equation}
    \frac{\dd}{\dd t}\left(\frac{f(r)}{\mathcal{L}} \dot{\phi}   \right) = \frac{\dd}{\dd t}\left[L A\left(\frac{
\dot{\varphi}_4 - \frac{A(r)}{4L}\dot{\phi}
}{\mathcal{L}}   \right)  \right] = \frac{\mathcal{J}_{4}}{4L m}\frac{\dd A(r)}{\dd t} \, .
\end{equation}
By comparing with equations in (\ref{eom4}), one can associate the charge $q$ of the particle in the lower dimensional theory to the R-symmetry charge $\mathcal{J}_{4} = -4L q$, which must be conserved throughout the motion. Hence (\ref{action}) is an effective action for the higher dimensional probe.

In this higher dimensional setup it is explicit that it is not consistent to take $\mathcal{J}_4$ to zero whilst keeping $q$ non-zero. This is because the charge of the particle $q$ is associated to the shrinking circle $\phi$, and after uplifting, $q$ mixes with the R-symmetry charge $\mathcal{J}_4$ associated to the internal angles due to the fibrations. The equations of motion for the embeddings are satisfied if only one of these fibered angles are given a profile, it need not require motion in all the directions of the fibered $\varphi_{i}$, here we consider without loss of generality $\varphi_4$ in 11d.
When the particle has an associated non-zero angular momentum, the particle does not reach the end of the space whilst retaining finite velocity, as its angular momentum has to be conserved. The particle can only reach the end of the space $r_0= \sqrt{QL}$ if $\mathcal{J}_4=0$ which implies $q = 0$. 

\subsection{Holographic Krylov Complexity in 4d}

We now aim to calculate the complexity for the particle in $4d$, described by \eqref{action}, and we do this using the Routhian formulation in order to remove the dependence on the coordinate $\phi$ and instead replace it by its associated Noether charge $\mathcal{J}$,
\begin{equation}
    \dot{\phi}=\frac{(\mathcal{J}-q A)\sqrt{f(r)r^{2}-L^{2}\dot{r}^{2}}}{L f(r)\sqrt{(\mathcal{J}-q A)^{2}+m^{2}f(r)}} \, ,\quad 
   \mathcal{R}(r, \dot{r}) =  \sqrt{\frac{(\mathcal{J}-q A)^{2}+m^{2}f(r)}{f(r)}\left(\frac{r^{2}}{L^{2}}-\frac{\dot{r}^{2}}{f(r)}   \right)} \, .
\end{equation}
From the Hamiltonian $\mathcal{H} = \frac{\partial \mathcal{R}}{\partial \dot{r}}\dot{r}- \mathcal{R}$, we obtain 
\begin{equation}\label{rdot AR 4d}
    \dot{r} = - \frac{r(t) \sqrt{f(r) \left(L^2 \mathcal{H}^2-m^2 r(t)^2\right)-r(t)^2 (\mathcal{J}-q A(r))^2}}{L^2 \mathcal{H}} \, .
\end{equation}
From the initial conditions $r(t = 0)= r_{\text{UV}}$ and $\dot{r}(t = 0)$, one can define the following dimensionless parameters
\begin{equation}
    \gamma = \left(\frac{L\mathcal{H}}{r_{\text{UV}}m}\right)^{2}, \quad \beta = \frac{1}{f(r_{\text{UV}})}\left(\frac{\mathcal{J}-q A(r_{\text{UV}})}{m}\right)^{2} \, .
\end{equation}
Subject to the constraint $\gamma = \beta + 1$. Moreover, by defining 
\begin{equation}
    \tilde{f}(x) = x^{2}-\frac{\tilde{\mu}}{x}-\frac{\tilde{Q}^{2}}{x^{2}} \, ,\quad \quad \tilde{A}(x) =  \frac{2q}{m} \tilde{Q}\left(\frac{1}{x}-\frac{1}{x_{0}}\right)\dd\phi 
\end{equation}
where $x = \frac{r}{r_{\text{UV}}}$ and the units of $[\tilde{Q}] = [\tilde{\mu}] = 1$. In this way, we define $f(r) = \frac{r_{\text{UV}}^{2}}{L^{2}}\tilde{f}(x)$ and $A(r) \equiv \frac{m r_{\text{UV}}}{q L} \tilde{A}(x)$.
Using this language equation \eqref{rdot AR 4d} can be written as
\begin{equation}
    \frac{\dd x}{\dd \tau} = -\frac{x(\tau)}{\gamma}\sqrt{\tilde{f}(x)[\gamma^{2} - \gamma x(\tau)^{2}]-\gamma \left(\tilde{A}(1)-\tilde{A}(x) + \sqrt{\beta \tilde{f}(1)}\right)^{2}x(\tau)^{2}} \, .
    \label{dotxAR}
\end{equation}
Where $x(\tau = 0) = 1 $ and the condition $\dot{x}(\tau = 0) = 0$ is implemented by the constraint $\gamma = \beta + 1$. A series expansion for \eqref{dotxAR} for early times is given by
\begin{equation}
    x(\tau) \approx 1 - \frac{2 (\beta +1) f(1)-2 A'(1) \sqrt{\beta  f(1)}-\beta  f'(1)}{4 (\beta +1)}\tau^{2} + \mathcal{O}(\tau^{3}) \quad \text{for} \quad \tau \to 0 \, ,
    \label{expAR0}
\end{equation}
where the prime indicates derivative with respect to $x$. For the supersymmetric case, meaning $\mu = 0$ the expansion is explicitly given by
\begin{equation}
     x(\tau) \approx 1- \frac{1-\tilde{Q} \left(-2 \sqrt{\beta -\beta  \tilde{Q}^2}+2 \beta  \tilde{Q}+\tilde{Q}\right)}{2 (\beta +1)}\tau^{2} + \mathcal{O}(\tau^{3}) \quad \text{for} \quad \tau \to 0 \, .
\end{equation}
In order to compute the complexity, we write the proper coordinate from the Routhian
\begin{equation}
   \mathcal{R} = \sqrt{g(r)^2 \frac{r^2}{L^2} - \dot{y}^2 } , \qquad \dot{y} = \sqrt{\frac{(\mathcal{J} - q \, A(r))^2 +m^2 \, f(r)}{f(r)}}\frac{\dot{r}}{\sqrt{f(r)}} = g(r) \frac{\dot{r}}{\sqrt{f(r)}}.
\end{equation}
The proper momentum is then given by
\begin{equation}
    P_{y} \equiv - \frac{\partial  \mathcal{R}}{\partial \dot{y}} = \frac{\dot{r}}{\sqrt{ \frac{f(r) r^{2}}{L^{2}}-\dot{r}^{2}}} = \frac{\dot{x}(\tau)}{\sqrt{\tilde{f}(x) x(\tau)^{2}-\dot{x}(\tau)^{2}}} .
\end{equation}
By using the expansion \eqref{expAR0} we can expand the proper momentum for early times, leading to 
\begin{equation}
    P_{y}(\tau) \approx \frac{2 A'(1) \sqrt{\beta  f(1)}+\beta  f'(1)-2 (\beta +1) f(1)}{2\sqrt{f(1)} (\beta +1)}\tau +\mathcal{O}\left(\tau^3\right) \quad \text{for} \quad \tau \to 0 \, ,
\end{equation}
which in the supersymmetric case where $\mu=0$, is 
\begin{equation}
    P_{y}(\tau) \approx \frac{1 + 2 \tilde{Q} \sqrt{\beta(1 - \tilde{Q}^2)}-(2 \beta +1) \tilde{Q}^2}{(\beta +1) \sqrt{1-\tilde{Q}^2}} \tau +\mathcal{O}\left(\tau^3\right) \quad \text{for} \quad \tau \to 0 \, .
\end{equation}

To summarise, we showed that the action in \eqref{action} effectively describes the probe in 11d after fixing $\mathcal{J}_{4} = -q$, which would be equivalent to applying the Routhian prescription for the probe in 11d after redefining $m$ and $\mathcal{J}$. This implies that, the complexity for both of the probes have the same behaviour, agreeing with the fact that the probes have the same quantum numbers from the field theory perspective.

\begin{figure}[h!]
    \centering
    \includegraphics[width=0.45\linewidth]{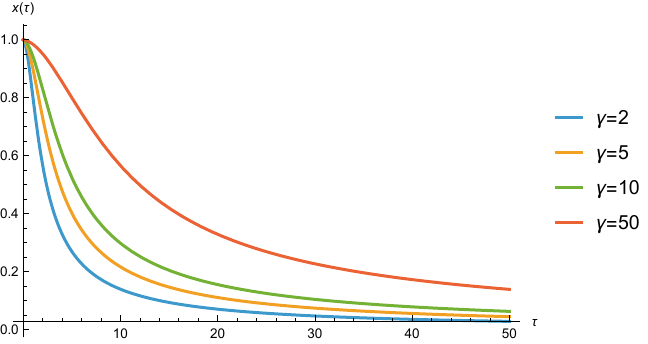}
    \quad \quad
        \includegraphics[width=0.45\linewidth]{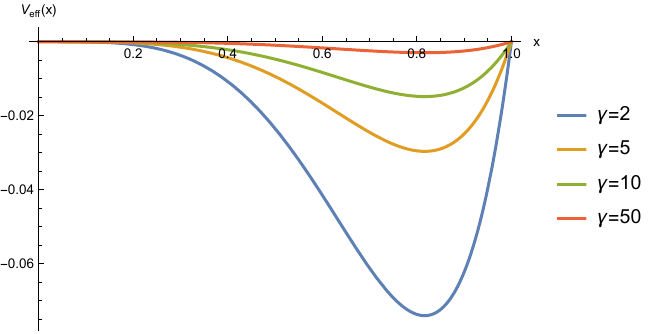}
    \caption{ [Left] plot of $x(\tau)$ for different values of $\gamma$ chosen as $\gamma=2,5,10,50$, with $Q=0.0001$. $\qquad \qquad$ [Right] Plot of the effective potential $V_{\text{eff}}(x)$, with $Q=0.0001$ for the supersymmetric case $\mu=0$.}
    \label{Veff AR}
\end{figure}

\bibliographystyle{JHEP}
\bibliography{main.bib}

@article{Imani:2025etp,
    author = "Imani, Hamid R. and Babaei Velni, Komeil and Mohammadi Mozaffar, M. Reza",
    title = "{Krylov complexity in Lifshitz-type Dirac field theories}",
    eprint = "2506.08765",
    archivePrefix = "arXiv",
    primaryClass = "hep-th",
    doi = "10.1140/epjc/s10052-025-14669-x",
    journal = "Eur. Phys. J. C",
    volume = "85",
    number = "9",
    pages = "958",
    year = "2025"
}

@article{Das:2026gko,
    author = "Das, Jayashish and Das, Suman and Pedraza, Juan F. and Qu, Le-Chen",
    title = "{Quantum chaos and late-time equipartition of symmetry-resolved Krylov complexity}",
    eprint = "2608.19346",
    archivePrefix = "arXiv",
    primaryClass = "hep-th",
    month = "8",
    year = "2026"
}

@article{Balasubramanian:2024ghv,
    author = "Balasubramanian, Vijay and Das, Rathindra Nath and Erdmenger, Johanna and Xian, Zhuo-Yu",
    title = "{Chaos and integrability in triangular billiards}",
    eprint = "2407.11114",
    archivePrefix = "arXiv",
    primaryClass = "hep-th",
    doi = "10.1088/1742-5468/adba41",
    journal = "J. Stat. Mech.",
    volume = "2025",
    number = "3",
    pages = "033202",
    year = "2025"
}

@article{Qu:2026dmv,
    author = "Qu, Le-Chen",
    title = "{A geodesic distance interpretation of Lanczos coefficients}",
    eprint = "2608.16982",
    archivePrefix = "arXiv",
    primaryClass = "hep-th",
    reportNumber = "IFT-UAM/CSIC-26-111",
    month = "8",
    year = "2026"
}

@article{Fan:2022xaa,
    author = "Fan, Zhong-Ying",
    title = "{Universal relation for operator complexity}",
    eprint = "2202.07220",
    archivePrefix = "arXiv",
    primaryClass = "quant-ph",
    doi = "10.1103/PhysRevA.105.062210",
    journal = "Phys. Rev. A",
    volume = "105",
    number = "6",
    pages = "062210",
    year = "2022"
}

@article{Huh:2023jxt,
    author = "Huh, Kyoung-Bum and Jeong, Hyun-Sik and Pedraza, Juan F.",
    title = "{Spread complexity in saddle-dominated scrambling}",
    eprint = "2312.12593",
    archivePrefix = "arXiv",
    primaryClass = "hep-th",
    reportNumber = "IFT-UAM/CSIC-23-178",
    doi = "10.1007/JHEP05(2024)137",
    journal = "JHEP",
    volume = "05",
    pages = "137",
    year = "2024"
}

@article{Nunez:2026vhw,
    author = "Nunez, Carlos and Pedraza, Juan F. and Subils, Javier G.",
    title = "{Complexity measures in holographic cascading theories with multiscale dynamics}",
    eprint = "2608.10060",
    archivePrefix = "arXiv",
    primaryClass = "hep-th",
    reportNumber = "IFT-UAM/CSIC-26-105",
    month = "8",
    year = "2026"
}

@article{Fatemiabhari:2026goj,
    author = "Fatemiabhari, Ali and Nunez, Carlos",
    title = "{Krylov Complexity, Confinement and Universality}",
    eprint = "2602.17757",
    archivePrefix = "arXiv",
    primaryClass = "hep-th",
    month = "2",
    year = "2026"
}

@article{Aharony:2008ug,
    author = "Aharony, Ofer and Bergman, Oren and Jafferis, Daniel Louis and Maldacena, Juan",
    title = "{N=6 superconformal Chern-Simons-matter theories, M2-branes and their gravity duals}",
    eprint = "0806.1218",
    archivePrefix = "arXiv",
    primaryClass = "hep-th",
    reportNumber = "WIS-12-08-JUN-DPP",
    doi = "10.1088/1126-6708/2008/10/091",
    journal = "JHEP",
    volume = "10",
    pages = "091",
    year = "2008"
}

@article{Nastase:2026lhz,
    author = "Nastase, Horatiu and Nunez, Carlos and Roychowdhury, Dibakar",
    title = "{Holographic Krylov Complexity for Charged, Composite and Extended Probes}",
    eprint = "2604.07432",
    archivePrefix = "arXiv",
    primaryClass = "hep-th",
    month = "4",
    year = "2026"
}

@article{Fatemiabhari:2026six,
    author = "Fatemiabhari, Ali and Nunez, Carlos and Santamaria, Ricardo T.",
    title = "{Complexity and Operator Growth in Holographic 6d SCFTs}",
    eprint = "2603.10106",
    archivePrefix = "arXiv",
    primaryClass = "hep-th",
    doi = "10.1016/j.nuclphysb.2026.117402",
    journal = "Nucl. Phys. B",
    volume = "1025",
    pages = "117402",
    year = "2026"
}

@article{Nunez:2026rgflow,
    author = "Nunez, Carlos and Roychowdhury, Dibakar",
    title = "{Krylov Complexity and $c$-function along RG Flows}",
    eprint = "2608.02715",
    archivePrefix = "arXiv",
    primaryClass = "hep-th",
    month = "8",
    year = "2026"
}

@article{Roychowdhury:2026eta,
    author = "Roychowdhury, Dibakar",
    title = "{Krylov complexity for $\eta$ deformed superstring backgrounds}",
    eprint = "2601.06555",
    archivePrefix = "arXiv",
    primaryClass = "hep-th",
    year = "2026"
}

@article{Zoakos:2026coulomb,
    author = "Zoakos, Dimitrios",
    title = "{Holographic Krylov complexity in the Coulomb branch of ${\cal N}=4$ SYM}",
    eprint = "2603.15435",
    archivePrefix = "arXiv",
    primaryClass = "hep-th",
    doi = "10.1007/JHEP06(2026)066",
    journal = "JHEP",
    volume = "06",
    pages = "066",
    year = "2026"
}

@article{Li:2025observer,
    author = "Li, Zhehan and Tian, Jia",
    title = "{The Holography of Spread Complexity: A Story of Observers}",
    eprint = "2506.13481",
    archivePrefix = "arXiv",
    primaryClass = "hep-th",
    year = "2025"
}

@article{Li:2026comments,
    author = "Li, Zhehan and Tian, Jia",
    title = "{Comments on holographic spread complexity}",
    eprint = "2607.18024",
    archivePrefix = "arXiv",
    primaryClass = "hep-th",
    month = "7",
    year = "2026"
}

@article{Qu:2025lanczos,
    author = "Qu, Le-Chen",
    title = "{Lanczos Meets Orthogonal Polynomials}",
    eprint = "2512.15857",
    archivePrefix = "arXiv",
    primaryClass = "hep-th",
    year = "2025"
}

@article{Roychowdhury:2026lin,
    author = "Roychowdhury, Dibakar",
    title = "{Krylov complexity for Lin--Maldacena geometries and their holographic duals}",
    eprint = "2604.16977",
    archivePrefix = "arXiv",
    primaryClass = "hep-th",
    doi = "10.1007/JHEP05(2026)197",
    journal = "JHEP",
    volume = "05",
    pages = "197",
    year = "2026"
}

@article{Roychowdhury:2026bmnstate,
    author = "Roychowdhury, Dibakar",
    title = "{Krylov state complexity for BMN matrix model}",
    eprint = "2605.10786",
    archivePrefix = "arXiv",
    primaryClass = "hep-th",
    month = "5",
    year = "2026"
}

@article{Roychowdhury:2026planewave,
    author = "Roychowdhury, Dibakar",
    title = "{Krylov Complexity for Plane Wave Matrix Model}",
    eprint = "2605.26055",
    archivePrefix = "arXiv",
    primaryClass = "hep-th",
    month = "5",
    year = "2026"
}

@article{Roychowdhury:2026bmnspectral,
    author = "Roychowdhury, Dibakar",
    title = "{Krylov complexity and spectral density of BMN matrix model}",
    eprint = "2607.24632",
    archivePrefix = "arXiv",
    primaryClass = "hep-th",
    month = "7",
    year = "2026"
}

@article{Graef:2026probe,
    author = "Graef, Eric L. and Murugan, Jeff and Nastase, Horatiu and Van Zyl, Hendrik J. R.",
    title = "{On the Universality of Probe Complexity in ${\cal N}=4$ SYM}",
    eprint = "2606.21662",
    archivePrefix = "arXiv",
    primaryClass = "hep-th",
    month = "6",
    year = "2026"
}

@article{Baume:2026chaos,
    author = "Baume, Florent and Cavusoglu, Atakan and Chakrabhavi, Vivek and Heckman, Jonathan J.",
    title = "{Controlled Chaos in 4D SCFTs}",
    eprint = "2606.23785",
    archivePrefix = "arXiv",
    primaryClass = "hep-th",
    month = "6",
    year = "2026"
}

@article{Fadafan:2026lifshitz,
    author = "Bitaghsir Fadafan, Kazem and Mohammadi Mozaffar, M. Reza",
    title = "{Holographic Krylov Complexity with Lifshitz Scaling and Hyperscaling Violation}",
    eprint = "2606.31724",
    archivePrefix = "arXiv",
    primaryClass = "hep-th",
    month = "6",
    year = "2026"
}

@article{Anabalon:2026moduli,
    author = "Anabalon, Andres and Nastase, Horatiu and Nunez, Carlos and Oyarzo, Marcelo and Stuardo, Ricardo",
    title = "{Moduli space of ${\cal N}=4$ Super Yang--Mills from AdS/CFT}",
    eprint = "2603.18141",
    archivePrefix = "arXiv",
    primaryClass = "hep-th",
    month = "3",
    year = "2026"
}

@article{Alfinito:2026vah,
    author = "Alfinito, Eleonora and Beccaria, Matteo",
    title = "{Krylov Correlators in $\mathfrak{sl}(2,\mathbb R)$ Models: Exact Results and Holographic Complexity}",
    eprint = "2605.17550",
    archivePrefix = "arXiv",
    primaryClass = "hep-th",
    month = "5",
    year = "2026"
}

@article{Muck:2026top,
    author = {M{\"u}ck, Wolfgang},
    title = "{Krylov complexity has it all}",
    eprint = "2605.28681",
    archivePrefix = "arXiv",
    primaryClass = "hep-th",
    month = "5",
    year = "2026"
}

@article{Anabalon:2021tua,
    author = "Anabalon, Andres and Ross, Simon F.",
    title = "{Supersymmetric solitons and a degeneracy of solutions in AdS/CFT}",
    eprint = "2104.14572",
    archivePrefix = "arXiv",
    primaryClass = "hep-th",
    doi = "10.1007/JHEP07(2021)015",
    journal = "JHEP",
    volume = "07",
    pages = "015",
    year = "2021"
}

@article{Nunez:2023nnl,
    author = "Nunez, Carlos and Oyarzo, Marcelo and Stuardo, Ricardo",
    title = "{Confinement in (1 + 1) dimensions: a holographic perspective from I-branes}",
    eprint = "2307.04783",
    archivePrefix = "arXiv",
    primaryClass = "hep-th",
    doi = "10.1007/JHEP09(2023)201",
    journal = "JHEP",
    volume = "09",
    pages = "201",
    year = "2023"
}

@article{Anabalon:2022aig,
    author = "Anabal\'on, Andr\'es and Gallerati, Antonio and Ross, Simon and Trigiante, Mario",
    title = "{Supersymmetric solitons in gauged $ \mathcal{N} $ = 8 supergravity}",
    eprint = "2210.06319",
    archivePrefix = "arXiv",
    primaryClass = "hep-th",
    doi = "10.1007/JHEP02(2023)055",
    journal = "JHEP",
    volume = "02",
    pages = "055",
    year = "2023"
}

@article{Nunez:2023xgl,
    author = "Nunez, Carlos and Oyarzo, Marcelo and Stuardo, Ricardo",
    title = "{Confinement and D5-branes}",
    eprint = "2311.17998",
    archivePrefix = "arXiv",
    primaryClass = "hep-th",
    doi = "10.1007/JHEP03(2024)080",
    journal = "JHEP",
    volume = "03",
    pages = "080",
    year = "2024"
}

@article{Anabalon:2024che,
    author = "Anabal\'on, Andr\'es and Nastase, Horatiu and Oyarzo, Marcelo",
    title = "{Supersymmetric AdS Solitons and the interconnection of different vacua of ${\cal N}=4$ Super Yang-Mills}",
    eprint = "2402.18482",
    archivePrefix = "arXiv",
    primaryClass = "hep-th",
    month = "2",
    year = "2024"
}

@article{Chatzis:2024kdu,
    author = "Chatzis, Dimitrios and Fatemiabhari, Ali and Nunez, Carlos and Weck, Peter",
    title = "{SCFT deformations via uplifted solitons}",
    eprint = "2406.01685",
    archivePrefix = "arXiv",
    primaryClass = "hep-th",
    month = "6",
    year = "2024"
}

@article{Benincasa:2011zu,
    author = "Benincasa, Paolo and Ramallo, Alfonso V.",
    title = "{Fermionic impurities in Chern-Simons-matter theories}",
    eprint = "1112.4669",
    archivePrefix = "arXiv",
    primaryClass = "hep-th",
    doi = "10.1007/JHEP02(2012)076",
    journal = "JHEP",
    volume = "02",
    pages = "076",
    year = "2012"
}

@article{Karch:2007pd,
    author = "Karch, Andreas and O'Bannon, Andy",
    title = "{Metallic AdS/CFT}",
    eprint = "0705.3870",
    archivePrefix = "arXiv",
    primaryClass = "hep-th",
    doi = "10.1088/1126-6708/2007/09/024",
    journal = "JHEP",
    volume = "09",
    pages = "024",
    year = "2007"
}

@article{OBannon:2007cex,
    author = "O'Bannon, Andy",
    title = "{Hall Conductivity of Flavor Fields from AdS/CFT}",
    eprint = "0708.1994",
    archivePrefix = "arXiv",
    primaryClass = "hep-th",
    doi = "10.1103/PhysRevD.76.086007",
    journal = "Phys. Rev. D",
    volume = "76",
    pages = "086007",
    year = "2007"
}

@article{Chatzis:2024top,
    author = "Chatzis, Dimitrios and Fatemiabhari, Ali and Nunez, Carlos and Weck, Peter",
    title = "{Conformal to confining SQFTs from holography}",
    eprint = "2405.05563",
    archivePrefix = "arXiv",
    primaryClass = "hep-th",
    month = "5",
    year = "2024"
}

@article{Craps:2024suj,
    author = "Craps, Ben and Evnin, Oleg and Pascuzzi, Gabriele",
    title = "{Multiseed Krylov Complexity}",
    eprint = "2409.15666",
    archivePrefix = "arXiv",
    primaryClass = "quant-ph",
    doi = "10.1103/PhysRevLett.134.050402",
    journal = "Phys. Rev. Lett.",
    volume = "134",
    number = "5",
    pages = "050402",
    year = "2025"
}

@article{Fatemiabhari:2024aua,
    author = "Fatemiabhari, Ali and Nunez, Carlos",
    title = "{From conformal to confining field theories using holography}",
    eprint = "2401.04158",
    archivePrefix = "arXiv",
    primaryClass = "hep-th",
    doi = "10.1007/JHEP03(2024)160",
    journal = "JHEP",
    volume = "03",
    pages = "160",
    year = "2024"
}

@article{Sfetsos:2010uq,
    author = "Sfetsos, Konstadinos and Thompson, Daniel C.",
    title = "{On non-abelian T-dual geometries with Ramond fluxes}",
    eprint = "1012.1320",
    archivePrefix = "arXiv",
    primaryClass = "hep-th",
    doi = "10.1016/j.nuclphysb.2010.12.013",
    journal = "Nucl. Phys. B",
    volume = "846",
    pages = "21--42",
    year = "2011"
}

@article{Lozano:2019zvg,
    author = "Lozano, Yolanda and Macpherson, Niall T. and Nunez, Carlos and Ramirez, Anayeli",
    title = "{Two dimensional ${\cal N}=(0,4)$ quivers dual to AdS$_3$ solutions in massive IIA}",
    eprint = "1909.10510",
    archivePrefix = "arXiv",
    primaryClass = "hep-th",
    doi = "10.1007/JHEP01(2020)140",
    journal = "JHEP",
    volume = "01",
    pages = "140",
    year = "2020"
}

@article{Anabalon:2024qhf,
    author = "Anabal\'on, A. and Astefanesei, D. and Gallerati, A. and Oliva, J.",
    title = "{Supersymmetric smooth distributions of M2-branes as AdS solitons}",
    eprint = "2402.00880",
    archivePrefix = "arXiv",
    primaryClass = "hep-th",
    doi = "10.1007/JHEP05(2024)077",
    journal = "JHEP",
    volume = "05",
    pages = "077",
    year = "2024"
}

@article{Macpherson:2024qfi,
    author = "Macpherson, Niall T. and Merrikin, Paul and Stuardo, Ricardo",
    title = "{Circle compactifications of Minkowski$_D$ solutions, flux vacua and solitonic branes}",
    eprint = "2412.15102",
    archivePrefix = "arXiv",
    primaryClass = "hep-th",
    month = "12",
    year = "2024"
}

@article{Balasubramanian:2025xkj,
    author = "Balasubramanian, Vijay and Caputa, Pawel and Sim{\'o}n, Joan",
    title = "{Variations on a theme of Krylov}",
    eprint = "2511.03775",
    archivePrefix = "arXiv",
    primaryClass = "hep-th",
    reportNumber = "YITP-25-171",
    doi = "10.1007/JHEP04(2026)172",
    journal = "JHEP",
    volume = "04",
    pages = "172",
    year = "2026"
}

@article{Chatzis:2025dnu,
    author = "Chatzis, Dimitrios and Hammond, Madison and Itsios, Georgios and Nunez, Carlos and Zoakos, Dimitrios",
    title = "{Universal Observables, SUSY RG-Flows and Holography}",
    eprint = "2506.10062",
    archivePrefix = "arXiv",
    primaryClass = "hep-th",
    month = "6",
    year = "2025"
}

@article{Macpherson:2025pqi,
    author = "Macpherson, Niall and Merrikin, Paul and Nunez, Carlos and Stuardo, Ricardo",
    title = "{Twisted-Circle Compactifications of SQCD-like Theories and Holography}",
    eprint = "2506.15778",
    archivePrefix = "arXiv",
    primaryClass = "hep-th",
    month = "6",
    year = "2025"
}

@article{Caputa:2024sux,
    author = "Caputa, Pawel and Chen, Bowen and McDonald, Ross W. and Sim{\'o}n, Joan and Strittmatter, Benjamin",
    title = "{Spread Complexity Rate as Proper Momentum}",
    eprint = "2410.23334",
    archivePrefix = "arXiv",
    primaryClass = "hep-th",
    reportNumber = "YITP-24-137",
    month = "10",
    year = "2024"
}

@article{Susskind:2019ddc,
    author = "Susskind, Leonard",
    title = "{Complexity and Newton's Laws}",
    eprint = "1904.12819",
    archivePrefix = "arXiv",
    primaryClass = "hep-th",
    doi = "10.3389/fphy.2020.00262",
    journal = "Front. in Phys.",
    volume = "8",
    pages = "262",
    year = "2020"
}

@article{Baiguera:2025dkc,
    author = "Baiguera, Stefano and Balasubramanian, Vijay and Caputa, Pawel and Chapman, Shira and Haferkamp, Jonas and Heller, Michal P. and Halpern, Nicole Yunger",
    title = "{Quantum complexity in gravity, quantum field theory, and quantum information science}",
    eprint = "2503.10753",
    archivePrefix = "arXiv",
    primaryClass = "hep-th",
    reportNumber = "YITP-25-39",
    month = "3",
    year = "2025"
}

@article{Balasubramanian:2022tpr,
    author = "Balasubramanian, Vijay and Caputa, Pawel and Magan, Javier M. and Wu, Qingyue",
    title = "{Quantum chaos and the complexity of spread of states}",
    eprint = "2202.06957",
    archivePrefix = "arXiv",
    primaryClass = "hep-th",
    doi = "10.1103/PhysRevD.106.046007",
    journal = "Phys. Rev. D",
    volume = "106",
    number = "4",
    pages = "046007",
    year = "2022"
}

@article{Rabinovici:2025otw,
    author = "Rabinovici, Eliezer and S{\'a}nchez-Garrido, Adri{\'a}n and Shir, Ruth and Sonner, Julian",
    title = "{Krylov Complexity}",
    eprint = "2507.06286",
    archivePrefix = "arXiv",
    primaryClass = "hep-th",
    reportNumber = "CERN-TH-2025-128",
    month = "7",
    year = "2025"
}

@article{Caputa:2025mii,
    author = "Caputa, Pawel and Di Giulio, Giuseppe and Loc, Tran Quang",
    title = "{Growth of block-diagonal operators and symmetry-resolved Krylov complexity}",
    eprint = "2507.02033",
    archivePrefix = "arXiv",
    primaryClass = "hep-th",
    reportNumber = "YITP-25-101",
    doi = "10.1103/9v9v-54zv",
    journal = "Phys. Rev. Res.",
    volume = "7",
    number = "4",
    pages = "043055",
    year = "2025"
}

@article{Caputa:2025ozd,
    author = "Caputa, Pawel and Di Giulio, Giuseppe and Loc, Tran Quang",
    title = "{Symmetry-Resolved Spread Complexity}",
    eprint = "2509.12992",
    archivePrefix = "arXiv",
    primaryClass = "hep-th",
    reportNumber = "YITP-25-146",
    month = "9",
    year = "2025"
}

@article{Caputa:2024xkp,
    author = "Caputa, Pawel and Kutak, Krzysztof",
    title = "{Krylov complexity and gluon cascades in the high energy limit}",
    eprint = "2404.07657",
    archivePrefix = "arXiv",
    primaryClass = "hep-ph",
    reportNumber = "YITP-24-49, IFJPAN-IV-2024-6",
    doi = "10.1103/PhysRevD.110.085011",
    journal = "Phys. Rev. D",
    volume = "110",
    number = "8",
    pages = "085011",
    year = "2024"
}

@article{Jiang:2025wpj,
    author = "Jiang, Xuhao and Halimeh, Jad C. and Srivatsa, N. S.",
    title = "{Krylov Complexity Meets Confinement}",
    eprint = "2511.03783", 
    archivePrefix = "arXiv",
    doi="10.1103/1gsg-zb8h",
    journal="Phys. Rev. D",
    primaryClass = "cond-mat.stat-mech",
    month = "11",
    year = "2025"
}

@article{Avdoshkin:2022xuw,
    author = "Avdoshkin, Alexander and Dymarsky, Anatoly and Smolkin, Michael",
    title = "{Krylov complexity in quantum field theory, and beyond}",
    eprint = "2212.14429",
    archivePrefix = "arXiv",
    primaryClass = "hep-th",
    doi = "10.1007/JHEP06(2024)066",
    journal = "JHEP",
    volume = "06",
    pages = "066",
    year = "2024"
}

@article{Nandy:2024evd,
    author = "Nandy, Pratik and Matsoukas-Roubeas, Apollonas S. and Mart{\'\i}nez-Azcona, Pablo and Dymarsky, Anatoly and del Campo, Adolfo",
    title = "{Quantum dynamics in Krylov space: Methods and applications}",
    eprint = "2405.09628",
    archivePrefix = "arXiv",
    primaryClass = "quant-ph",
    reportNumber = "RIKEN-iTHEMS-Report-24",
    doi = "10.1016/j.physrep.2025.05.001",
    journal = "Phys. Rept.",
    volume = "1125-1128",
    pages = "1--82",
    year = "2025"
}

@article{Caputa:2021sib,
    author = "Caputa, Pawel and Magan, Javier M. and Patramanis, Dimitrios",
    title = "{Geometry of Krylov complexity}",
    eprint = "2109.03824",
    archivePrefix = "arXiv",
    primaryClass = "hep-th",
    doi = "10.1103/PhysRevResearch.4.013041",
    journal = "Phys. Rev. Res.",
    volume = "4",
    number = "1",
    pages = "013041",
    year = "2022"
}

@article{Fan:2024iop,
    author = "Fan, Zhong-Ying",
    title = "{Momentum-Krylov complexity correspondence}",
    eprint = "2411.04492",
    archivePrefix = "arXiv",
    primaryClass = "hep-th",
    month = "11",
    year = "2024"
}

@article{He:2024pox,
    author = "He, Peng-Zhang",
    title = "{Revisit the relationship between spread complexity rate and radial momentum}",
    eprint = "2411.19172",
    archivePrefix = "arXiv",
    primaryClass = "hep-th",
    month = "11",
    year = "2024"
}

@article{Heller:2024ldz,
    author = "Heller, Michal P. and Papalini, Jacopo and Schuhmann, Tim",
    title = "{Krylov Spread Complexity as Holographic Complexity beyond Jackiw-Teitelboim Gravity}",
    eprint = "2412.17785",
    archivePrefix = "arXiv",
    primaryClass = "hep-th",
    doi = "10.1103/spcr-jgm6",
    journal = "Phys. Rev. Lett.",
    volume = "135",
    number = "15",
    pages = "151602",
    year = "2025"
}

@article{Fu:2025kkh,
    author = "Fu, Yichao and Jeong, Hyun-Sik and Kim, Keun-Young and Pedraza, Juan F.",
    title = "{Toward Krylov-based holography in double-scaled SYK}",
    eprint = "2510.22658",
    archivePrefix = "arXiv",
    primaryClass = "hep-th",
    reportNumber = "IFT-UAM/CSIC-25-105, APCTP Pre2025 - 020",
    month = "10",
    year = "2025"
}

@article{Rabinovici:2023yex,
    author = "Rabinovici, E. and S{\'a}nchez-Garrido, A. and Shir, R. and Sonner, J.",
    title = "{A bulk manifestation of Krylov complexity}",
    eprint = "2305.04355",
    archivePrefix = "arXiv",
    primaryClass = "hep-th",
    doi = "10.1007/JHEP08(2023)213",
    journal = "JHEP",
    volume = "08",
    pages = "213",
    year = "2023"
}

@article{Fatemiabhari:2025cyy,
    author = "Fatemiabhari, Ali and Nastase, Horatiu and Roychowdhury, Dibakar",
    title = "{Holographic Krylov complexity in ${\cal N}=4$ SYM}",
    eprint = "2511.19286",
    archivePrefix = "arXiv",
    primaryClass = "hep-th",
    month = "11",
    year = "2025"
}

@article{Chatzis:2025hek,
    author = "Chatzis, Dimitrios and Hammond, Madison and Itsios, Georgios and Nunez, Carlos and Zoakos, Dimitrios",
    title = "{Supersymmetric AdS Solitons, Coulomb Branch Flows and Twisted Compactifications}",
    eprint = "2511.18128",
    archivePrefix = "arXiv",
    primaryClass = "hep-th",
    month = "11",
    year = "2025"
}

@article{Lozano:2019emq,
    author = "Lozano, Yolanda and Macpherson, Niall T. and Nunez, Carlos and Ramirez, Anayeli",
    title = "{AdS$_3$ solutions in Massive IIA with small $\mathcal{N}=(4,0)$ supersymmetry}",
    eprint = "1908.09851",
    archivePrefix = "arXiv",
    primaryClass = "hep-th",
    doi = "10.1007/JHEP01(2020)129",
    journal = "JHEP",
    volume = "01",
    pages = "129",
    year = "2020"
}

@article{Lozano:2019ywa,
    author = "Lozano, Yolanda and Macpherson, Niall T. and Nunez, Carlos and Ramirez, Anayeli",
    title = "{AdS$_3$ solutions in massive IIA, defect CFTs and T-duality}",
    eprint = "1909.11669",
    archivePrefix = "arXiv",
    primaryClass = "hep-th",
    doi = "10.1007/JHEP12(2019)013",
    journal = "JHEP",
    volume = "12",
    pages = "013",
    year = "2019"
}

@article{Fatemiabhari:2025usn,
    author = "Fatemiabhari, Ali and Nastase, Horatiu and Nunez, Carlos and Roychowdhury, Dibakar",
    title = "{Holographic Krylov complexity in confining gauge theories}",
    eprint = "2511.22717",
    archivePrefix = "arXiv",
    primaryClass = "hep-th",
    month = "11",
    year = "2025"
}

@article{Ambrosini:2024sre,
    author = "Ambrosini, Marco and Rabinovici, Eliezer and S{\'a}nchez-Garrido, Adri{\'a}n and Shir, Ruth and Sonner, Julian",
    title = "{Operator K-complexity in DSSYK: Krylov complexity equals bulk length}",
    eprint = "2412.15318",
    archivePrefix = "arXiv",
    primaryClass = "hep-th",
    reportNumber = "CERN-TH-2025-040",
    doi = "10.1007/JHEP08(2025)059",
    journal = "JHEP",
    volume = "08",
    pages = "059",
    year = "2025"
}

@article{Susskind:2020gnl,
    author = "Susskind, Leonard and Zhao, Ying",
    title = "{Complexity and Momentum}",
    eprint = "2006.03019",
    archivePrefix = "arXiv",
    primaryClass = "hep-th",
    doi = "10.1007/JHEP03(2021)239",
    journal = "JHEP",
    volume = "03",
    pages = "239",
    year = "2021"
}

@article{Susskind:2018tei,
    author = "Susskind, Leonard",
    title = "{Why do Things Fall?}",
    eprint = "1802.01198",
    archivePrefix = "arXiv",
    primaryClass = "hep-th",
    month = "2",
    year = "2018"
}

@article{Barbon:2020uux,
    author = "Barbon, J. L. F. and Martin-Garcia, J. and Sasieta, M.",
    title = "{A Generalized Momentum/Complexity Correspondence}",
    eprint = "2012.02603",
    archivePrefix = "arXiv",
    primaryClass = "hep-th",
    doi = "10.1007/JHEP04(2021)250",
    journal = "JHEP",
    volume = "04",
    pages = "250",
    year = "2021"
}

@article{Das:2024tnw,
    author = "Das, Rathindra Nath and Demulder, Saskia and Erdmenger, Johanna and Northe, Christian",
    title = "{Spread complexity for the planar limit of holography}",
    eprint = "2412.09673",
    archivePrefix = "arXiv",
    primaryClass = "hep-th",
    doi = "10.1007/JHEP06(2025)166",
    journal = "JHEP",
    volume = "06",
    pages = "166",
    year = "2025"
}

@article{Fatemiabhari:2025poq,
    author = "Fatemiabhari, Ali and Nastase, Horatiu and Nunez, Carlos and Roychowdhury, Dibakar",
    title = "{Holographic Krylov Complexity for Conformal Quiver Gauge Theories}",
    eprint = "2512.14812",
    archivePrefix = "arXiv",
    primaryClass = "hep-th",
    month = "12",
    year = "2025"
}

@article{Parker:2018yvk,
    author = "Parker, Daniel E. and Cao, Xiangyu and Avdoshkin, Alexander and Scaffidi, Thomas and Altman, Ehud",
    title = "{A Universal Operator Growth Hypothesis}",
    eprint = "1812.08657",
    archivePrefix = "arXiv",
    primaryClass = "cond-mat.stat-mech",
    doi = "10.1103/PhysRevX.9.041017",
    journal = "Phys. Rev. X",
    volume = "9",
    number = "4",
    pages = "041017",
    year = "2019"
}

@article{Jeong:2026iac,
    author = "Jeong, Hyun-Sik",
    title = "{Krylov Subspace Dynamics as Near-Horizon AdS$_2$ Holography}",
    eprint = "2602.11627",
    archivePrefix = "arXiv",
    primaryClass = "hep-th",
    reportNumber = "APCTP Pre2026 - 003",
    month = "2",
    year = "2026"
}

@article{Thompson:2019ipl,
    author = "Thompson, Daniel C.",
    editor = "Anagnostopoulos, Konstantinos and others",
    title = "{An Introduction to Generalised Dualities and their Applications to Holography and Integrability}",
    eprint = "1904.11561",
    archivePrefix = "arXiv",
    primaryClass = "hep-th",
    doi = "10.22323/1.347.0099",
    journal = "PoS",
    volume = "CORFU2018",
    pages = "099",
    year = "2019"
}

@article{Chatzis:2026ekd,
    author = "Chatzis, Dimitrios and Hammond, Madison and Nunez, Carlos and Ramallo, Alfonso V. and Santamaria, Ricardo T.",
    title = "{Holographic Spread Complexity from Branes and Strings}",
    eprint = "2607.00074",
    archivePrefix = "arXiv",
    primaryClass = "hep-th",
    month = "6",
    year = "2026"
}

\end{document}